\documentclass[final,authoryear,5p,times,twocolumn]{elsarticle}

\usepackage{epsfig}

\usepackage{amssymb}

\usepackage{lineno}
\usepackage{rotating}
\usepackage{color,soul}
\usepackage{longtable}

\usepackage[bookmarks = true, bookmarksnumbered = true, pdfpagemode =None, pdfstartview = FitH, pdfpagelayout = SinglePage, colorlinks = true, urlcolor = blue, citecolor = monbleu]{hyperref}

\newcommand{\pderiv}[2]{\frac{\partial #1}{\partial #2}}

\newcommand{\beq}{\bigskip\begin{equation}}
\newcommand{\eeq}{\bigskip\end{equation}}

\journal{Icarus}

\begin{document}

\begin{frontmatter}



\title{Mid-Infrared Mapping of Jupiter's Temperatures, Aerosol Opacity and Chemical Distributions with IRTF/TEXES}


\author[le]{Leigh N. Fletcher}
\ead{leigh.fletcher@leicester.ac.uk}
\author[swri]{T.K. Greathouse}
\author[jpl]{G.S. Orton}
\author[jpl]{J.A. Sinclair}
\author[ox]{R.S. Giles}
\author[ox]{P.G.J. Irwin}
\author[meu]{T. Encrenaz}


%

\address[le]{Department of Physics \& Astronomy, University of Leicester, University Road, Leicester, LE1 7RH, UK}
\address[swri]{Southwest Research Institute, Division 15, 6220 Culebra Road, San Antonio, Texas 78228, USA}
\address[jpl]{Jet Propulsion Laboratory, California Institute of Technology, 4800 Oak Grove Drive, Pasadena, CA, 91109, USA}
\address[ox]{Atmospheric, Oceanic \& Planetary Physics, Department of Physics, University of Oxford, Clarendon Laboratory, Parks Road, Oxford, OX1 3PU, UK}
\address[meu]{LESIA, Observatoire de Paris, CNRS, UPMC, Univ. Paris Diderot, 92195 Meudon, France.}


\linenumbers

\begin{abstract}

Global maps of Jupiter's atmospheric temperatures, gaseous composition and aerosol opacity are derived from a programme of 5-20 $\mu$m mid-infrared spectroscopic observations using the Texas Echelon Cross Echelle Spectrograph (TEXES) on NASA's Infrared Telescope Facility (IRTF).   Image cubes from December 2014 in eight spectral channels, with spectral resolutions of $R\sim2000-12000$ and spatial resolutions of $2-4^\circ$ latitude, are inverted to generate 3D maps of tropospheric and stratospheric temperatures, 2D maps of upper tropospheric aerosols, phosphine and ammonia, and 2D maps of stratospheric ethane and acetylene.  The results are compared to a re-analysis of Cassini Composite Infrared Spectrometer (CIRS) observations acquired during Cassini's closest approach to Jupiter in December 2000, demonstrating that this new archive of ground-based mapping spectroscopy can match and surpass the quality of previous investigations, and will permit future studies of Jupiter's evolving atmosphere.  The visibility of cool zones and warm belts varies from channel to channel, suggesting complex vertical variations from the radiatively-controlled upper troposphere to the convective mid-troposphere.  We identify mid-infrared signatures of Jupiter's 5-$\mu$m hotspots via simultaneous M, N and Q-band observations, which are interpreted as temperature and ammonia variations in the northern Equatorial Zone and on the edge of the North Equatorial Belt (NEB).  Equatorial plumes enriched in NH$_3$ gas are located south-east of NH$_3$-desiccated `hotspots' on the edge of the NEB.  Comparison of the hotspot locations in several channels across the 5-20 $\mu$m range indicate that these anomalous regions tilt westward with altitude. Aerosols and PH$_3$ are both enriched at the equator but are not co-located with the NH$_3$ plumes.  The equatorial temperature minimum and PH$_3$/aerosol maxima have varied in amplitude over time, possibly as a result of periodic equatorial brightenings and the fresh updrafts of disequilibrium material.  Temperate mid-latitudes display a correlation between mid-IR aerosol opacity and the white albedo features in visible light (i.e., zones).  We find hemispheric asymmetries in the distribution of tropospheric PH$_3$, stratospheric hydrocarbons and the 2D wind field (estimated via the thermal-wind equation) that suggest a differing efficiency of mechanical forcing (e.g., vertical mixing and wave propagation) between the two hemispheres that we argue is driven by dynamics rather than Jupiter's small seasonal cycle.  Jupiter's stratosphere is notably warmer at northern mid-latitudes than in the south in both 2000 and 2014, although the latter can be largely attributed to strong thermal wave activity near $30^\circ$N that dominates the 2014 stratospheric maps and may be responsible for elevated C$_2$H$_2$ in the northern hemisphere.  A vertically-variable pattern of temperature and windshear minima and maxima associated with Jupiter's Quasi Quadrennial Oscillation (QQO) is observed at the equator in both datasets, although the contrasts were more subdued in 2014.  Large-scale equator-to-pole gradients in ethane and acetylene are superimposed on top of the mid-latitude mechanically-driven maxima, with C$_2$H$_2$ decreasing from equator to pole and C$_2$H$_6$ showing a polar enhancement, consistent with a radiatively-controlled circulation from low to high latitudes.  Cold polar vortices beyond $\sim60^\circ$ latitude can be identified in the upper tropospheric and lower stratospheric temperature maps, suggesting enhanced radiative cooling from polar aerosols.  Finally, compositional mapping of the Great Red Spot confirms the local enhancements in PH$_3$ and aerosols, the north-south asymmetry in NH$_3$ gas and the presence of a warm southern periphery that have been noted by previous authors.  

\end{abstract}


\end{frontmatter}


\section{Introduction}
\label{intro}

Thermal infrared sounding of Jupiter provides a rich resource for investigation of the dynamical, chemical and cloud-forming processes shaping the three-dimensional structure of the planet's atmosphere.  The 5-25 $\mu$m region provides access to a host of spectral absorption and emission features, superimposed onto a continuum of hydrogen-helium emission and aerosol opacity, from which we can determine the horizontal and vertical distributions of temperature, composition and aerosol structures from the churning cloud tops to the overlying stratosphere.   Spatially-resolved thermal mapping from Voyager, Galileo and Cassini allowed us to explore the connection between the dynamic activity observed in the cloud-forming region and the relatively unexplored circulation and chemistry of the middle atmosphere (upper troposphere and stratosphere).    However, instruments to exploit this spectral range are absent from future missions to Jupiter, including the upcoming Juno spacecraft.  In this study we report on a regular programme of spectroscopic mapping observations from NASA's Infrared Telescope Facility (IRTF), aiming to match and surpass the capabilities of previous spacecraft thermal-IR observations to provide a new database for investigators studying Jovian climate, dynamics and chemistry.  Our aim is to bridge the observational gap in IR spectroscopy between the Cassini and Juno epochs (2000 and 2016, respectively).

Multi-wavelength imaging in narrow-band filters covering the 5-25 $\mu$m spectral range \citep[including those from the Galileo photopolarimeter-radiometer instrument,][]{96orton} have proven highly effective in constraining atmospheric temperatures at discrete pressure levels, and data amassed over several decades have revealed: (i) tropical variability associated with stratospheric wind and temperature oscillations \citep[analogous to Earth's quasi-biennial oscillation,][]{91orton,91leovy,94orton,99friedson,06simon}; (ii) belt/zone variability caused by the life cycle of jovian `global upheavals' \citep{95rogers}, particularly the fade and revival cycle of the South Equatorial Belt \citep[SEB,][]{11fletcher_fade}; (iii) a characterisation of the Galileo probe entry site as a region of uniquely powerful atmospheric subsidence and desiccation \citep{98orton, 98ortiz, 05friedson}; (iv) understanding of the thermal aftermath of large impact events \citep[e.g.,][and references therein]{04harrington}; and (v) the thermal structure and variability of Jupiter's large anticyclones like the Great Red Spot \citep{10fletcher_grs}.  Despite these successes, temperatures derived from thermal imaging observations are subject to large degeneracies with chemical composition and cloud opacity \citep{09fletcher_imaging}, rendering the quantitative results highly uncertain.  Spatial mapping of tropospheric and stratospheric gases, in particular, requires us to spectrally resolve the forest of absorption and emission features to derive abundances. It is this deficiency in spectroscopy that our IRTF spectral programme seeks to address.

Spatially resolved spectral maps of Jupiter have been provided by Voyager/IRIS \citep[Infrared Radiometer and Spectrometer,][]{77hanel} and Cassini/CIRS \citep[Composite Infrared Spectrometer,][]{04flasar}, but these were limited to snapshots during brief flybys, so they failed to explore the temporal variability of the thermal emission.  Voyager-1 and -2 spectra (March and July 1979, respectively) have been presented as zonally averaged spectra for interpretation \citep{83conrath, 84conrath, 81flasar, 92griffith, 92carlson, 96sada, 98conrath, 00simon}, although sparse longitudinally-resolved coverage was available and has been used to investigate the Great Red Spot \citep{92griffith, 96sada, 02simon, 06read_grs} and the spatial distribution of water ice signatures \citep{00simon}.  Only the spectral maps of Cassini/CIRS between December 2000 - January 2001 can claim to have provided near-global coverage by sweeping its detectors from north to south to generate multiple maps over approximately two weeks.  The Cassini datasets have provided us with tropospheric and stratospheric temperature maps \citep{04flasar, 06li}, distributions of the disequilibrium species phosphine \citep[PH$_3$,][]{04irwin, 09fletcher_ph3,10fletcher_grs}, distributions of ammonia, the key condensible in Jupiter's upper troposphere \citep{06achterberg}, cloud opacity \citep{04wong,05matcheva,09fletcher_ph3} and stratospheric hydrocarbons \citep{04kunde, 07nixon, 10nixon, 13zhang}.  Temporal variability of temperatures was observed during this close flyby \citep{04flasar, 06li}, but no orbital mission has ever provided a long-term database to study this fourth dimension.  

The best hope for characterisation of the variability of the thermal and chemical environment is therefore ground-based spectroscopy, albeit limited to regions free of terrestrial contamination (the M, N and Q bands near 5, 10 and 20 $\mu$m, respectively).  Ground-based spectroscopy permits the high spectral resolutions required to resolve spectral line shapes.  However, previous studies have focussed on discrete regions so that spatio-spectral mapping is rare and global coverage has not been previously published \citep{87kostiuk,93livengood,98sada,11fast,11fletcher_trecs}.  Observations from the IRSHELL spectrometer on the IRTF \citep[achieving spectral resolutions of $R\approx10,000$,][]{89lacy} were employed to map Jupiter's temperatures, clouds and distributions of phosphine and ammonia in the $10-36^\circ$S domain \citep{98lara} in 1991, although only zonal-mean cross-sections are shown.  IRSHELL was subsequently used in 1994 to map emission surrounding the Shoemaker-Levy 9 impact sites \citep{97griffith,97bezard_nh3}. IRSHELL was retired in 1994 as a successor, TEXES \citep[the Texas Echelon Cross Echelle Spectrograph,][]{02lacy} was developed as a visitor instrument for the IRTF.  TEXES has been previously employed to trace the fate of HCN and H$_2$O related to the Shoemaker-Levy 9 impact \citep{04griffith, 13cavalie} and to determine Jupiter's ammonia isotopologue ratios \citep{14fletcher_texes}.  The current programme of TEXES spatio-spectral mapping is described in Section \ref{obs}, and the development of a data reduction and spectral inversion pipeline for TEXES data is described in Section \ref{analysis}.  Having established the methodology, we present global spatio-spectral maps of Jupiter's temperatures, tropospheric disequilibrium species and condensible volatiles, tropospheric aerosol opacity and stratospheric hydrocarbon distributions in Section \ref{results}.  The results are compared to similar maps of Jupiter's composition from the Cassini flyby in 2000, showing that ground-based scan mapping of Jupiter can now match, and in some cases surpass, spacecraft flyby observations.  

\section{Data}
\label{obs}

\begin{figure*}
\begin{centering}
\centerline{\includegraphics[angle=0,scale=.80]{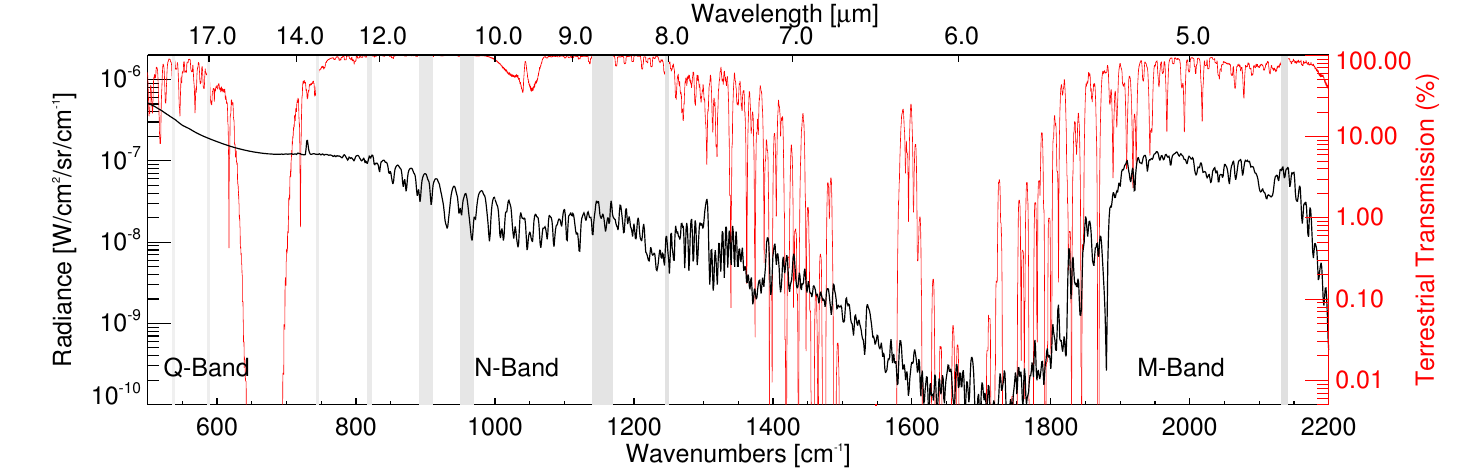}}
\caption{Synthetic spectrum of Jupiter calculated at 1 cm$^{-1}$ spectral resolution compared to the Earth's transmission spectrum from ATRAN \citep{92lord} (right hand axis, red line, calculated for Mauna Kea using an airmass of unity and precipitable water vapour column of 3 mm).  The nine TEXES channels are shown as grey vertical bars, selected for their high telluric transmission and sensitivity to key atmospheric gases.  The M, N and Q bands are labelled.}
\label{jupspx}
\end{centering}
\end{figure*}

The TEXES instrument \citep[Texas Echelon Cross Echelle Spectrograph,][]{02lacy} is a cross-dispersed grating spectrograph able to record spatially-resolved spectra throughout the M (5 $\mu$m), N (7-13 $\mu$m) and Q (17-24 $\mu$m) bands.  Fig. \ref{jupspx} compares a synthetic spectrum of Jupiter to the Earth's transmission windows:  the Q band is shaped by the collision-induced absorption of H$_2$ and He from which we can determine upper tropospheric temperatures; the N-band features broad absorption features of ammonia and phosphine, plus emission features of methane (a probe of stratospheric temperatures), ethane and acetylene (products of methane photolysis); and the M band senses thermal emission from the mid-troposphere attenuated by overlying clouds, hazes, PH$_3$, NH$_3$, CH$_3$D and other minor species.  We focus on the N and Q bands in this study, with initial results from the M band to be presented elsewhere \citep{16encrenaz_texes}.

Given that our primary targets are lines formed in the upper troposphere and lower stratosphere, pressure broadening dominates and the maximum TEXES spectral resolution ($R\approx80,000$ in cross-dispersed mode) is not required.  Our programme uses medium ($R\sim15,000$) and low ($R\sim2000$) spectral resolutions, employing the $1.4\times45$-arcsecond slit to cover the entire jovian disc at a series of distinct wavelength settings, as described below.  The lower spectral resolutions bypass the echelon grating and simply use the echelle or first-order grating as the disperser \citep{02lacy}, allowing the full slit length to be imaged onto the $256\times256$ SiAs detector array.  

TEXES was used in `scan mode,' whereby we aligned the slit along the celestial north-south and stepped from west to east across the planet in 0.7" increments (Nyquist sampling the 1.4" slit width), with 2-second integrations at each step.  We did not align directly with Jupiter's central meridian so that any row/column defects on the detector could be readily distinguished from the banded structure of Jupiter.  These scans started and finished on blank sky to permit background subtraction from the on-source measurements.  Scans in a particular setting were repeated 2-4 times in quick succession to build up signal-to-noise and minimise the risks of data loss due to cosmic rays or detector defects, before moving on to the next spectral setting.  Unlike mid-infrared imaging, we do not use the chopping secondary, using the off-target scan steps instead of nodded pairs to remove the background.  Given that Jupiter's 10-hour rotation is longer than its visibility from the IRTF, we requested groups of 2-3 consecutive nights in order to cover as many jovian longitudes as possible in a short space of time, building up near-complete maps of the planet.  Typically 5-10 individual scan maps were obtained for each of the nine spectral settings used in this study.  This combination of the long TEXES slit, efficient scan-mapping and calibration routines developed by the TEXES instrument scientists, and carefully-selected spectral settings permits the global temperature and composition mapping described in Section \ref{results}.

This global mapping programme has, to date, provided maps in February 2013, October and December 2014, March and November 2015, and January and April 2016.  Each observing run used a standard set of TEXES settings at low and medium spectral resolutions, detailed in Table \ref{settings}.  Settings were chosen based on their sensitivity to temperatures in a particular altitude range or the presence of absorption/emission features in relatively clear regions of the telluric transmission spectrum.  Two channels of the February 2013 dataset sensing tropospheric NH$_3$ were previously published by \citet{14fletcher_texes}.  In this study we use the full December 2014 dataset of nine channels, detailed in Table \ref{data1}-\ref{data2} in Appendix \ref{datarecord}, acquired over two nights (December 8th and 9th), as an excellent example of the quality of the maps that can be derived from TEXES data.   A suite of 9 channels took approximately 70 minutes to acquire, and was cycled repeatedly over approximately 6-8 hours.  December 8th focussed on $0-180^\circ$W and December 9th focussed on $180-360^\circ$W.  

\begin{table*}[htb]
\caption{Nine spectral channels considered in this study, showing the spectral resolution, coverage, diffraction-limited spatial resolution and key spectral features.}
\begin{center}
\begin{tabular}{|c|c|c|c|c|c|}
\hline
Central Wavenumber & Resolving Power & Coverage & Resolution & Diffraction Limit & Key Features/Objectives \\
\hline
538 cm$^{-1}$ & 7907 & 537-541 cm$^{-1}$ & 0.068 cm$^{-1}$ & 1.56" & H$_2$-He tropospheric T \\
586 cm$^{-1}$ & 5836 & 584-589 cm$^{-1}$ & 0.101 cm$^{-1}$ & 1.43" & H$_2$-He tropospheric T \\
744 cm$^{-1}$ & 10292 & 742-747 cm$^{-1}$ & 0.072 cm$^{-1}$ & 1.12" & C$_2$H$_2$ \\
819 cm$^{-1}$ & 7724 & 815-823 cm$^{-1}$ & 0.106 cm$^{-1}$ & 1.02" & C$_2$H$_6$ \\
901 cm$^{-1}$ & 2896 & 885-915 cm$^{-1}$ & 0.311 cm$^{-1}$ & 0.93" & NH$_3$ \\
960 cm$^{-1}$ & 2664 & 945-975 cm$^{-1}$ & 0.360 cm$^{-1}$ & 0.87" & NH$_3$ \& PH$_3$ \\
1161 cm$^{-1}$ & 2157 &1138-1170 cm$^{-1}$ & 0.538 cm$^{-1}$ & 0.72" & PH$_3$, CH$_3$D and Aerosols \\
1248 cm$^{-1}$ & 12358 & 1243-1252 cm$^{-1}$ & 0.101 cm$^{-1}$ & 0.67" & CH$_4$ stratospheric T \\
2137 cm$^{-1}$ & 12366 & 2131-2142 cm$^{-1}$ & 0.173 cm$^{-1}$ & 0.39" & Deep cloud opacity\\
\hline
\end{tabular}
\end{center}
\label{settings}
\end{table*}%

\subsection{TEXES Data Processing}

\subsubsection{Radiometric and wavelength calibration}
Target spectra were radiometrically calibrated and flat-fielded using two observations of the sky emission and two observations of a room-temperature black body (a high-emissivity metal chopper blade just above the entrance to the Dewar, temperature $T_{black}$), observed immediately prior to each scan.  If we assume that $T_{black}$ is approximately equal to the sky ($T_{sky}$) and telescope ($T_{tel}$) temperatures, then the difference between the black body and the sky observations can be used as the flat field to account for both the telluric and instrument emission.   The calibrated target intensity $I_{\nu}(\mbox{target})$ is therefore given by \citep{02lacy}:
\begin{equation}
I_{\nu}(\mbox{target})=S_{\nu}(\mbox{target-sky})\frac{B_{\nu}(T_{tel})}{S_{\nu}(\mbox{black - sky})}
\end{equation}
where $S_{\nu}(\mbox{target-sky})$ is the measured flux difference between the target and the sky, $S_{\nu}(\mbox{black - sky})$ is the measured flux difference between the black and the sky, and $B_{\nu}(T_{tel})$ is the black body flux at the ambient temperature of the telescope.  As the black body fills the TEXES field of view, we need not account for the FOV-filling corrections that are typically required if standard divisors (e.g., mid-IR bright stars or asteroids) are used, providing a highly efficient calibration scheme that has been found to match the accuracy of more standard absolute calibration techniques.  This sky subtraction cannot remove the telluric absorption completely, doing a better job with gases in Earth's warmer troposphere (e.g., CO$_2$ and H$_2$O) than those in the cold and high stratosphere (e.g., O$_3$).  Variable water vapour and clouds (especially thin cirrus clouds) between each step of the scan are partially accounted for using the small portions of sky available at the ends of the slit away from the target.  However, as we shall see in Section \ref{analysis}, the calibration becomes less accurate in regions where $T_{sky}$ and $T_{black}$ differ substantially (i.e., where the sky emission is low), and where the TEXES system response \citep[Fig. 5 of][]{02lacy} becomes small. Given the high sensitivity of spectral inversions to this radiometric accuracy, we still require cross-calibration with space-based measurements for the purpose of this study.

The TEXES data reduction package \citep{02lacy} performs the required sky subtraction, flat fielding and radiometric calibration, as well as corrections for optical distortions within the instrument and the removal of dead pixels on the detector.  The measured sky scans were correlated with a model for the Earth's transmission spectrum to assign wavelengths to each pixel, although this too required fine tuning prior to spectral inversion.  A custom-designed IDL pipeline was created to assign latitudes, longitudes, Doppler shifts and emission angles (observing zenith angles) to each pixel using a visual fit to the location of the planetary limb.    Each individual scan map was then interpolated onto a regular $1\times1^\circ$ grid and radiances were Doppler shifted back to the rest frame for subsequent analysis (i.e., removing redshifts from the dusk limb and blueshifts from the dawn limb).  To improve further on the wavelength calibration in each spectral setting, a forward model based on Cassini/CIRS determinations of temperatures, composition and aerosol opacity \citep{09fletcher_ph3} was used to identify spectral features.  This was compared to a TEXES spectrum averaged within $\pm30^\circ$ of latitude and longitude of the sub-observer point for every individual scan map, and any differences were used to improve the accuracy of the spectral calibration via a shift-and-stretch method \citep{14fletcher_texes}.   The average sky transmission (using the same Doppler shift as the Jupiter data) was used to identify contaminated regions of each spectrum.  

\subsubsection{Inter-cube variability}

\begin{figure*}
\begin{centering}
\centerline{\includegraphics[angle=0,scale=.85]{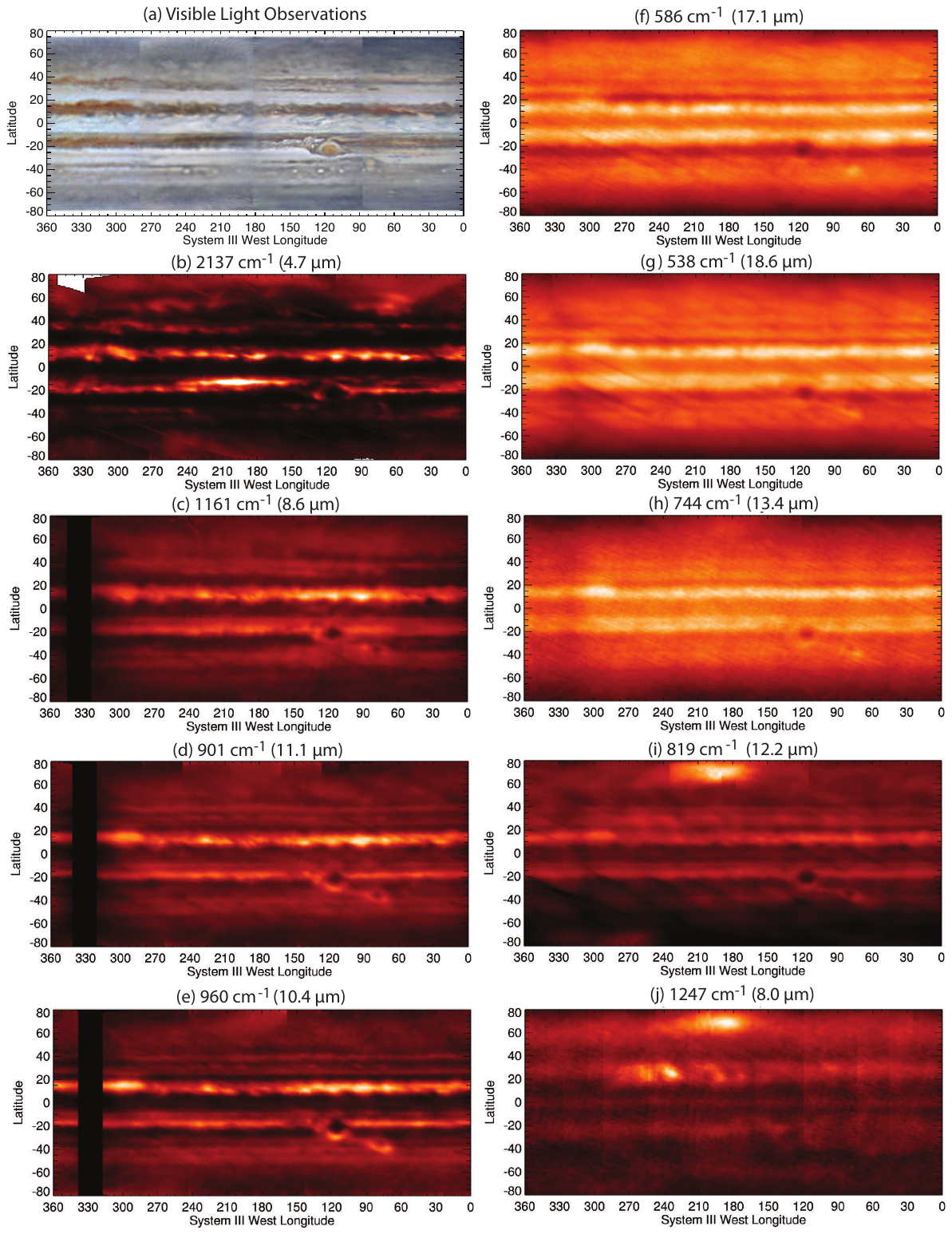}}
\caption{Examples of raw intensity maps assembled by averaging the spectral radiances from each channel over all wavelengths where transmission exceeded 80\% of the clearest region of the channel.  All latitudes are planetographic. Emission angle effects have been corrected using the airmass for display purposes, and some artefacts (vertical striping) remain visible.  The thermal maps are compared to a visible-light map assembled by M. Vedovato for the same dates, using images from Ian Sharp (04:24UT, Dec 8), F. Fortunato (06:45UT, Dec 8), H. Einaga (19:40UT, Dec 8) and T. Horiuchi (17:51UT, Dec 9), from right to left, respectively.  Panels are organised so that those in the left column sense clouds and deep ($p>500$ mbar) temperatures, those in the upper right sense upper-tropospheric temperatures via the Q-band, and those in the lower right sense the stratosphere (or a mixture of tropospheric and stratospheric emission). }
\label{texesmaps} 
\end{centering}
\end{figure*}

The individual wavelength-corrected and absolutely-calibrated scan maps were combined into global maps, with the raw data for each spectral setting shown in Fig. \ref{texesmaps}.  To create these maps, we averaged all data at a particular latitude/longitude with a zenith angle within $10^\circ$ of the minimum (i.e., as close to nadir as possible for each location).  Although empirically corrected using the zenith angle, the maps sometimes show discontinuities in radiance as vertical stripes, due to the mismatch of zenith angles between adjacent longitudes.  Upon initial inspection, we discovered small radiance offsets from cube to cube in a particular setting, potentially correlated with changes to the sky background during the observing run.  Variable water humidity or cirrus cloud over the course of the two nights would change the effectiveness of the absolute calibration, and produced stark steps in the absolute radiance in the global maps of the order 5-15\% depending on the specific setting.  Whilst this level of variability is within the conservative 20\% uncertainty envelope usually quoted for calibration of ground-based data, it is insufficiently accurate to permit spatially-resolved retrievals.  

\begin{figure*}
\begin{centering}
\centerline{\includegraphics[angle=0,scale=.80]{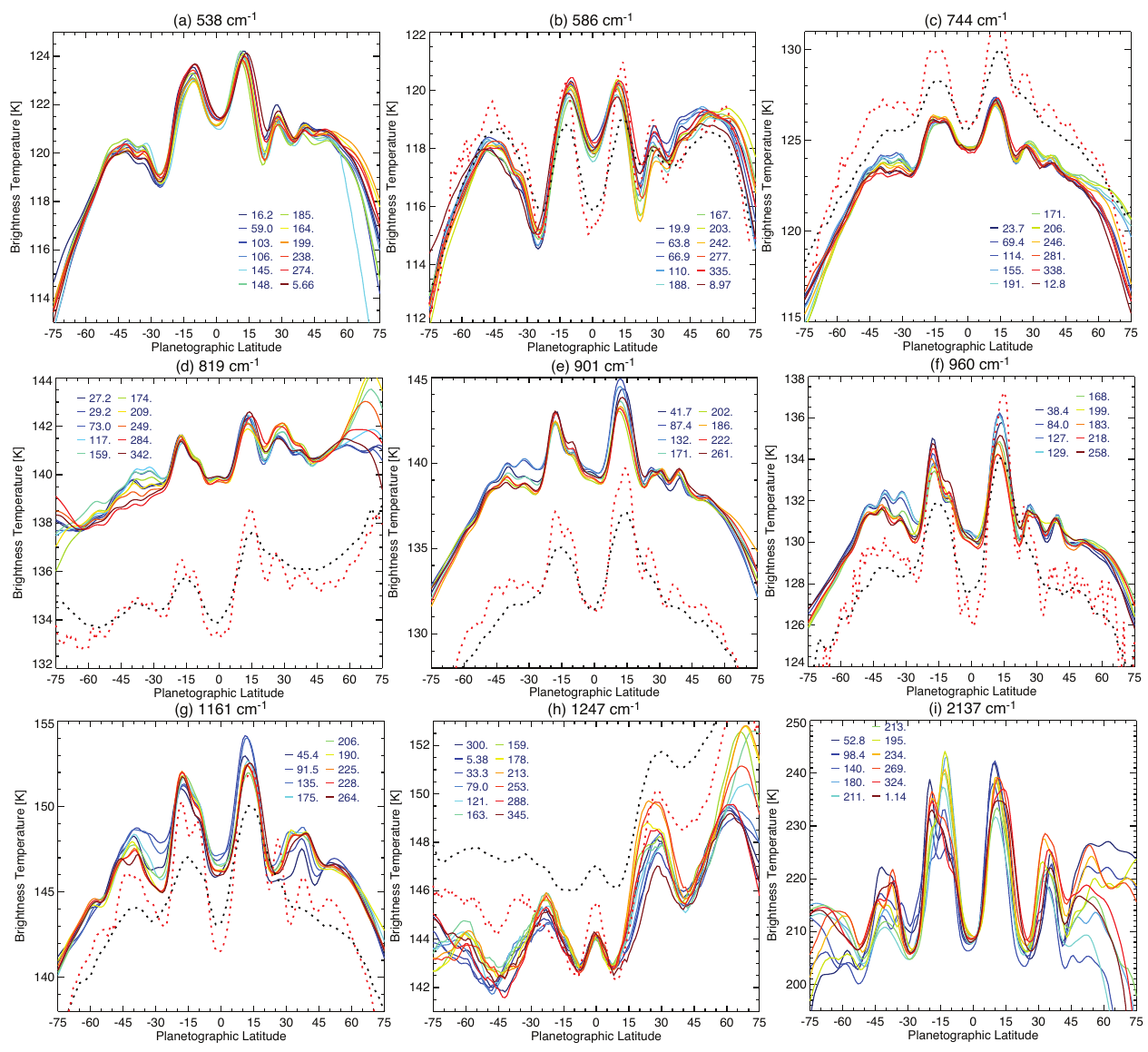}}
\caption{Central-meridian averaged radiances extracted from each TEXES spectral cube and averaged over the wavelength range of the setting (where the transmission exceeds 80\% of the clearest region in the setting).  Each cube has been normalised to a specific latitude range as described in the main text, and the colours correspond to the central meridian longitudes shown in the inset key.  These are compared to Cassini/CIRS spectra averaged over the same spectral range for (a) 0.5 cm$^{-1}$ data averaged from November 15th 2000 to February 15th 2001 with a low spatial resolution (black dotted); and (b) 2.5 cm$^{-1}$ data acquired at closest approach on December 31st 2000 (red dotted).  This provides an indication of the spatial resolution of the TEXES data when compared to CIRS, and also highlights systematic differences between CIRS and TEXES that will be explored in Section \ref{analysis}.  Note that CIRS spectra do not cover the 538 cm$^{-1}$ and 2137 cm$^{-1}$ channels, so are not shown in panels (a) and (i).}
\label{meridianspx}
\end{centering}
\end{figure*}

We therefore extracted averaged radiances from within $\pm10^\circ$ longitude of the central meridian for each scan map, averaged over the spectral channel, and normalised them all to the median value within a specific latitude range.  We chose latitude ranges that are relatively unaffected by Jupiter's intrinsic longitudinal variability - equatorial regions for stratospheric channels (CH$_4$, C$_2$H$_2$ and C$_2$H$_6$) and limb-darkened high latitudes for tropospheric channels.  These corrected central meridian radiances are shown for each channel in Fig. \ref{meridianspx}, compared to Cassini/CIRS zonally-averaged radiances, averaged over the same spectral range as each TEXES channel.  This is not a quantitatively accurate comparison, given that CIRS radiances have a lower spectral resolution (both 0.5 cm$^{-1}$ and 2.5 cm$^{-1}$ observations are shown) and are not affected by terrestrial contamination.  Nevertheless, they reveal that large-scale offsets between the TEXES and CIRS absolute calibrations are present, which will be dealt with in Section \ref{analysis}.

One unfortunate feature of the TEXES image cubes is a `column noise' due to blemishes on the filter, which manifests as a vertical stripe on the images that appears to have a lower brightness than the rest of the image.  As Jupiter's central meridian was at an angle to the detector rows and columns (the slit was aligned along the celestial north-south), this translates to diagonal striping in the cylindrical maps.  Such stripes can be seen in the Q-band images (Fig. \ref{texesmaps}f-g) but are also present at 819 cm$^{-1}$ (Fig. \ref{texesmaps}i).  The combination of the blemishes, Jupiter's intense brightness at these wavelengths, and the relative clarity of the telluric atmosphere (i.e., very little sky flux) means that we have no direct means to remove them from the data via flat fielding.  This adds additional uncertainty to the retrieved products which will be assessed in Section \ref{analysis}.

\subsubsection{Spatial resolution}

The highest spatial resolution of the Cassini/CIRS maps of Jupiter was 2700 km from 137 R$_J$ using its $273\times273$ $\mu$rad detectors, equivalent to $2.2^\circ$ at Jupiter's equator.   The TEXES observations occurred when Jupiter was at a distance of 4.83 AU ($7.2\times10^8$ km), so that the spatial resolution varies between 2400-5500 km (0.67-1.56" for diffraction-limited observations from the 3-m IRTF between the longest and shortest wavelengths, 1248 cm$^{-1}$ and 538 cm$^{-1}$), equivalent to $1.9-4.4^\circ$ latitude at Jupiter's equator.  A 0.75" seeing corresponds to the same spatial resolution as the best CIRS dataset, with wavelengths larger than 10$\mu$m being diffraction limited.  The spatial resolution of the two datasets is therefore comparable at wavelengths below 10 $\mu$m.  For the remainder of this paper we explore both zonal-mean spectra and spatially resolved spectra from both TEXES and CIRS.  

\subsection{Inspection of Images}
\label{image_inspection}

Before proceeding with inversion of the TEXES, we describe some of the features revealed in Fig. \ref{texesmaps} in comparison with a montage of visible light images (Fig. \ref{texesmaps}a), kindly provided by M. Vedovato based on observations by amateur observers.  Images of the $0-180^\circ$W were acquired approximately 10 hours (one rotation) before the TEXES maps, whereas images of $180-360^\circ$W coincided with the TEXES maps.  It is important to note that the TEXES maps of a particular region were all acquired within approximately 70 minutes of one another (Tables \ref{data1}-\ref{data2}), so longitudinal motions would have been negligible during this interval.  The reader is referred to Table \ref{beltzone} for nomenclature for the belt/zone structure used in the text that follows.  All latitudes in this study are planetographic.

The dominant features of the maps are the cool, cloudy zones and warm, cloud-free belts, punctuated by dramatic wave activity and large anticyclonic vortices (the Great Red Spot near $120^\circ$W and Oval BA near $80^\circ$W).  The visibility of the warm emission from the belts varies as a function of wavelength (and therefore altitude), with tropospheric belts in the North Temperate Domain ($20-50^\circ$N) being most prominent in NH$_3$-sensitive channels (10.4, 11.1, 12.2 $\mu$m in Fig. \ref{texesmaps}d,e and i) and hard to distinguish in aerosol-sensitive channels (4.7 and 8.6 $\mu$m,  Fig. \ref{texesmaps}b-c).  In particular, the warm band at $27^\circ$N (the North Temperate Belt (NTB), bordered by a prograde jet at $24^\circ$N and a retrograde jet at $31^\circ$N, Table \ref{beltzone}) that is visible in Figs. \ref{texesmaps}d and e does not appear to have a readily distinguishable counterpart in the visible light image - a thermal anomaly potentially masked by overlying aerosols.  

This warm NTB near $27^\circ$N and another belt near $40^\circ$N (the North North Temperate Belt, NNTB) straddle a colder zone (the North Temperate Zone, NTZ), within which we see several warm patches near $30^\circ$N that coincide with dark albedo structures in the visible (known as `brown barges'). These barges are at the limit of the resolution of the IRTF, but can be seen as bright patches at 4.7, 8.7, 10.4 and 11.1 $\mu$m (Figs. \ref{texesmaps}b-e), indicating that they are depleted in both ammonia and aerosols \citep{15orton_dps}.  They cannot be seen in the upper-troposphere sensitive filters from $13-18\mu$m, suggesting deep-seated features.  The NH$_3$-sensitive channels also reveal up to three distinct temperate belts in the southern hemisphere between $30-50^\circ$S, the most equatorward of which is partially disrupted by the passage of the GRS and Oval BA.  A chain of anticyclonic white ovals (AWOs) can be seen in the visible-light image in the South South Temperate Belt (SSTB), but are at the limit of the spatial resolution of the TEXES observations - they can be seen as darker patches in the 10-11 $\mu$m maps.  These same filters reveal non-uniformity within the equatorial zone, where regions of brighter emission coincide with visibly-dark albedo structures, suggesting small gaps in the otherwise thick reflective clouds.

Fig. \ref{texesmaps}i-j (12.2 and 8.0 $\mu$m) show the most sensitivity to stratospheric temperatures via emission from ethane and methane, respectively.  The 8-$\mu$m map is unlike any other, showing banded structures (a warm equator and cool neighbouring latitudes; warm mid-latitude bands) that have no counterpart in the deeper troposphere.  The mid-latitude stratospheric bands exhibit dramatic wave activity, particularly in the northern hemisphere in the $180-270^\circ$W region.  This stratospheric wave impacts both the temperature and composition of the mid-stratosphere, and will be discussed in Section \ref{results}.  Heating associated with the northern auroral oval is evident between $180-210^\circ$W \citep[as observed previously in ground-based observations, e.g.,][]{93livengood, 93kostiuk}, although high-spectral resolution TEXES observations \citep{15sinclair_dps} are required to determine the vertical structure of this energy deposition (from a combination of Joule heating in response to currents flowing downwards from the homopause level and direct deposition by precipitating electrons).  There is no evidence of heating associated with the southern aurora, but given the timing of the TEXES observations (northern summer) this may be due to a poor observing geometry for southern high latitudes.  Furthermore, the southern auroral oval occurs at a higher latitude ($\sim75^\circ$S) than that in the north.

\begin{figure*}
\begin{centering}
\centerline{\includegraphics[angle=0,scale=.90]{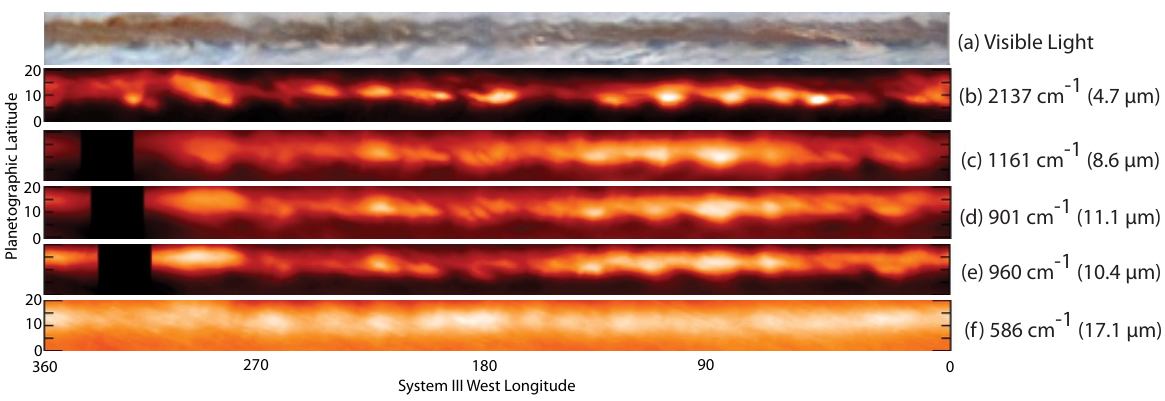}}
\caption{Focusing on dynamic activity in the North Equatorial Belt (NEB) between 0 and 20$^\circ$N, covering all longitudes.  Visible-light images show the `hotspots' as features of extremely low albedo on the southern edge of the red-brown belt on the prograding jet at $7^\circ$N.  The visible-light images are the same as those in Fig. \ref{texesmaps}:  images between $0-180^\circ$W we're taken approximately 10 hours prior to the TEXES images; the images between $180-360^\circ$W coincided with TEXES.   The TEXES observations of a particular longitude were taken within ~70 minutes of one another.}
\label{texesNEB}
\end{centering}
\end{figure*}

\begin{figure*}
\begin{centering}
\centerline{\includegraphics[angle=0,scale=.90]{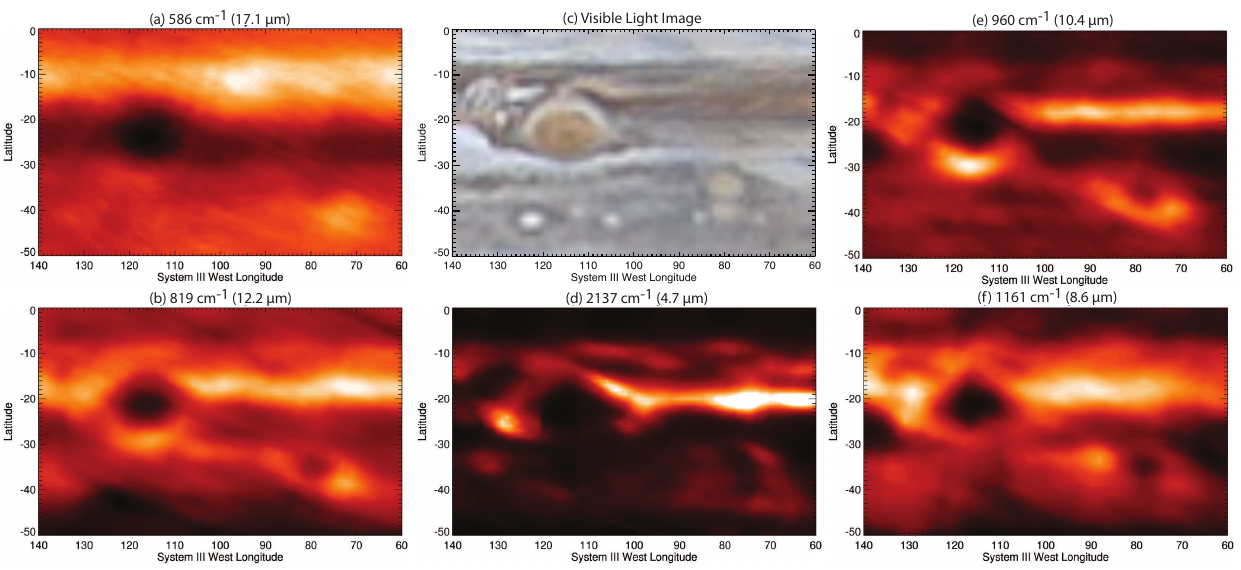}}
\caption{TEXES maps focusing on the Great Red Spot and Oval BA, demonstrating the co-alignment of the images from 5-18 $\mu$m.  The visible light images are there same as those in Fig. \ref{texesmaps}. }
\label{texesGRS}
\end{centering}
\end{figure*}

Besides the large-scale banded structures, the TEXES dataset also allows us to probe the vertical structure of smaller-scales.  Two examples are shown in Figs. \ref{texesNEB} and \ref{texesGRS}, for the North Equatorial Belt (NEB) and the region surrounding the GRS and Oval BA.  The brightness variations in the NEB are related to Rossby wave activity on the prograding jet at $7^\circ$N.  Visibly-dark structures in the visible images are associated with cloud-free regions at 4.7 $\mu$m, where the dearth of aerosol opacity permits emission from deeper, warmer layers.  These `5-$\mu$m hotspots' are in fact visible throughout the M and N-bands, showing that they are perturbing the temperature, aerosol and possibly the composition field in the 400-600 mbar region.  They are harder to observe in the Q-band, although this may simply be related to the lower spatial resolution.  The most interesting feature of Fig. \ref{texesNEB} is the offsets observed in the hotspot locations as a function of wavelength, primarily in the eastern hemisphere (observations acquired on December 8th 2014).  From Table \ref{data1}, we see that TEXES scans at different wavelengths were taken in a strict sequence, so that the same spatial locations on the planet would have been covered with no more than 70 minutes separation between one wavelength and the next, and it was often much faster - for example, images at 901, 960, 1161 and 2137 cm$^{-1}$ focussed on $90^\circ$W longitude were acquired within 30 minutes.   Could this represent a real tilt of the hotspots westward with height, from the deepest sensing 2137-cm$^{-1}$ filter (Fig. \ref{texesNEB}b) to the highest sensing 960 cm$^{-1}$ channel (Fig. \ref{texesNEB}f)?  This tilt is not observed everywhere within the NEB, with hotspots in the western hemisphere generally more co-aligned as a function of depth, and we speculate that this could be due to the differences in the thickness of NH$_4$SH clouds between the eastern and western hotspots.  The offset with respect to the visible light observations near $90^\circ$W (Fig. \ref{texesNEB}a) may be a temporal offset due to ten hours separation between the TEXES and amateur images, during which features on the NEBs jet could move east by $\sim3.4^\circ$ longitude.  Nevertheless, there is a closer alignment of the albedo patterns with the N-band observations than there is with the M-band observations, supporting the idea that the M-band probes levels beneath the top-most cloud decks.

Could the observed longitudinal offsets in Fig. \ref{texesNEB} simply be caused by inaccuracies in the spatial registration in the TEXES cubes?  While uncertainties in latitude and longitude of around $1^\circ$ are certainly possible, the images of the Great Red Spot and other features in Fig. \ref{texesGRS} are part of exactly the same image cubes as those in Fig. \ref{texesNEB}.  Here we have no evidence for longitudinal shifts between any of the filters presented - the GRS and Oval BA are aligned in each of these images.  The structure of the two giant vortices matches that previously presented \citep{10fletcher_grs}, namely:  high cloud opacity observed at both 1161 and 2137 cm$^{-1}$, the cloud covering a broader area in the short-wavelength channel; a warm southern periphery in the 400-600 mbar region observed at 819 and 960 cm$^{-1}$ (coinciding with high cloud opacity) but not in the 200-400 mbar region observed at 586 cm$^{-1}$; and a warm and aerosol-free SEB (particularly associated with rifting activity in the northwest wake of the GRS) contrasted against a cold and cloudy South Tropical Zone (STropZ).  Fig. \ref{texesGRS} shows a superior spatial resolution to those maps of the GRS acquired by Cassini/CIRS \citep[see Fig. 4 of][]{10fletcher_grs}.  Recalling that each pixel in these images represents a full TEXES spectrum of eight channels, Figs. \ref{texesNEB} and \ref{texesGRS} highlight the capability for temperature, composition and aerosol sounding within the giant vortices and other regions of interest on Jupiter.

\begin{table*}[htb]
\caption{Nomenclature for the belt/zone structure observable in IRTF images between $\pm40^\circ$ latitude, using the zonal winds of \citet{03porco} and following \citet{95rogers}.  Acronyms followed by a lower case `n' or `s' are the names of the jets, with the latitudes in planetographic coordinates.}
\begin{center}
\begin{tabular}{|c|c|c|c|c|c|}
\hline
Acronym & Name & Southern Jet & Southern Jet Speed [m/s] & Northern Jet & Northern Jet Speed [m/s] \\
\hline
SSTB & South South Temperate Belt & SSTBs, $-39.6^\circ$ & $-8.4\pm8.4$ & SSTBn, $-36.2^\circ$ & $34.2\pm3.7$ \\
STZ & South Temperate Zone & SSTBn, $-36.2^\circ$ & $34.2\pm3.7$ & STBs, $-32.4^\circ$ & $-16.1\pm5.6$ \\
STB & South Temperate Belt & STBs, $-32.4^\circ$ & $-16.1\pm5.6$ & STBn, $-27.1^\circ$ & $48.1\pm2.3$  \\
STropZ & South Tropical Zone & STBn, $-27.1^\circ$ & $48.1\pm2.3$ & SEBs, -19.7$^\circ$ & $-62.0\pm18.4$ \\
SEB & South Equatorial Belt & SEBs, -19.7$^\circ$ & $-62.0\pm18.4$ & SEBn, -7.2$^\circ$ & $136.9\pm6.5$ \\
EZ & Equatorial Zone & SEBn, -7.2$^\circ$ & $136.9\pm6.5$ & NEBs, $6.9^\circ$ & $113.9\pm20.5$ \\
NEB & North Equatorial Belt & NEBs, $6.9^\circ$ & $113.9\pm20.5$ & NEBn, $17.4^\circ$ & $-20.1\pm5.1$ \\
NTropZ & North Tropical Zone & NEBn, $17.4^\circ$ & $-20.1\pm5.1$ & NTBs, $24.2^\circ$ & $136.2\pm3.3$ \\
NTB & North Temperate Belt & NTBs, $24.2^\circ$ & $136.2\pm3.3$ & NTBn, $31.4^\circ$ & $-24.8\pm5.1$ \\
NTZ & North Temperate Zone & NTBn, $31.4^\circ$ & $-24.8\pm5.1$ & NNTBs, $35.4^\circ$ & $31.8\pm2.6$ \\
NNTB & North North Temperate Belt & NNTBs, $35.4^\circ$ & $31.8\pm2.6$ & NNTBn, $39.6^\circ$ & $-14.7\pm9.2$ \\

\hline

\hline
\end{tabular}
\end{center}
\label{beltzone}
\end{table*}

\section{TEXES retrieval pipeline}
\label{analysis}

\begin{figure*}
\begin{centering}
\centerline{\includegraphics[angle=0,scale=.75]{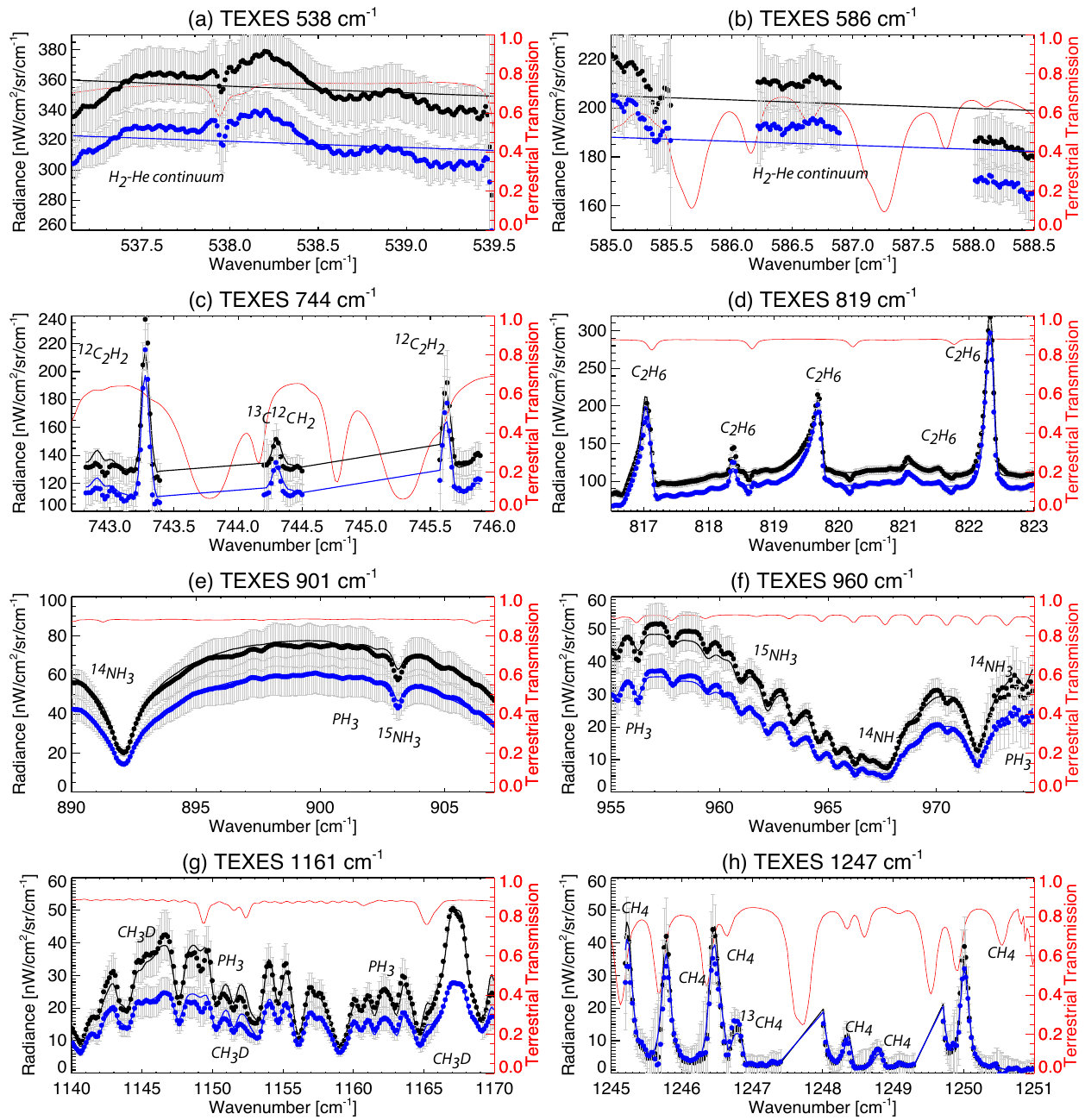}}
\caption{Two examples of zonally-averaged TEXES spectra at the equator (blue) and the North Equatorial Belt (black).  The data are shown as circles with error bars, the best-fitting model is shown as the solid line of the same colour.  The red line denotes the TEXES transmission spectrum (taking both the sky emission and losses within the telescope into account), indicating the presence of telluric contamination.   Gaps in the spectrum show regions that were not used in the spectral inversions.  Approximate locations of key features are indicated, but the features shown in panels f and g are blends of multiple lines of NH$_3$, PH$_3$ and CH$_3$D.  Q-band spectra in panels a and b should be flat (as shown by the solid lines), so variations are artefacts related to the poor telluric transmission.}
\label{spectrum}
\end{centering}
\end{figure*}

\subsection{Spectral Model and Inversion}

\begin{figure*}
\begin{centering}
\centerline{\includegraphics[angle=0,scale=.65]{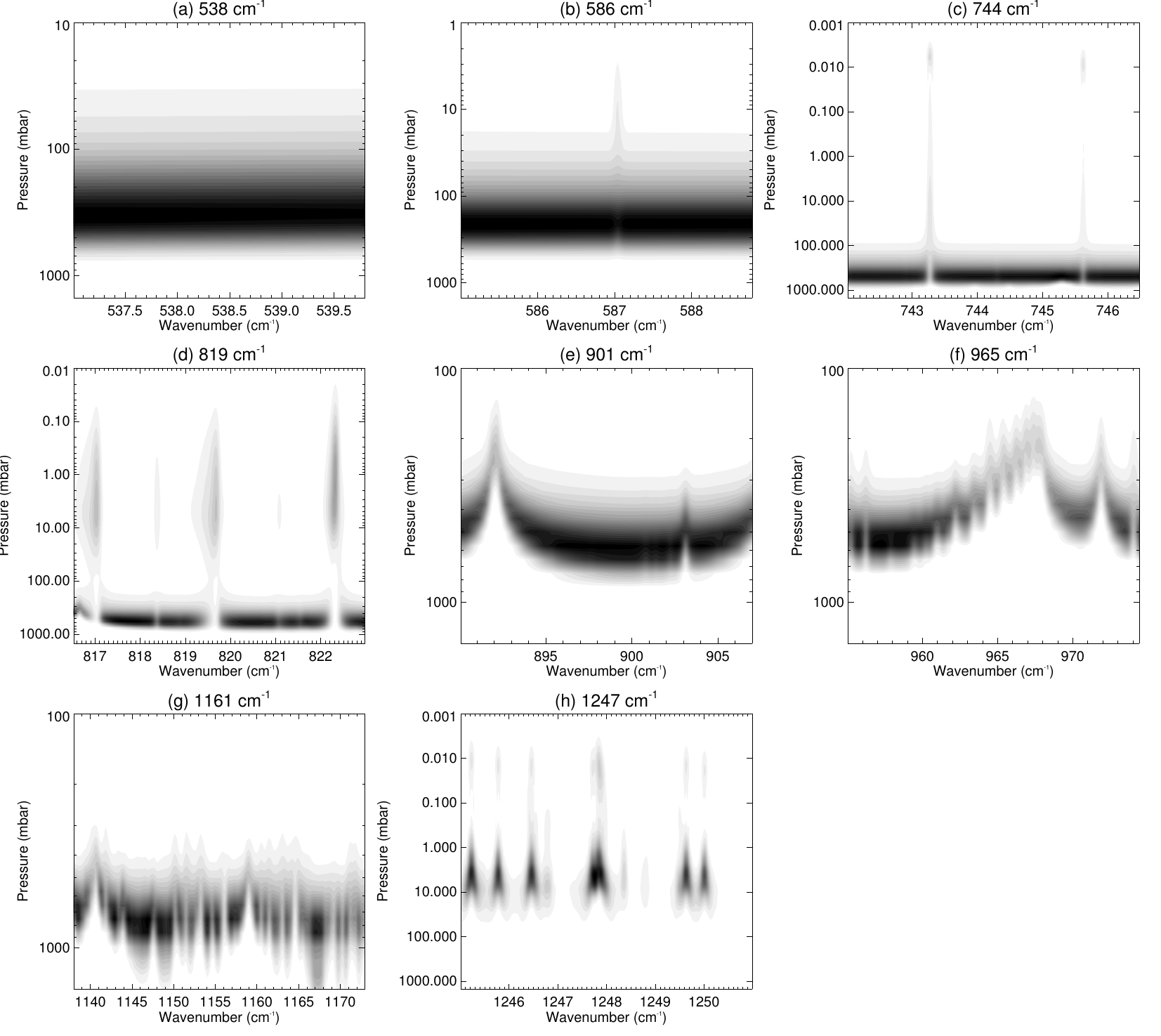}}
\caption{Contribution functions (Jacobians of temperature) computed for the TEXES channels used in this study.  The vertical axis changes from panel to panel, depending on how much of the stratosphere is being sampled.  Grey contours are given in steps of 0.05 from zero (white) to one (black), normalised to the strongest contribution for this spectral channel. }
\label{contrib}
\end{centering}
\end{figure*}

Fig. \ref{spectrum} shows the eight TEXES N and Q-band channels considered in this work, with key spectral features labelled.  The corresponding vertical sensitivity is shown in Fig. \ref{contrib}.  Zonal-mean TEXES spectra at 901 and 960 cm$^{-1}$ were previously analysed by \citet{14fletcher_texes} using the radiative-transfer and spectral-retrieval algorithm, NEMESIS \citep{08irwin}.  This work extends that previous analysis to include a further six spectral settings at 538, 586, 744, 819, 1161 and 1247 cm$^{-1}$ as shown in Table \ref{settings}, performing simultaneous retrievals from all eight channels.  We developed the retrieval pipeline one channel at a time, starting from the troposphere-sensing N-band channels and subsequently adding in capabilities for stratospheric temperatures (CH$_4$), upper tropospheric temperatures (538 an 586 cm$^{-1}$) and stratospheric composition (C$_2$H$_2$ and C$_2$H$_6$).  The addition of M-band channels to sound the mid-troposphere will be the subject of future work.  At each stage we performed tests to determine the retrieval sensitivity to the prior and the implications of adding the new channels. 

The forward model calculation uses the correlated-$k$ method \citep{89goody_ck,91lacis}, which required the pre-tabulation of smooth $k$-distributions (ranking absorption coefficients $k$ according to their frequency distributions) based on a variety of sources of spectral line data (Table \ref{tab:linedata}).  Isotopologues for methane ($^{12}$CH$_4$, CH$_3$D and $^{13}$CH$_4$) and ammonia ($^{14}$NH$_3$ and $^{15}$NH$_3$) were treated separately, but hydrocarbon isotopologues were combined into single tables. The $k$-distributions for each channel are pre-convolved with an instrument function with the spectral resolutions shown in Table \ref{settings}, calculated directly from the grating equation depending on the grating angle and the angular size of the TEXES slit.  These distributions are then combined into a single tabulation for each of the species listed in Table \ref{tab:linedata}.  The TEXES instrument function is expected to be a convolution of a Gaussian and a Lorenztian, but testing of a variety of instrument functions by \citet{14fletcher_texes} showed that the use of a simple Gaussian was sufficient for analysis of the TEXES data at low and moderate resolutions.  These $k$-distributions (with different spectral resolutions for each of the eight channels), combined with both the collision-induced absorption in Table \ref{tab:linedata} and aerosol absorption described below, constitute the forward model.

\begin{table*}[htp]
\caption{Sources of spectroscopic linedata (all isotopologues taken from the same sources).  Exponents for temperature dependence $T^n$ given in the final column.}
\begin{center}
\begin{tabular}{|p{2 cm}|p{7cm}|p{5cm}|p{3cm} |}
\hline
{\bf Gas} & {\bf Line Intensities} & {\bf Broadening Half Width} & {\bf Temperature Dependence $T^n$} \\
\hline

Collision-induced absorption (CIA) & H$_2$-H$_2$ opacities from \citet{07orton}, plus additional H$_2$-He, H$_2$-CH$_4$ and CH$_4$-CH$_4$ opacities from \citet{88borysow}, \citet{86borysow} and \citet{87borysow}, respectively.   & - & - \\
\hline

CH$_4$, CH$_3$D & \citet{03brown} & H$_2$ broadened using a half-width of 0.059 cm$^{-1}$atm$^{-1}$ at 296 K & $n=0.44$ \citep{93margolis} \\
\hline

C$_2$H$_6$& \citet{07vander} \citep[also found in GEISA 2009,][]{11geisa} & 0.11 cm$^{-1}$atm$^{-1}$ at 296 K \citep[][for H$_2$ and He, respectively]{88halsey,87blass} &  $n=0.94$ \citep{88halsey} \\
\hline

C$_2$H$_2$ & GEISA 2003 \citep{05geisa} \citep[unchanged in GEISA 2009 at 13.6 $\mu$m,][]{11geisa} & Fits to data in \citet{92varanasi} & - \\
\hline

PH$_3$ & \citet{03kleiner} & Broadened by both H$_2$ and He using $\gamma_{H_2}=0.1078-0.0014J$ cm$^{-1}$atm$^{-1}$ and $\gamma_{He}=0.0618-0.0012J$ cm$^{-1}$atm$^{-1}$ \citep{93levy, 04bouanich} &   $n=0.702-0.01J$  ($J$ is the rotational quantum number) \citep{04salem} \\
\hline

NH$_3$ & \citet{03kleiner} \citep[also found in GEISA 2009,][]{11geisa} & Empirical model of \citet{94brown} & Empirical model of \citet{94brown} \\
\hline

C$_2$H$_4$ & GEISA 2003 \citep{05geisa} & Fits to data in \citet{03bouanich_c2h4, 04bouanich_c2h4} (B. Bezard, \textit{pers. comm.}) & $n=0.73$ \citep{04bouanich_c2h4} \\
\hline

H$_2$ Quad. & HITRAN 2012 \citep{13rothman} & 0.0017 cm$^{-1}$atm$^{-1}$ \citep{94reuter} & $n=0.75$ \citep{13rothman} \\

\hline
\end{tabular}
\end{center}
\label{tab:linedata}
\end{table*}

The NEMESIS optimal estimation retrieval algorithm allows us to fit the TEXES spectra via a Levenburg-Marquardt iterative scheme, whilst using smooth \textit{a priori} state vectors to ensure physically-realistic solutions \citep[see][for a full discussion of this technique]{00rodgers,08irwin}.  The \textit{a priori} jovian atmosphere was specified on 120 levels from 10 bar to 1 $\mu$bar, using reference profiles of temperature, ammonia, phosphine, ethane and acetylene from a low-latitude mean of Cassini/CIRS results \citep{07nixon,09fletcher_ph3}.   The CIRS-derived temperature profile originally used the $T(p)$ from the Galileo Atmospheric Structure Instrument \citep[ASI,][]{98seiff} as a prior.  The deep helium and methane mole fractions were set to 0.136 and $1.81\times10^{-3}$, respectively, based on the Galileo probe measurements of \citet{98niemann} that were used to constrain the photochemical model of \citet{05moses_jup}.    Methane then decreased with altitude following the diffusive photochemical model of \citet{05moses_jup}, which was also used as the prior for the C$_2$H$_2$ and C$_2$H$_6$ measurements of \citet{07nixon}.  Ethylene (C$_2$H$_4$) is included based on the photochemical model of \citet{96romani} as it may have a minor effect near 950 cm$^{-1}$.  We assumed isotopologue ratios of D/H$_{(CH_4)}=2\times10^{-5}$ \citep[the value in the protosolar cloud,][]{03geiss}, a terrestrial ratio of $^{13}$C/$^{12}$C and a $^{15}$N/$^{14}$N ratio of $2.3\times10^{-3}$ \citep{01owen} that was previously confirmed by modelling of the TEXES 901- and 960-cm$^{-1}$ spectra \citep{14fletcher_texes}.  


Combining these priors with the $k$-distributions for each TEXES channel, we show contribution functions (Jacobians for temperature, or the rate of change of radiance with respect to the $T(p)$ profile) in Fig. \ref{contrib} to show how the vertical sensitivity of the TEXES data varies as a function of wavelength.  Note that these were computed for a nadir geometry - the greater atmospheric path at higher zenith angles would cause these contribution functions to move to higher altitudes.  This figure introduces some of the complexity of modelling the TEXES spectra.  Firstly, the contribution functions associated with the fine hydrocarbon emissions are often multi-lobed, with sensitivity in the 1-10 mbar range and a tail of sensitivity in the line cores probing the 5-15 $\mu$bar range.  The relative weight of these two regions is a complex function of the vertical temperature and composition structure, with observations at higher spectral resolution providing more data points (and hence more retrieval sensitivity) for the lowest pressures sensed in the line cores.  Cassini/CIRS 2.5-cm$^{-1}$ resolution spectra, by contrast, do not provide sufficient sensitivity to probe $p<1$ mbar in this nadir geometry. The ethane line cores, for example, sense a broad range from 0.1-20 mbar, making inferences of vertical $T(p)$ and composition gradients extremely degenerate with the limited data available.  

Secondly, Fig. \ref{contrib} shows that accurate radiometric calibration will be essential when attempting to combine multiple channels, because there are regions of significant overlap in vertical sensitivity.  The deepest tropospheric pressures probed are $\sim800$ mbar at 1161 cm$^{-1}$, $\sim600$ mbar at 819 cm$^{-1}$ and 901 cm$^{-1}$, $\sim550$ mbar at 965 cm$^{-1}$, $\sim500$ mbar at 744 cm$^{-1}$, $\sim300$ mbar at 538 cm$^{-1}$ and $\sim200$ mbar at 586 cm$^{-1}$.  The cores of the NH$_3$ lines probe up towards the 150-300 mbar level in Fig. \ref{contrib}e, where temperature constraint must come from the Q-band channels (e.g., Fig. \ref{contrib}b).  The continuum between the C$_2$H$_2$ features at 744 cm$^{-1}$ senses the 400-700 mbar level, significantly overlapping the continuum in Figs. \ref{contrib}(d-g).  Any inconsistencies between these continuum radiances would result in difficulties in selecting representative temperatures for these altitude levels, as we shall see below.

The information content of the TEXES spectra is such that we derive a full profile of atmospheric temperature, but we retrieve a single scaling factor for the hydrocarbons and a parameterised profile for NH$_3$ and PH$_3$ (a constant mole fraction up to a transition pressure $p_0$, above which the abundance declines due to condensation and/or photolytic destruction with a fractional scale height, $f$), for reasons we discuss in Section \ref{sensitivity}.  The abundance of NH$_3$ and PH$_3$ is forced to zero for altitudes above the tropopause.  We also derive a scale factor for the optical depth of a single aerosol layer, modelled as a simple grey absorber at the 800-mbar level with a compact scale height $0.2\times$ the gas scale height \citep{04wong, 05matcheva, 06achterberg, 09fletcher_ph3}, and later test the TEXES sensitivity to different cloud parameterisations.  Each spectral inversion therefore provides estimates of the 3D thermal profile and 2D distributions of PH$_3$, NH$_3$, C$_2$H$_6$, C$_2$H$_2$ and $\sim800$mbar aerosol opacity.

\subsection{Radiometric Comparison to Cassini}
\label{cirscomp}

The comparison of central-meridian averages between the CIRS dataset in 2000 and the TEXES dataset in 2014 (Fig. \ref{meridianspx}) suggested that systematic radiometric offsets might be present.  If these offsets had been confined to a single region of Jupiter, such as those associated with the most dramatic changes over time like the NEB and SEB, then we might have considered them to be real.  But these offsets are seen globally, which strongly suggests a defect in the radiometric calibration.  In their analysis of TEXES spectra of Saturn's stratospheric vortex, \citet{16fouchet_texes} found TEXES-derived stratospheric temperatures to be systematically cooler than those derived from CIRS.  They attributed this to the significant difference in spatial resolution found when convolving Cassini's high-spatial-resolution thermal maps with a seeing-limited FWHM that was reasonable for the IRTF at the time of their measurements.  However, as the TEXES and CIRS Jupiter datasets have a comparable spatial resolution, we cannot attribute the radiometric offsets observed in Fig. \ref{meridianspx} to the same effect.  

\begin{figure*}
\begin{centering}
\centerline{\includegraphics[angle=0,scale=.75]{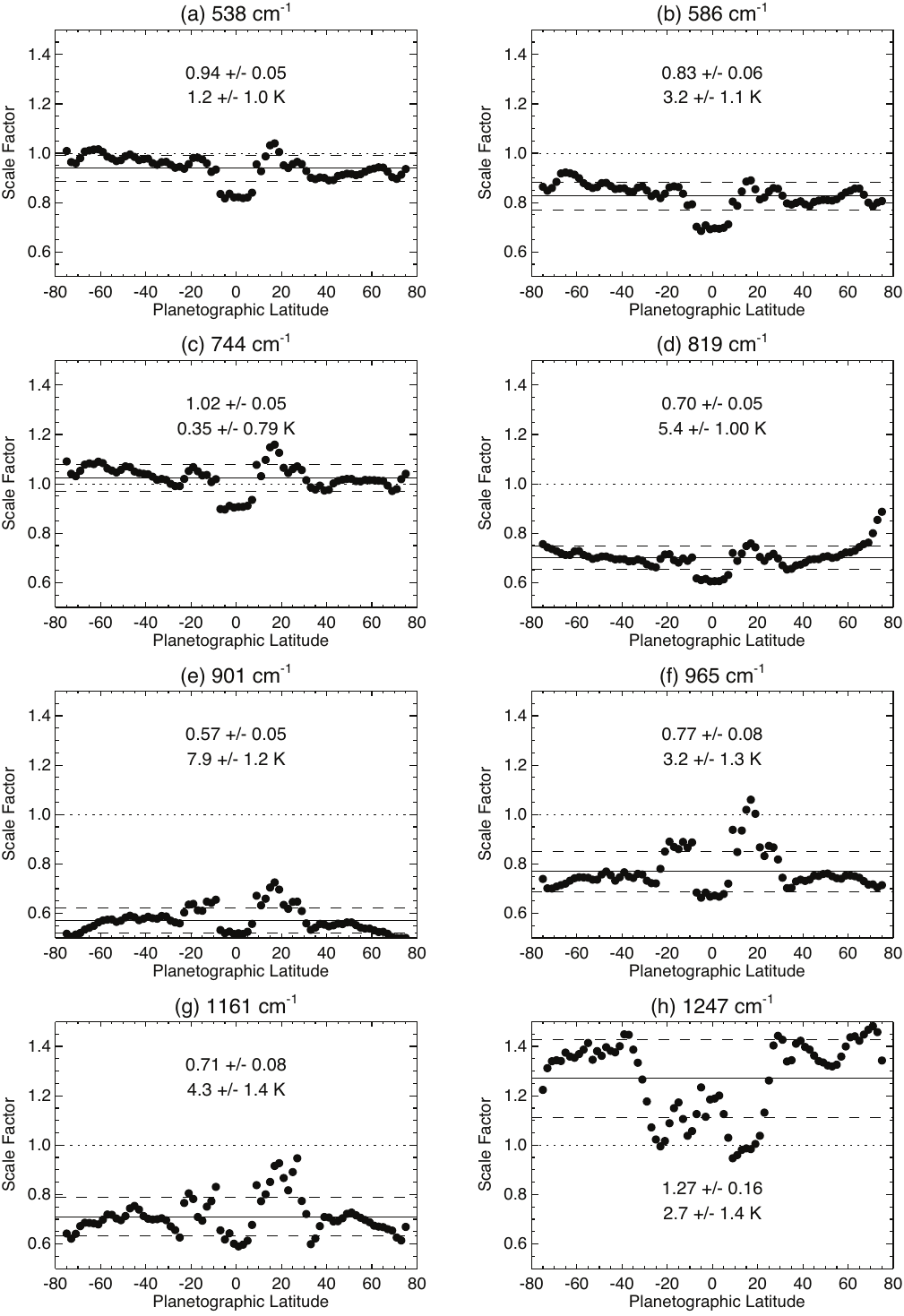}}
\caption{The latitudinal variation of the multiplicative scale factor (circles) that must be applied to the TEXES spectral radiances to match the forward-modelled radiances based on Cassini/CIRS atmospheric retrievals.  The solid and dashed horizontal line show the mean scale factor and its standard deviation, respectively (also given as the topmost number in each panel).  Overlapping with the dotted horizontal line would imply that no scaling is necessary, as is the case for the 744 cm$^{-1}$ channel.  The bottom number in each panel is the equivalent change in the temperature of a black body required to reproduce these radiance offsets.}
\label{scalefactors}
\end{centering}
\end{figure*}

\begin{figure*}
\begin{centering}
\centerline{\includegraphics[angle=0,scale=.80]{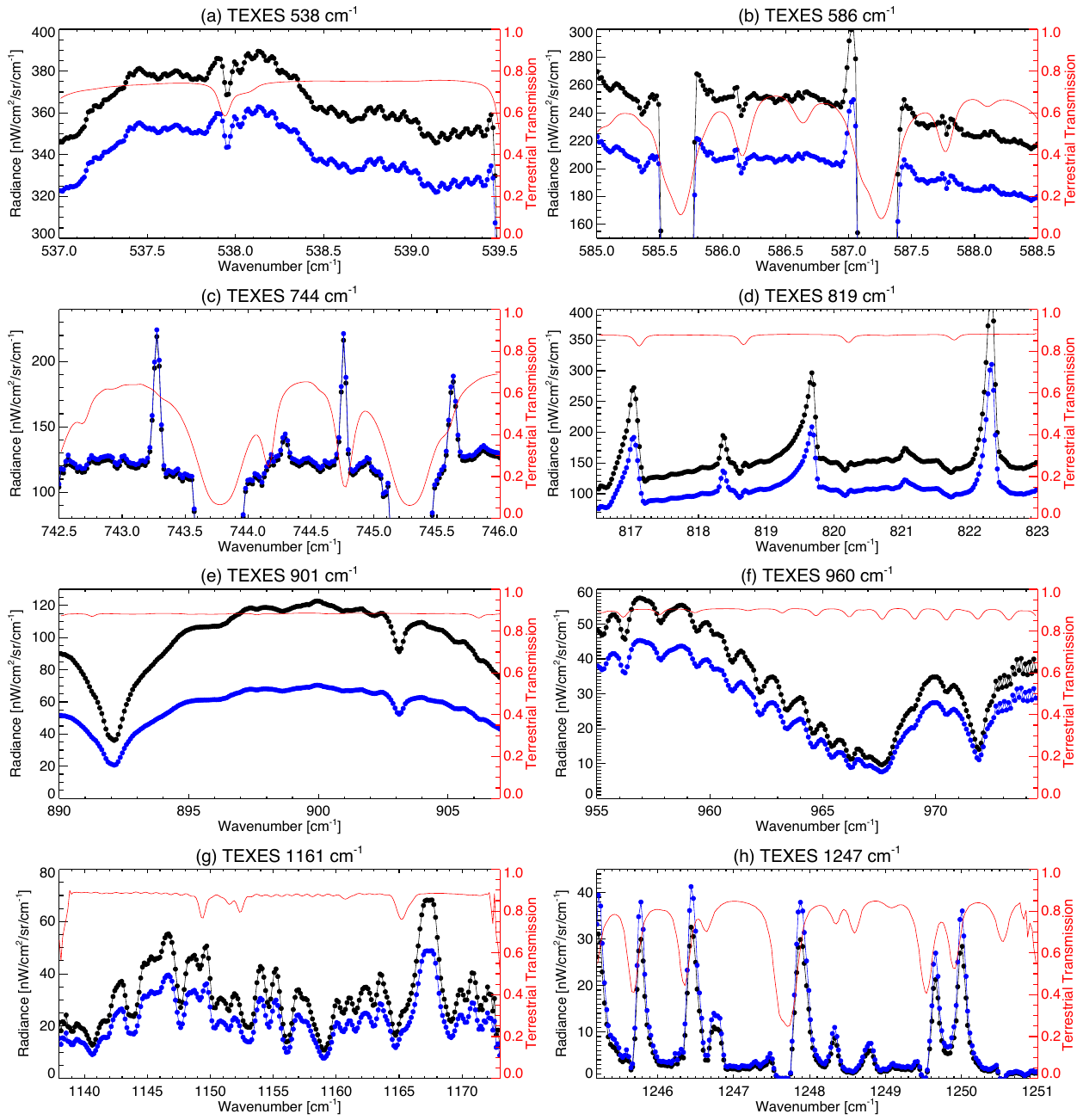}}
\caption{TEXES zonally-averaged spectra before (black) and after (blue) application of a scaling factor (Fig. \ref{scalefactors}) to correct the radiometric calibration to match a synthetic model based on Cassini (this example is for a spectrum at $15^\circ$N).  The red line shows the sky transparency (accounting for both sky transmission and losses in the telescope).  Differences are most notable in the centre of the N-band (panels d, e, f) where the sky transparency was largest.}
\label{cirs_fmodel}
\end{centering}
\end{figure*}

To assess the magnitude of the CIRS-TEXES discrepancy, we compare zonally-averaged TEXES spectra to Cassini-based forward models at every latitude.  CIRS spectra at 2.5-cm$^{-1}$ spectral resolution were extracted from the latest calibration of the CIRS database (version 4.2) for the ATMOS02A map acquired on December 31st, 2000. We replicated the work of \citet{09fletcher_ph3}, fitting temperatures, PH$_3$, NH$_3$ and aerosols as described above.  This was extended by simultaneously fitting scale factors for the hydrocarbon distributions (C$_2$H$_6$ and C$_2$H$_2$).  The resulting zonal-mean temperature, composition and aerosol opacity will be presented in Section \ref{results}, but these were used to forward model the TEXES channels, using the observing geometry (latitudes and emission angles) of the TEXES spectra themselves.  Fig. \ref{scalefactors} shows that the required radiance scale factor for each channel is largely constant as a function of latitude (the mean offsets and their standard deviations are also shown), with the exception of the tropical region (NEB, EZ and SEB) that displays the largest atmospheric variability as a function of time, and thus the largest differences between the Cassini (2000) and TEXES (2014) observations.  Fig. \ref{cirs_fmodel} compares the raw spectra both with and without the multiplicative scaling factor applied.   

The mean scale factors for the radiance in each channel are broadly consistent with the offsets observed in the zonal-mean spectral averages in Fig. \ref{meridianspx}, but are more accurate because we are able to compare data and models at the same spectral resolution.  Only those regions unaffected by telluric contamination and away from narrow Jovian emission features (i.e., the continuum between hydrocarbon lines) are used to estimate the scale factor, and we find that systematic offsets between TEXES and the CIRS forward model are detected for almost every channel.  Intriguingly, the majority of the TEXES observations overpredict the flux and need to be scaled downwards, whereas those at 1247 cm$^{-1}$ need to be scaled up to match CIRS.  We note that previously-reported TEXES observations of Jupiter \citep{14fletcher_texes} and Saturn \citep{16fouchet_texes} have shown offsets in the same direction as those found here.  

Given that Jupiter's temperature and composition are intrinsically variable, these globally-averaged scale factors are highly uncertain - indeed, low-latitude 1247-cm$^{-1}$ observations are actually consistent with CIRS-derived temperatures.  Although the scale factors may look sizeable (in one case up to 45\%), it is more meaningful to view these as brightness temperature offsets - the equivalent change in black body temperature required to reproduce the difference, and an estimate of the atmospheric temperature at the altitude probed by a particular channel.  We find that Q-band temperatures only need to be decreased by $\sim3$ K, 700-1200 cm$^{-1}$ temperatures need to be decreased by $\sim3$ to $\sim8$ K, and the 1248-cm$^{-1}$ spectrum needs to be increased by $\sim3$ K (all values given in Fig. \ref{scalefactors}).  Indeed, the largest discrepancy between CIRS and TEXES occurs where the sky is at its most transparent (i.e., $T_{sky}$ is at its smallest), at 819 and 901 cm$^{-1}$, and we would expect to derive global tropospheric temperatures some 5-8 K warmer from the TEXES data than from the CIRS data.  The lack of sky flux at these wavelengths means that the TEXES flat (the difference between the reference black body card and the sky emission) is subject to larger uncertainties in the most transparent regions.  Furthermore, the TEXES system response is smallest in the regions showing the largest offsets \citep[Fig. 5 of ][]{02lacy}.  Future observations have been scheduled to better characterise the systematic offsets in TEXES Jupiter spectra over consecutive nights as a resource for other users of TEXES.

In summary, we find systematic offsets between CIRS and TEXES that are difficult to explain without invoking global changes to Jupiter's temperatures, which we deem unlikely.  A more likely explanation is that the TEXES absolute calibration scheme becomes less accurate in regions of high terrestrial transmission, and we must account for these offsets in subsequent modelling.  Note that smaller-scale latitudinal differences between CIRS and TEXES are real, and are investigated in Section \ref{results}.

\subsection{Error Handling}

TEXES spectra are affected by sources of both random and systematic uncertainty, with the latter being the hardest to quantify.  The inspection of central-meridian radiances from individual TEXES cubes (Section \ref{obs}) revealed a high level of precision from cube to cube and night to night, with radiances reproducible from cube to cube at the 5-15\% level, although some of this can be attributed to Jupiter's own intrinsic longitudinal variability.  However, cross-comparison with CIRS observations in Section \ref{cirscomp} suggest a radiometric accuracy that varies with wavelength by up to 50\%.  The implications for this accuracy on spectral inversions will be discussed in Section \ref{results}.  

Precision uncertainties on TEXES spectra were explored in \citet{14fletcher_texes}, providing several approaches to estimating the measurement noise.  The uncertainty in a particular spectral channel varies with time (due to variable sky emission and stability during a night) and wavelength (with larger values close to telluric features).  For each cube considered in this study, we calculate the standard deviation of the radiance for each wavelength in $10\times10$ pixel squares from the four corners of the array (i.e., away from Jupiter).  We then average this over the wavelength range and compare to the radiance in the centre of the cube (i.e., the centre of Jupiter), and finally average this over all cubes in a particular spectral setting.  This allows us to estimate the background flux variation as follows:  1.1\% at 538 cm$^{-1}$, 1.5\% at 586 cm$^{-1}$, 3.8\% at 744 cm$^{-1}$, 0.6\% at 819 cm$^{-1}$, 0.3\% at 901 cm$^{-1}$, 1.2\% at 960 cm$^{-1}$, 0.9\% at 1161 cm$^{-1}$ and 8.2\% at 1248 cm$^{-1}$.  As expected, this standard deviation is smallest where the atmosphere is most transparent.  

When TEXES spectra were zonally or spatially averaged, we compare these `background uncertainties' to the standard deviation of the mean spectrum and take the most conservative as our initial estimate of the random uncertainty.  However, if this were to be applied uniformly across the TEXES spectrum, we would be assigning equal weight to both clear and telluric-contaminated regions in the inversions.  Following \citet{14fletcher_texes}, we therefore weight our measurement uncertainty using the measured sky emission spectra \citep[rather than a modelled telluric transmission following][]{05greathouse}, accounting for the Doppler shift that was applied to each pixel of the Jupiter cubes to bring the wavelengths to their rest states.  The wavelength-dependent standard deviation was inflated by a factor of two in the vicinity of strong telluric features so that they would be effectively ignored in the spectral inversion.  Furthermore, the worst-affected spectral regions were removed from the fit entirely, producing gaps in the spectral coverage of a single channel.  

Finally, although the calibration pipeline attempts to correct for the variable transmission across each channel, we found that artificial slopes were present in the 538, 586 and 744 cm$^{-1}$ spectra that were being misinterpreted by the spectral inversion algorithm.  Forward models suggest that the continuum should effectively be flat away from the H$_2$ S(1) and C$_2$H$_2$ emission features, so we empirically corrected the data by dividing through by a $\tau^n$, where $\tau$ is a normalised sky spectrum to preserve the absolute flux calibration and $n$ is a tuning parameter to flatten the spectrum.  The measurement uncertainty was increased in these regions by the difference between the original and flattened spectrum.  These random precision uncertainties were fixed for the remainder of the spectral inversions, and the influence of systematic uncertainties in accuracy are considered in the following sections.

\section{Retrieval Sensitivity}
\label{sensitivity}

Before applying the NEMESIS spectral retrieval algorithm to zonal averages and spatially-resolved spectra, we first created a zonal-mean spectrum of Jupiter's tropical domain ($\pm20^\circ$ latitude) from the TEXES cubes to demonstrate the influence of the radiometric accuracy and \textit{a priori} temperature, gas and cloud distributions on the robustness of the retrievals.  We compared our retrieved properties to those from the Cassini/CIRS ATMOS2A map (December 31, 2000) at 2.5-cm$^{-1}$ spectral resolution, which was spatially averaged in the same way as the TEXES cubes.  

\subsection{Influence of calibration uncertainties}

Assuming initially that the TEXES radiometric calibration was accurate, we attempted to fit the eight channels simultaneously by varying $T(p)$, parameterised NH$_3$ and PH$_3$ distributions, scale factors for C$_2$H$_2$, C$_2$H$_6$ and the opacity of a 800-mbar grey cloud.  The quality of the resulting fit was extremely poor.  Despite sensing similar atmospheric levels, the continuum regions of the 744 cm${-1}$ and 901 cm$^{-1}$ channels showed such inconsistency that the tropospheric temperatures that were required to match the 901 cm$^{-1}$ channel caused significant overestimation of the continuum at 744 cm$^{-1}$.  Fitting these two channels independently, we found that the 744 cm$^{-1}$ continuum required 440-mbar temperatures of 132-135 K, whereas the 901-cm$^{-1}$ channel required temperatures between 141-143K at the same altitude.  These $\sim10$-K temperature differences have an order-of-magnitude effect on the retrieved abundances of NH$_3$, which dominates the N-band absorption spectrum.  The higher the tropospheric temperature, the more ammonia was required to reproduce the spectrum.  We note that 440-mbar temperatures in the 130-135-K range (i.e., those from the 744 cm$^{-1}$ continuum) are more consistent with the independent CIRS analysis of \citep{06achterberg} and the $\sim130$-K temperature derived from the Voyager radio science experiment and Galileo probe Atmospheric Structure instrument for this pressure level \citep{92lindal_nep,98seiff}.  Furthermore, our estimates of the NH$_3$ abundance are an order of magnitude higher than those of \citet{06achterberg} if we assume the TEXES calibration to be accurate.  This qualitatively confirms the need to scale the 901-cm$^{-1}$ radiance downwards. 

As a second example, the stratospheric temperatures required to fit the 819 cm$^{-1}$ channel significantly overestimated the flux in the CH$_4$ lines at 1247 cm$^{-1}$.  Stratospheric temperature fits to the 819 cm$^{-1}$ channel (while also permitting ethane to vary) suggested 1-mbar temperatures in the 167-178 K range, depending on the latitude.  Conversely, fits to the 1247-cm$^{-1}$ CH$_4$ lines suggested 1-mbar temperatures in the 159-169K range.  The Cassini/CIRS results derived from a full 600-1400 cm$^{-1}$ spectrum favoured 1-mbar temperatures between these two extremes, qualitatively supporting a decrease in the 819-cm$^{-1}$ channel and an increase in the 1248-cm$^{-1}$ channel to make things consistent.  Although individual channels can be reproduced in isolation, this difficulty in fitting all channels simultaneously was evident everywhere,  despite our best attempts to do so after a thorough exploration of the priors.  

\begin{figure*}
\begin{centering}
\centerline{\includegraphics[angle=0,scale=.60]{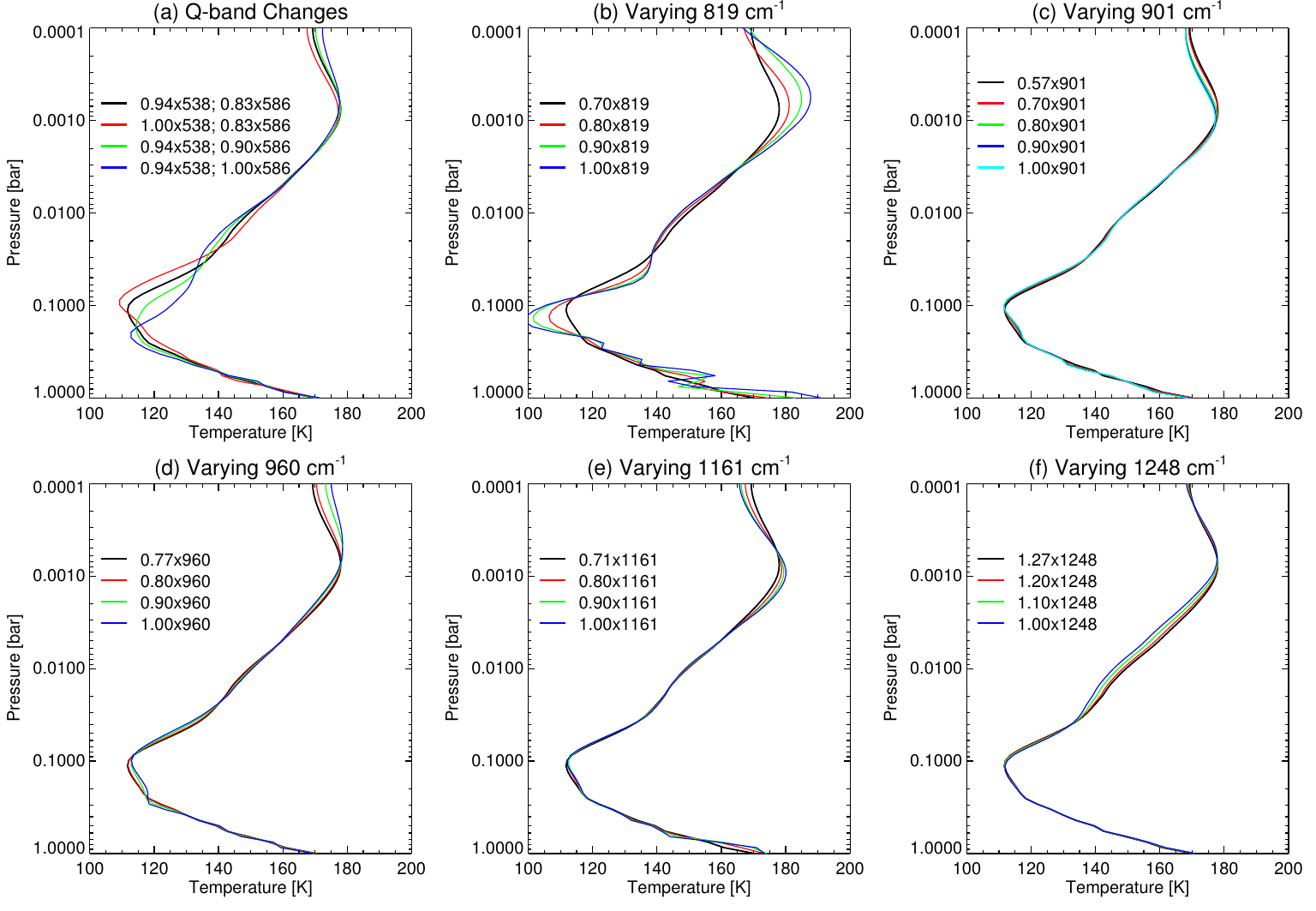}}
\caption{Demonstration of the effects on the $T(p)$ of scaling the TEXES channels relative to one another.  Panel a shows the tropopause effects of scaling the Q-band channels; panels b-f show the results from scaling the 819, 901, 960, 1161 and 1248-cm$^{-1}$ channels.  The  effects on the composition and the quality of the spectral fit are discussed in Section \ref{sensitivity}. }
\label{scaletest}
\end{centering}
\end{figure*}

Although the radiometric accuracy of Cassini/CIRS is itself subject to uncertainty and would certainly benefit from independent ground-based confirmation, the quantitative similarity between CIRS and Voyager observations of Jupiter \citep[e.g.,][]{06simon} gives us cause to trust the CIRS calibration.  We therefore apply the global-mean scale factors shown in Fig. \ref{scalefactors} globally to the TEXES data (equivalent to brightness temperature differences of 8-K in the worst case) and rerun the inversion.  This allows us to reproduce the eight TEXES channels with a temperature structure that looks reasonable and is quantitatively similar to that derived from Cassini.  In Fig. \ref{scaletest} we demonstrate how the retrieved $T(p)$ structure is modified by changing the scale factor for each channel, one at a time, to return it to the original TEXES values (i.e., a scale factor of one).  In some channels the effects are rather straightforward - scaling the 1248-cm$^{-1}$ from 1.0 to 1.3 (Fig. \ref{scaletest}f) has the effect of warming the mid-stratospheric temperatures by $\sim4$ K whilst improving the quality of the fit to the data, consistent with the expected changes in black body temperature from a 30\% change in radiance (Fig. \ref{scalefactors}). Changing the scale factor for the 901, 960 and 1161 cm$^{-1}$ channels has a subtle effect on the $T(p)$ (Fig. \ref{scaletest}c-e) but a dramatic effect on the retrieved ammonia and cloud abundances and the ability to fit the spectrum.  

More complicated effects occur when the contribution functions for a TEXES channel overlap the upper troposphere and lower stratosphere in Figs. \ref{scaletest}a-b.  The 819 cm$^{-1}$ channel senses both the troposphere and stratosphere, but retaining the unscaled TEXES data causes a failure of our model to converge to a reasonable solution - the highly oscillatory structure in Fig. \ref{scaletest}b is an example of a retrieval struggling to fit inconsistent data, resulting in the overall goodness-of-fit $\chi^2/N$ (where $N$ is the number of spectral points) increasing from $\sim0.8$ in the best case to $\sim4.7$ for the worst case.  The shape of the tropopause region demonstrates a high sensitivity to the scalings applied in the Q-band due to the limited constraint provided by the contribution functions from the other channels - hence temperatures vary wildly here in an attempt to improve the quality of the spectral fit with limited success (the $\chi^2/N$ varies between 0.8 and 0.9 for the four cases shown).   This presents a significant problem for the robustness of retrievals in this region, particularly as the Q-band spectra suffer from significant telluric contamination.  

In summary, the uncertainty in the TEXES radiometric calibration impacts the quantities that can be derived from these data, particularly given the degeneracies inherent in spectral inversion.  We proceed with the CIRS-derived scaling factors (Fig. \ref{scalefactors}), which allow us to fit the TEXES data with the best goodness-of-fit and a smooth $T(p)$ structure.


\subsection{Sensitivity to the temperature prior}
\begin{figure}
\begin{centering}
\includegraphics[angle=0,scale=.70]{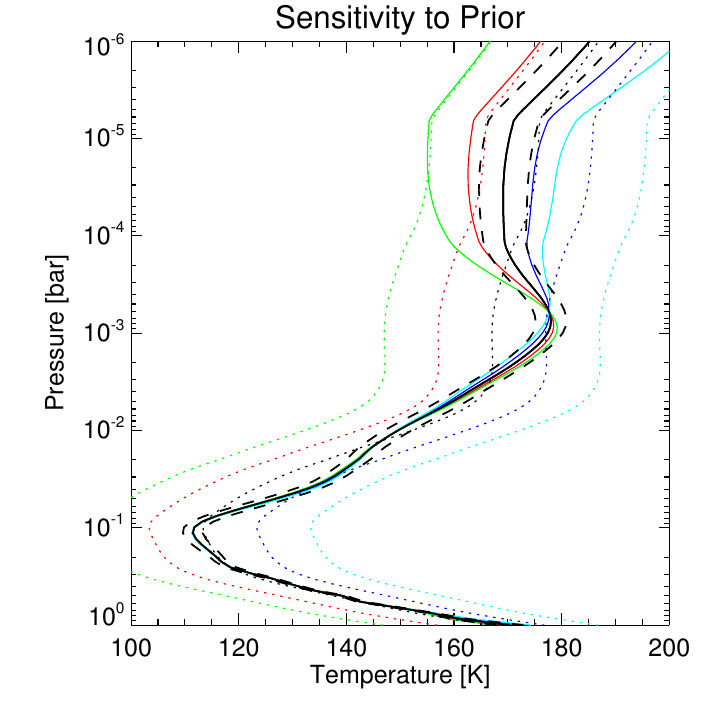}
\caption{Testing the retrieval sensitivity to the $T(p)$ prior.  Five different priors were used, varying $\pm20$ K from our reference atmosphere (dotted lines).  The corresponding retrieved $T(p)$ (solid lines) are remarkably consistent until $p<0.5$ mbar, when their deviations begin to exceed the formal uncertainty on the optimal estimate (dashed lines). }
\label{Tprior}
\end{centering}
\end{figure}

Fig. \ref{Tprior} shows the sensitivity of the retrieved temperature structure in Jupiter's tropics to changes in our \textit{a priori} thermal structure.  Five different priors were used, varying $\pm20$ K from the reference atmosphere described in Section \ref{analysis}.  The retrieved $T(p)$ remains consistent throughout the troposphere and lower stratosphere, only beginning to differ significantly in the mid-stratosphere near 1 mbar.  For $p<0.5$ mbar the retrieved $T(p)$ differences begin to exceed the uncertainty on our nominal retrievals, implying that even with the high spectral resolution of TEXES we still have a significant sensitivity to the prior in the upper stratosphere.  Note that the uncertainties shown in Fig. \ref{Tprior} are the formal errors on the optimal estimate derived from the measurement, smoothing and \textit{a priori} error covariance matrices \citep[equation 22 of][]{08irwin}, but they underestimate the true uncertainty shown in Fig. \ref{Tprior} for the upper stratosphere.  

As ethane and acetylene emission bands both have multi-lobed contribution functions that probe these low pressures, this will have implications for our ability to determine hydrocarbon abundances.  For the five cases in Fig. \ref{Tprior}, the scaling factors for the prior C$_2$H$_2$ and C$_2$H$_6$ abundances vary by 40\% and 6\%, respectively.  Acetylene's strong dependence on the upper atmospheric temperature is unsurprising given the high-altitude peaks of the contribution functions in Fig. \ref{contrib}, whereas ethane senses the deeper stratosphere where temperatures are better constrained.  The quality of the fits varies from $\chi^2/N=0.79-0.81$ for this particular tropical spectrum, with marginally better fits for the warmest upper stratosphere.  Repeating this test for a tropical-mean of the CIRS 2.5 cm$^{-1}$ observations shows exactly the same problem, with $T(p)$ profiles becoming dependent on the prior for $p<0.2$ mbar, and corresponding uncertainties in the C$_2$H$_2$ and C$_2$H$_6$ abundances of 25\% and 3\%, respectively.  The hydrocarbon abundances are strongly sensitive to this upper atmospheric temperature uncertainty, so we must identify some way to constrain the prior using previous measurements. We note that our nominal prior has a 1-mbar temperature of 167 K \citep[consistent with the estimate of 168-K from][]{92lindal_nep} and that \citet{98seiff} showed a quasi-isothermal structure up to $p<2 \mu$bar, consistent with the warmer retrievals in Fig. \ref{Tprior}.  However, this is only representative of one location on Jupiter, and higher spectral resolutions (with more high-altitude information content) will be needed to properly constrain temperatures and acetylene abundances in Jupiter's upper atmosphere for $p<1$ mbar.

\subsection{Sensitivity to the hydrocarbon priors}

\begin{figure*}
\begin{centering}
\centerline{\includegraphics[angle=0,scale=.70]{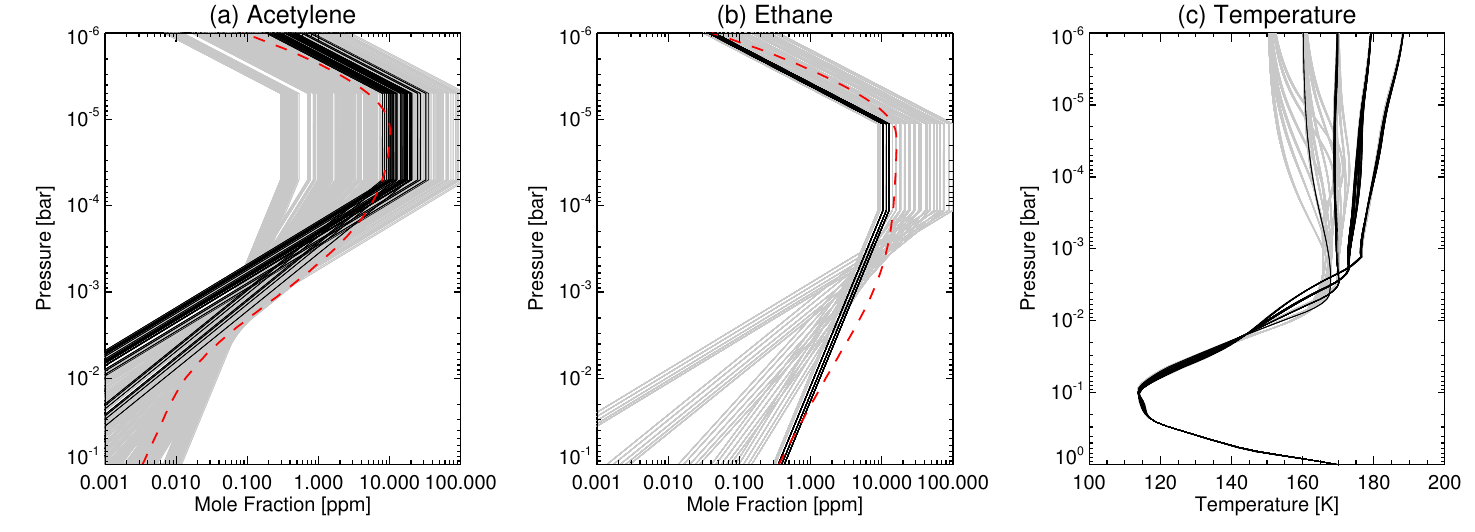}}
\caption{Testing sensitivity of TEXES inversions to the chosen priors for (a) acetylene, (b) ethane and (c) stratospheric temperature, as described in the main text.  Grey lines in panels (a) and (b) show vertical profiles scaled during the retrieval to fit the tropical-mean spectrum; black lines show those models that reproduce the respective emission feature to within a $\Delta\chi^2/N=1$. The red dashed lines are low-latitude averages of the results of \citet{07nixon} for comparison.  Grey lines in panel (c) are the corresponding temperature profiles derived from the data for five different priors; the black lines are those models which reproduce the full spectrum within a $\Delta\chi^2/N=1$. }
\label{stratos_test}
\end{centering}
\end{figure*}

With this uncertainty in the vertical $T(p)$, are we particularly sensitive to the choice of prior for the hydrocarbon distributions?  Based on \citet{07nixon}, the abundances of C$_2$H$_2$ and C$_2$H$_6$ peak in the 15-20 $\mu$bar region (as shown in Fig. \ref{stratos_test}), where our thermal uncertainties are extremely large.  To explore these uncertainties, we constructed a grid of stratospheric temperature and hydrocarbon priors in the following way:  (a) using isotherms in the upper stratosphere between 150 and 190 K, smoothly connecting to our nominal temperature prior in the lower stratosphere; (b) an ethane distribution with zero abundance for $p>100$ mbar, a range of fractional scale heights for $100>p>0.1$ mbar; constant mole fractions of 0.5-5.0 ppm between $0.1>p>0.01$ mbar; and a decline with altitude for $p<0.01 mbar$; and (c) an acetylene distribution with similar parameters but with a higher-altitude region of constant abundance between $0.05>p>0.005$ mbar.  With five variables defining the temperature and hydrocarbon priors, we reran the tropical retrievals for both CIRS and TEXES over a thousand times, recording the best-fitting 2-mbar abundance and the goodness-of-fit to the 744, 819 and 1247-cm$^{-1}$ channels.  The scaled profiles are shown in Fig. \ref{stratos_test}.

This experiment demonstrated that both CIRS and TEXES spectra contain information on the vertical distributions of temperature and hydrocarbons, with some priors leading to better fits than others.  The scatter in the best-fitting C$_2$H$_2$ profiles (Fig. \ref{stratos_test}a) is larger than those for C$_2$H$_6$ (Fig. \ref{stratos_test}b), given that only three narrow acetylene lines are observed in the TEXES channels.  The retrieval is still able to converge on a vertical hydrocarbon profile even though the upper atmospheric temperatures are uncertain (Fig. \ref{stratos_test}c).  All of the profiles, despite radically different abundances in the upper atmosphere, converge in the 0.1-5.0 mbar region where TEXES is most sensitive.  In this tropical-mean spectrum, we found 2-mbar acetylene abundances between 0.02 ppm and 0.07 ppm (column densities integrated for $p<10$ mbar of $2.5-4.5\times10^{15}$ molecules/m$^2$) and 2-mbar ethane abundances between 1.5-2.5 ppm ($1.9-2.3\times10^{16}$ molecules/m$^2$ at $p<10$ mbar), depending on the vertical profile and retaining only those models that reproduced the data within $1\sigma$.   These abundances fall within the range of previous studies \citep[see the excellent summary in Fig. 1 of][]{13zhang}.  The most radical deviations of the abundance profiles from previous work failed to reproduce the data satisfactorily.   

Unfortunately, even these large error ranges are underestimates for TEXES, given the lack of $T(p)$ information at high altitude and the uncertainty in the radiometric scaling.  If the TEXES calibration were completely accurate, then the spectral resolution used in this study is sufficient to provide information on the vertical stratospheric structure.  But further experimentation showed that the favoured prior was extremely sensitive to our choice of radiometric scaling.  To make the problem tractable in the absence of other constraints on the upper atmosphere, we chose to use hydrocarbon priors based on photochemical modelling \citep{05moses_jup}, which were themselves based on $T(p)$ profiles similar to those used in our notional prior.  These have been previously validated against CIRS spectra at higher spectral resolution \citep{07nixon, 13zhang}.  We caution the reader that alternative vertical distributions also produce acceptable fits to the TEXES and CIRS data, and that our technique of scaling these profiles would be unable to distinguish between a change in abundance at one altitude, a change in the vertical abundance gradient, or a change in upper atmospheric temperature.  Our inversions therefore assume that horizontal temperature changes at microbar pressures mirror those at millibar pressures (from smooth relaxation to the upper atmospheric prior set by Voyager and Galileo radio science experiments) and that the hydrocarbon profiles have the same vertical shapes everywhere.  Ultimately a combination of radiative and photochemical models will be required to set better priors for temperature and hydrocarbons in the upper atmosphere to break this extremely challenging degeneracy.




\subsection{Sensitivity to tropospheric priors}

\begin{figure*}
\begin{centering}
\centerline{\includegraphics[angle=0,scale=.70]{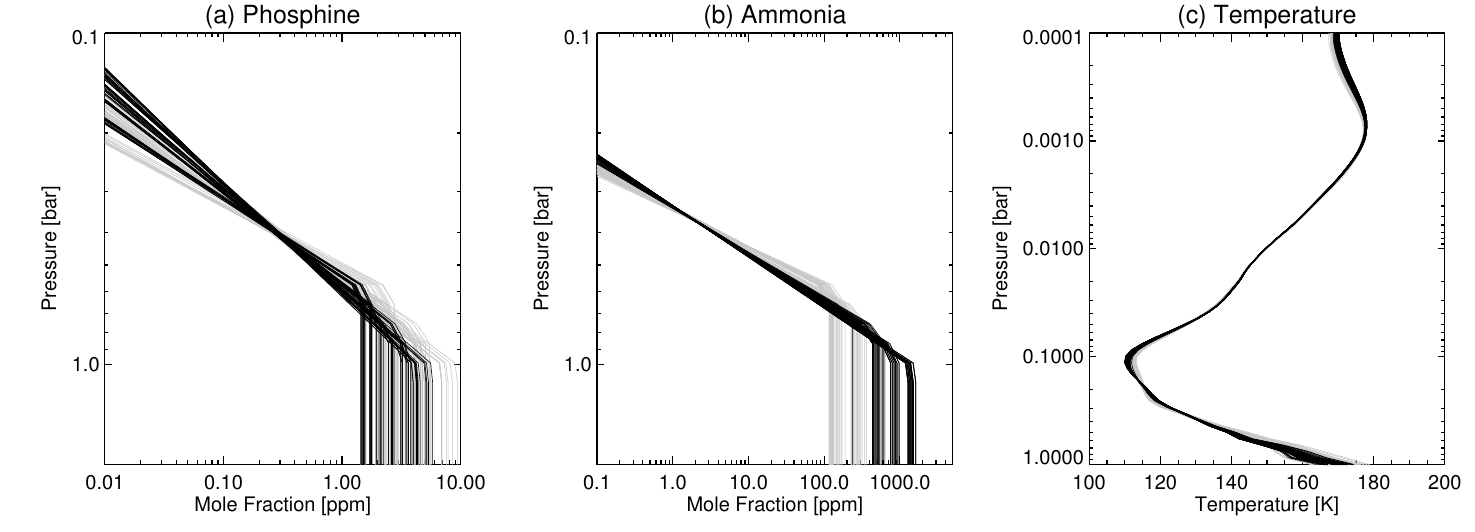}}
\caption{Testing the sensitivity of TEXES inversions to the chosen priors for vertical distributions of (a) PH$_3$ and (b) NH$_3$, and (c) the corresponding effects on the $T(p)$ retrieval. Models fitting the data within a $\Delta\chi^2/N=1$ are shown as black lines, models with poorer fits are shown as the grey lines.  }
\label{tropos_test}
\end{centering}
\end{figure*}

Degenerate solutions also exist when we consider the various contributors - temperature, aerosols and the vertical distributions of ammonia and phosphine - that shape spectra in the 819, 901, 960 and 1161 cm$^{-1}$ channels.  Adopting a similar strategy to those employed in the stratosphere, we explored this degeneracy via a grid, allowing the PH$_3$ and NH$_3$ transition pressures $p_0$ and the cloud base pressure to vary between 600 and 1000 mbar; the cloud scale height to vary between 0.1 and 0.5 (i.e., compact or extended); and repeated this for two cloud cross-section models: (i) a spectrally-uniform absorber, and (ii) a cloud of spherical NH$_3$ ice particles \citep{84martonchik} with a standard gamma distribution of mean particle radii 10 $\mu$m and variance 5 $\mu$m \citep[i.e., large particles consistent with][]{09fletcher_ph3}.  We then allowed the NH$_3$ and PH$_3$ deep abundances and fractional scale heights, plus the temperatures and opacity of our aerosol layer, to vary freely during the tropical-mean retrieval, and the resulting NH$_3$, PH$_3$ and temperature profiles are shown in Fig. \ref{tropos_test}.

We found negligible difference in the fitting quality between the compact and extended aerosol layers, limited sensitivity to the base pressure of the clouds themselves, and no difference in fitting quality between the grey cloud or large NH$_3$ ice particles.  The use of the extended cloud favoured $T(p)$ profiles (Fig. \ref{tropos_test}c) with warmer 1-bar temperatures ($\sim172$ K) whereas the compact cloud favoured $\sim161$ K \citep[the Galileo probe measured $\sim166$ K at the same altitude,][]{98seiff}.  Given the broad-band effect of aerosol opacity on Jupiter's spectrum, a wider spectral range would be needed to provide new constraints on the upper tropospheric aerosols, so we use a $p=800$ mbar compact cloud of large NH$_3$ ice spheres for the remainder of this study, representative of previous literature in this range \citep{04wong,05matcheva,06achterberg,09fletcher_ph3}.  This choice has very little effect on the 440-mbar temperatures and gas abundances, but a more substantial impact near 800 mbar - temperatures here are $\sim2$K warmer for higher-altitude cloud bases, but this is smaller than the $\sim10$ K scatter in temperatures due to the poor constraint on the deep ammonia abundance.  This reflects the substantial degeneracy between all three parameters (temperatures, aerosols and gas abundances) at $p>800$ mbar.

Retrieved vertical profiles of ammonia and phosphine (Fig. \ref{tropos_test}a-b) overlap in the 400-600 mbar range despite large differences in the location of the profile transition pressures $p_0$ and the deep mole fraction $q_0$.  The sensitivity to the deep abundances is limited in the TEXES data, resulting in the scatter of results spanning an order of magnitude for both species.  For ammonia, the best-fitting profiles had $p_0$ near 800 mbar, which is also consistent with the altitude of the putative NH$_3$ condensation cloud.  At higher pressures, the Galileo probe indicated a deep NH$_3$ mole fraction of $570\pm220$ ppm for $p>8$ bar \citep{04wong}, but Jupiter observations at microwave wavelengths support a depletion for $p<4$ bar to reach 100-200 ppm levels in the 1-2 bar region \citep{01depater,05showman}.  The deep abundances estimated by TEXES fall between these two extremes.  At higher altitudes, the TEXES fits support a steep decline in NH$_3$ to $\sim5$ ppm near 440 mbar, consistent with the $2-10$ ppm range reported by \citep{06achterberg}.  For the remainder of this study, we fix the NH$_3$ $p_0$ to 800 mbar and vary both the deep abundance and fractional scale height to fit the data.

The deep abundance of PH$_3$ is poorly constrained by the TEXES data.  The PH$_3$ profiles in Fig. \ref{tropos_test}a demonstrate that the fitting quality is only weakly sensitive to $p_0$, with values in the range 600-800 mbar reproducing the data within $1\sigma$ and a best-fit for $p_0=750$ mbar.  PH$_3$ abundances for the best-fitting models all overlap near 400 mbar where TEXES has the most sensitivity, with mole fractions in the range 0.35-0.45 ppm depending on the choice of vertical profile.  If we fix $p_0$ to 750 mbar, we derive deep abundances that are consistent with the $q_0\sim2$ ppm estimated for $p>1$ bar by previous mid-infrared studies \citep[see][and references therein]{09fletcher_ph3}, but larger than estimates of $q_0\sim0.7$ ppm using the deeper-sounding 5-$\mu$m window \citep[][using the same spectral inversion techniques]{15giles}.  Resolving this discrepancy requires simultaneous modelling of both the 5- and 10-$\mu$m PH$_3$ bands and is beyond the scope of the current study, so we fix the PH$_3$ to $q_0=2$ ppm for $p>750$ mbar for the remainder of this study.  

In summary, despite the excellent spatial and spectral resolution of the TEXES Jupiter dataset, one significant challenge hampers its analysis - the radiometric calibration.  If the calibration of the eight channels were accurate, then the exploration of parameter space described above would have provided some insight into the vertical distributions of temperature, hydrocarbons, tropospheric gases and aerosols.  Instead, we have systematically tuned the \textit{absolute} abundances and temperatures to broadly reflect previous investigations.  We are now able to explore \textit{relative} spatial variability in each of these properties in the next Section, but with the caveat that systematic uncertainties are large.

\section{Results and Discussion}
\label{results}

In this Section, we present a comparison of Jupiter's temperatures, composition and aerosol opacity from both CIRS (2000, $L_s=110^\circ$) and TEXES observations (2014 $L_s=175^\circ$).  Zonal-mean spectra were computed from all TEXES and CIRS data on a $2^\circ$ latitudinal grid with a $4^\circ$ width.  Spatially resolved spectra are computed on the same latitudinal grid, but with a longitudinal step of $2^\circ$ and a width of $4^\circ$, resulting in approximately 11,000 spectra for a global map between $60^\circ$N and $60^\circ$S.  For the zonal-mean spectra we retrieved vertical temperature profiles at every location along with (i) the optical depth of the 800-mbar compact cloud of 10-$\mu$m radius NH$_3$ ice spheres; (ii) the scale height for PH$_3$ above a well-mixed mole fraction of 2 ppm for $p>750$ mbar; (iii) the deep mole fraction and fractional scale height for NH$_3$ with a transition pressure of $p=800$ mbar; and (iv) scale factors for low-latitude mean profiles of C$_2$H$_2$ and C$_2$H$_6$ from \citet{07nixon}.  The retrieval strategy for the spatially-resolved maps was similar, except that we simply scaled a low-latitude mean of the PH$_3$, NH$_3$, C$_2$H$_2$ and C$_2$H$_6$ profiles derived from the zonal-mean spectra.  We caution the reader that the choice of temperature and gaseous priors does indeed influence the retrieval, and that alternative distributions are often able to reproduce the data equally well.  In particular, the hydrocarbon vertical gradients are held constant and all variations are assumed to be horizontal.  This is the first time that global maps of these species have been presented from mid-infrared spectroscopy.

\subsection{Temperatures}
\label{res_temp}
\begin{figure*}
\begin{centering}
\includegraphics[angle=0,scale=.70]{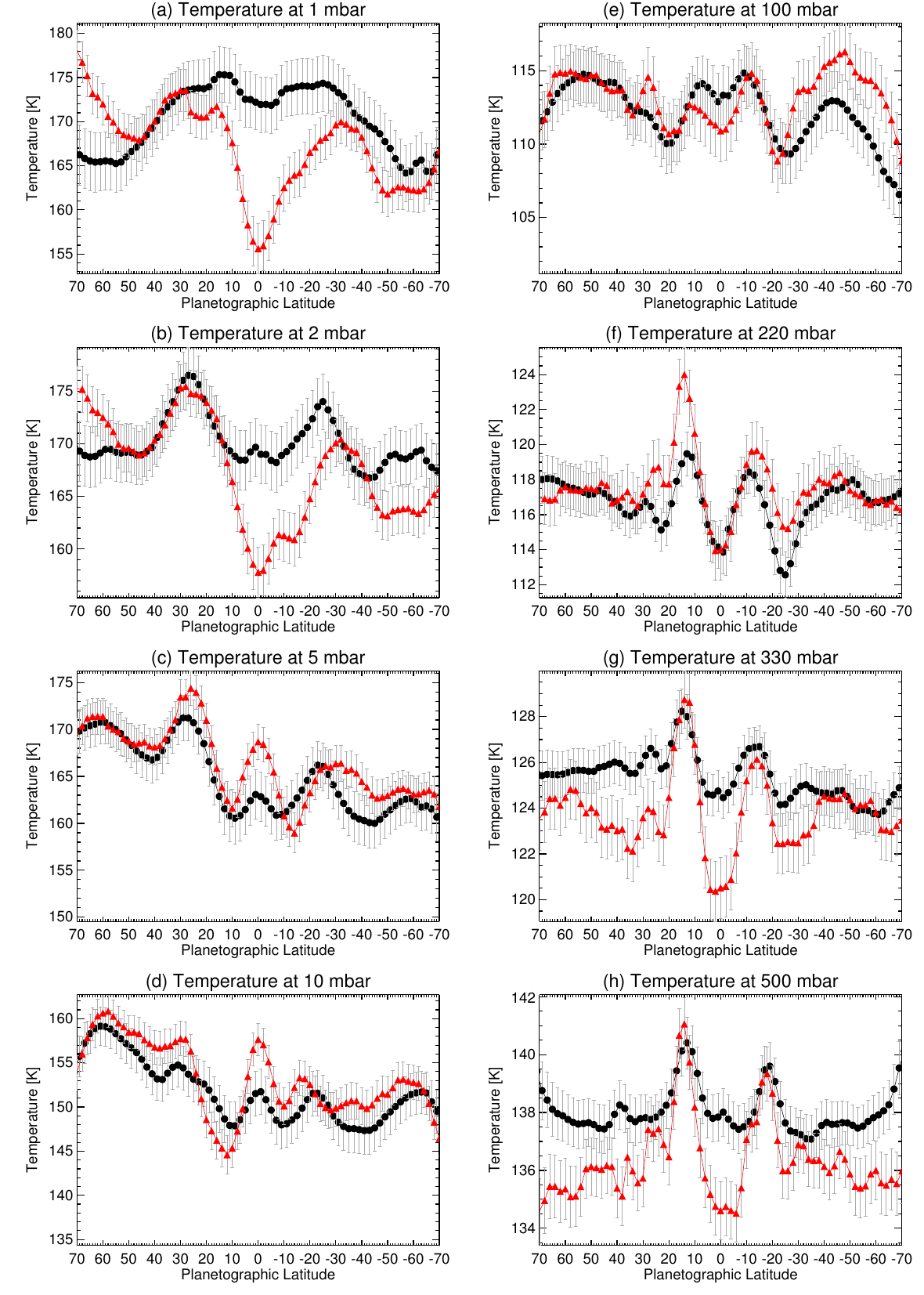}
\caption{Comparison of the zonal-mean temperatures between TEXES (black circles) and CIRS (red triangles) at pressure levels in the stratosphere (left) and troposphere (right).  The grey error bars represent retrieval uncertainties, comprising random measurement error, smoothing error and degeneracies.  Systematic calibration uncertainties are not included in these error bars.  }
\label{zonalT}
\end{centering}
\end{figure*}

\begin{figure}
\begin{centering}
\includegraphics[angle=0,scale=.75]{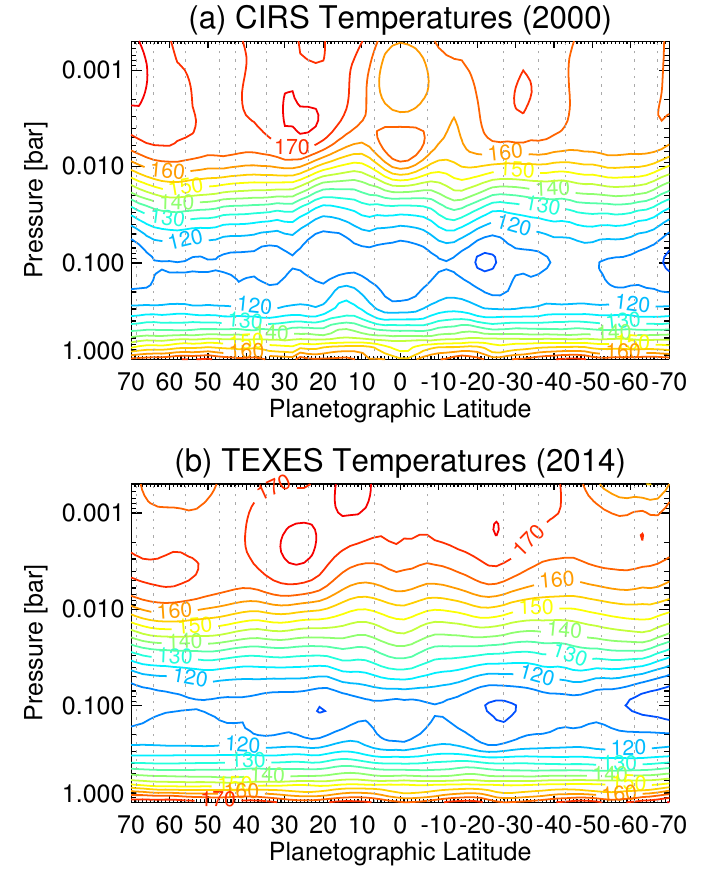}
\caption{Comparison of the zonal-mean temperature contours between TEXES and CIRS.  Vertical dotted lines show the location of prograde jets as measured by Cassini \citep{03porco}. Contours have a 5-K separation. }
\label{Tcontour}
\end{centering}
\end{figure}

\begin{figure*}
\begin{centering}
\centerline{\includegraphics[angle=0,scale=.9]{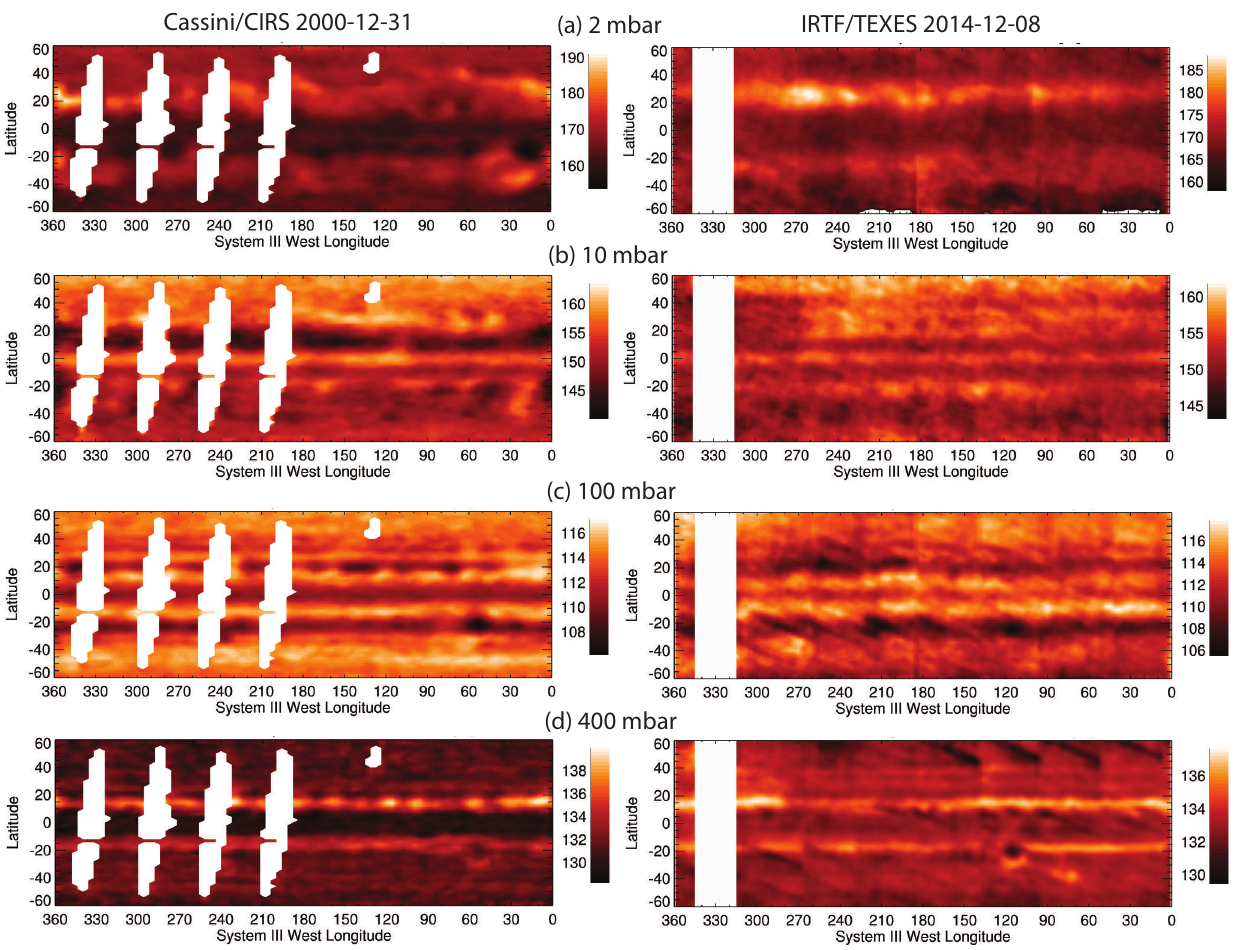}}
\caption{Global maps of Jupiter's temperatures at four pressure levels: 2 and 10 mbar in the stratosphere; 100 and 400 mbar in the troposphere.  We compared Cassini/CIRS measurements from December 2000 on the left to IRTF/TEXES measurements from December 2014 on the right.  Mean retrieval uncertainties are 2.3 K, 2.0K, 1.9 K and 1.0 K for the four pressure levels from top to bottom, excluding systematic uncertainties due to the radiometric calibration of TEXES.  Gaps in spatial coverage are shown as white.  Diagonal striations, particularly in the 400-mbar TEXES map, are due to variable striping on the detector that could not be fully removed.}
\label{mapT}
\end{centering}
\end{figure*}

Although several previous authors have used the December 2000 CIRS 2.5 cm$^{-1}$ observations of Jupiter to determine the zonal-mean temperatures \citep[e.g.,][]{04flasar_jup,05matcheva,06simon,06achterberg,09fletcher_ph3}, we repeated these measurements to ensure that the retrieval parameterisations were identical between the TEXES and CIRS inversions.  Fig. \ref{zonalT} compares Jupiter's zonal-mean temperatures in 2000 and 2014 at five different pressures; Fig. \ref{Tcontour} shows latitude-pressure zonal cross-sections of the temperature field; and Fig. \ref{mapT} compares global temperature maps.  These results clearly demonstrate that ground-based TEXES observations are able to match the large-scale structures observed by Cassini.

\subsubsection{Tropical and Temperate Domains}
The tropical domain shows the largest temperature contrasts between the warm NEB and SEB and the cold equatorial zone.  The NEB generally appears warmer than the SEB at 250-500 mbar, but the contrast between the two belts in Fig. \ref{zonalT}f-g has varied with time, from $\sim2-4$ K at 250 mbar in 2000 to $\sim1$ K in 2014.  The shifting contrasts are unsurprising given the longitudinal variability observed in both belts in Fig. \ref{mapT}.  Rossby waves are thought to modulate the temperatures of these belts \citep{89deming, 89magalhaes, 94orton, 96harrington, 97deming}, although their connection to the equatorial Rossby wave responsible for the `5-$\mu$m hotspots' \citep{90allison, 98ortiz, 00showman} has so far remained unclear.   By combining TEXES observations in the M, N and Q-bands, Fig. \ref{texesNEB} clearly indicates that the 5-$\mu$m hotspots are associated with thermal perturbations throughout the upper troposphere, although the correspondence is hardest to see at the Q-band wavelengths.   We return to this problem, and the temperature distribution in the vicinity of the Great Red Spot, in Section \ref{closeup}.

The temperature at the equator, and therefore the contrast between the EZ and neighbouring belts, has also varied with time.  The TEXES data indicate a warmer EZ at all longitudes and tropospheric pressures (the reduced latitudinal contrast can be seen in Fig. \ref{mapT}), $\sim2-4$ K warmer than that observed by CIRS in Fig. \ref{zonalT}g.  Similar levels of equatorial variability near 250 mbar were observed by \citet{94orton} and \citet{06simon} prior to Cassini's flyby, and this variability was tentatively associated with equatorial brightening events, suggestive of fresh updrafts and cooling.  Indeed, the EZ was the coldest region on the planet in 2000 (115 K at 250 mbar), whereas the STropZ was coldest in 2014. Away from the tropics, the thermal data lacks the spatial resolution to identify temperature contrasts across the numerous jets in the temperate domain ($\pm25-65^\circ$) observed by Cassini \citep{03porco}.  The narrow belts and zones of the temperate domain are visible in the raw data (Fig. \ref{texesmaps}), as well as the variability associated with the brown barges described in Section \ref{image_inspection}, but these are not captured on our $2\times2^\circ$ spatial grid in Fig. \ref{mapT}.

\subsubsection{Polar Vortices}
There is no notable change in the deep temperatures at $p>200$ mbar as we enter the polar domains (beyond the highest-latitude prograde jets at $66.8^\circ$S and $68.7^\circ$N).  However, if we move higher into the tropopause region (80-100 mbar) and the stratosphere $p>5$ mbar), we find that the temperature drops significantly poleward of $\pm60^\circ$ in both the CIRS and TEXES data.  Large cold anomalies sit over both polar regions in the 5-100 mbar range, implying a strong negative shear on the prograde polar jets in both hemispheres.  This behaviour was also evident in thermal retrievals from Voyager and Cassini by \citet{06simon}, who discussed the presence of cold polar vortices and their implications for middle atmospheric circulation and composition.  Jupiter's polar regions are characterised by the regular belt/zone structure giving way to smaller-scale turbulence, and by a sharp rise in the number density and optical thickness of stratospheric aerosols in the 10-20 mbar region \citep{13zhang_aer}.  These aerosols contribute to the radiative budget \citep{15zhang} and could serve to enhance radiative cooling over the poles without the need to invoke large-scale upwelling and adiabatic cooling.  Moving into the mid-stratosphere ($p<5$ mbar) there is no evidence for the cold polar vortices in Fig. \ref{zonalT}a-b - instead, temperatures appear to rise poleward of $\pm60^\circ$, potentially in association with the regions of auroral heating \citep[see][for an exploration of the temperatures at these high latitudes]{15sinclair_dps}. 

\subsubsection{Equatorial Stratosphere and QQO}
The equatorial tropopause and stratosphere are strongly influenced by Jupiter's quasi-quadrennial oscillation (QQO), a regular $\sim4.2$-year cycle of changing tropical temperatures that has been compared to the Earth's quasi-biennial (26-month) oscillation (QBO) \citep{91orton, 91leovy, 99friedson, 06simon}.  These temperature changes are the result of zonal-wind reversals due to stresses imparted by upward-propagating waves, although \citet{06simon} showed that the temperature oscillations were rather more complex (a superposition of many different periods) and that the amplitude varied with time, particularly in response to the 1994 Shoemaker-Levy 9 collision.  The QQO can be seen in our CIRS inversions as a vertical chain of warm and cool airmasses, but this is much less apparent in the TEXES inversions.  In 2000, a large cool airmass at 1 mbar sat above a warm airmass at 5 mbar.  Fourteen years (3.5 QQO cycles) later the latitudinal temperature contrasts at both altitudes were much more subdued in Fig. \ref{zonalT}a-d, showing a small equatorial maximum near 5 mbar and a small equatorial minimum at 1 mbar.   This is broadly consistent with a long-term record of the QQO phase (Greathouse et al., in prep), which indicates that the equatorial stratosphere should have a local maximum at 10 mbar and a local minimum at 0.4 mbar in December 2014.  However, our moderate-resolution TEXES spectra at 1247 cm$^{-1}$ are insufficient to fully resolve the vertical structure of the QQO  - either broadband spectral coverage (like CIRS) or higher TEXES spectral resolutions are required.   On Saturn, Cassini has observed the stratospheric airmasses associated with its quasi-periodic oscillation to sink towards the tropopause over time \citep{08fouchet, 08orton_qxo, 10fletcher_seasons, 11guerlet, 11schinder, 16fletcher}, and Jupiter's QQO may be responsible for modulating the equatorial temperatures near the tropopause in Fig. \ref{zonalT}e.

\subsubsection{Mid-Latitude Stratosphere and Waves}
\label{stratos_wave}
Jupiter's mid-latitude stratospheric temperatures are largely symmetric about the equator for $p>50$ mbar, but become significantly asymmetric at higher altitudes.  In 2000, the northern stratosphere at 5 mbar was $\sim10$ K warmer at $30^\circ$N than at $30^\circ$S (Fig. \ref{zonalT}-\ref{Tcontour}), a situation also found 20 years earlier by Voyager (1979) \citep{06simon}.  The 5-mbar contrast was smaller in 2014 but still indicated warmer northern mid-latitudes.  Furthermore, the peak southern temperatures moved from $\sim30^\circ$S to $\sim20^\circ$S between 2000 and 2014.  This substantial variability is consistent with the 1979-2001 record of stratospheric temperatures from the ground \citep{94orton, 06simon}.  Given that the northern stratosphere was warmer than the south in 1979 ($L_s=190^\circ$, early northern autumn), 2000 ($L_s=110^\circ$, just after northern summer solstice) and 2014 ($L_s=175^\circ$, late northern summer), it is plausible that this is a seasonal effect due to Jupiter's $3^\circ$ obliquity.   However, the 10-mbar time series of \citet{06simon} implied that simple radiative heating and cooling could not explain the phasing of the stratospheric temperature changes.    Alternatively, the number density of stratospheric aerosols is higher in Jupiter's northern hemisphere poleward of $30^\circ$N \citep{13zhang_aer}, which could produce additional stratospheric radiative heating that contributes to the asymmetry (note that radiative simulations without stratospheric aerosols produce negligible north-south asymmetries in stratospheric temperature, S. Guerlet, \textit{pers. comms.}).   However, there appear to be no notable stratospheric aerosol enhancements at the mid-latitudes \citep{13zhang_aer}, so an alternative explanation for the $\pm15-40^\circ$ warm bands is required.

\begin{figure}
\begin{centering}
\includegraphics[angle=0,scale=.65]{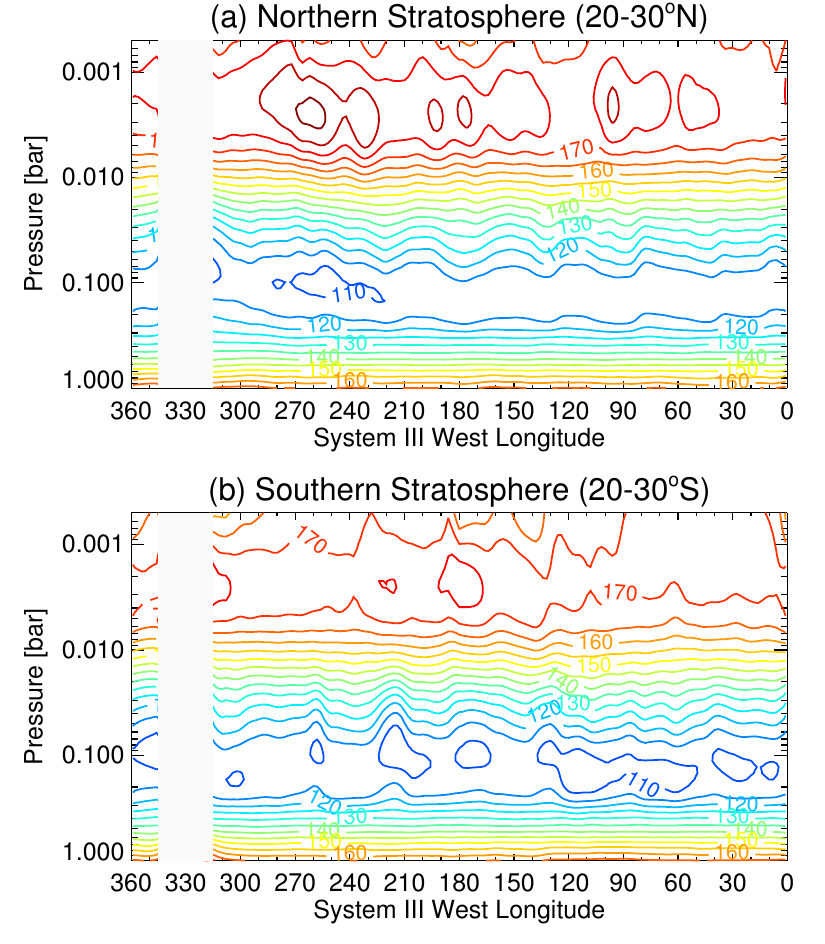}
\caption{Longitudinal cross-section of temperatures through stratospheric wave activity in Jupiter's northern and southern hemispheres observed by TEXES in December 2014.  Contours are spaced every 5 K, and we see that northern and southern 2-mbar temperatures would be near-equal without the prominent northern wave.}
\label{stratoswave}
\end{centering}
\end{figure}

The stratospheric temperature maps (Fig. \ref{mapT}) provide further insights, showing how the stratosphere is dominated by broad bands of warmer temperatures in the $\pm15-40^\circ$ range, with the north being warmer than the south in 2014.  We note, however, that this is not true at all longitudes in Fig. \ref{mapT}, with some longitudes having approximately equal temperatures at the northern and southern mid-latitudes.  This is better shown by the longitudinal temperature cross-sections in 2014 (Fig. \ref{stratoswave}), where we compare east-west cuts at $\pm20-30^\circ$ latitude.  2-mbar temperatures vary from $\sim170$ K in the quiescent stratosphere in both hemispheres up to $\sim185$ K at the peaks of the prominent northern stratospheric wave in Fig. \ref{stratoswave}a.  Both Fig. \ref{mapT} and \ref{stratoswave} demonstrate the high degree of longitudinal variability evident at Jupiter's mid-latitudes.  The temperature asymmetry appears to be driven by the wave activity present primarily in the northern hemisphere, with the warmest temperatures near $25^\circ$N (i.e., above the North Tropical Zone and North Temperate Belt) and $240-270^\circ$W in the TEXES maps. This wave is not continuous around the planet, but is evident between $150-270^\circ$W with a $15-20^\circ$ wavelength.  The CIRS stratospheric maps show similar levels of longitudinal variability in both hemispheres, but with a less well-defined wavelength. 

Ground-based observations over multiple years must be used to constrain Jupiter's stratospheric wave properties (e.g., common latitudes, wavelengths, phase speed and variability), but we speculate that the temperature asymmetry between northern and southern mid-latitudes is driven mechanically, rather than radiatively.  The ephemeral wave activity could be more common and stronger in the northern hemisphere (as it is in 2014 in Fig. \ref{mapT}), in response to asymmetries in the strength of mechanical forcing from meteorological activity in the troposphere.  Vertically propagating waves can transport energy from the deeper atmosphere, but the ability of these waves to reach the stratosphere depends on the vertical and latitudinal curvature of the background zonal-wind field (see below), which is indeed asymmetric between the two hemispheres \citep[e.g.,][]{90conrath} and could result in hemispheric differences in the prevalence of stratospheric wave activity.   We note, however, that no large-scale plume activity was spotted in visible images of either the NTropZ or NTB that would obviously be coupling tropospheric weather to stratospheric wave activity in December 2014. Alternatively, we can postulate a large-scale stratospheric circulation pattern, with subsidence and adiabatic heating being stronger at northern mid-latitudes than at southern mid-latitudes in 2000 and 2014 \citep[vertical velocities of order $5\times10^{-6}$ m/s are required based on the work of][]{06simon}.  Another alternative is that jovian mid-latitudes are controlled by an extratropical QQO, varying in step with the changes seen at the equator.  Earth's extratropical variations appear to be correlated with the equatorial QBO \citep[Section 12.5 of][]{87andrews}, although this does not necessarily imply a direct dynamical connection.  Distinguishing between large-scale meridional overturning, radiative heating and cooling, and mixing and heating via wave activity, will be a challenge for future observations and models.

\subsection{Ammonia}

\begin{figure*}
\begin{centering}
\includegraphics[angle=0,scale=.70]{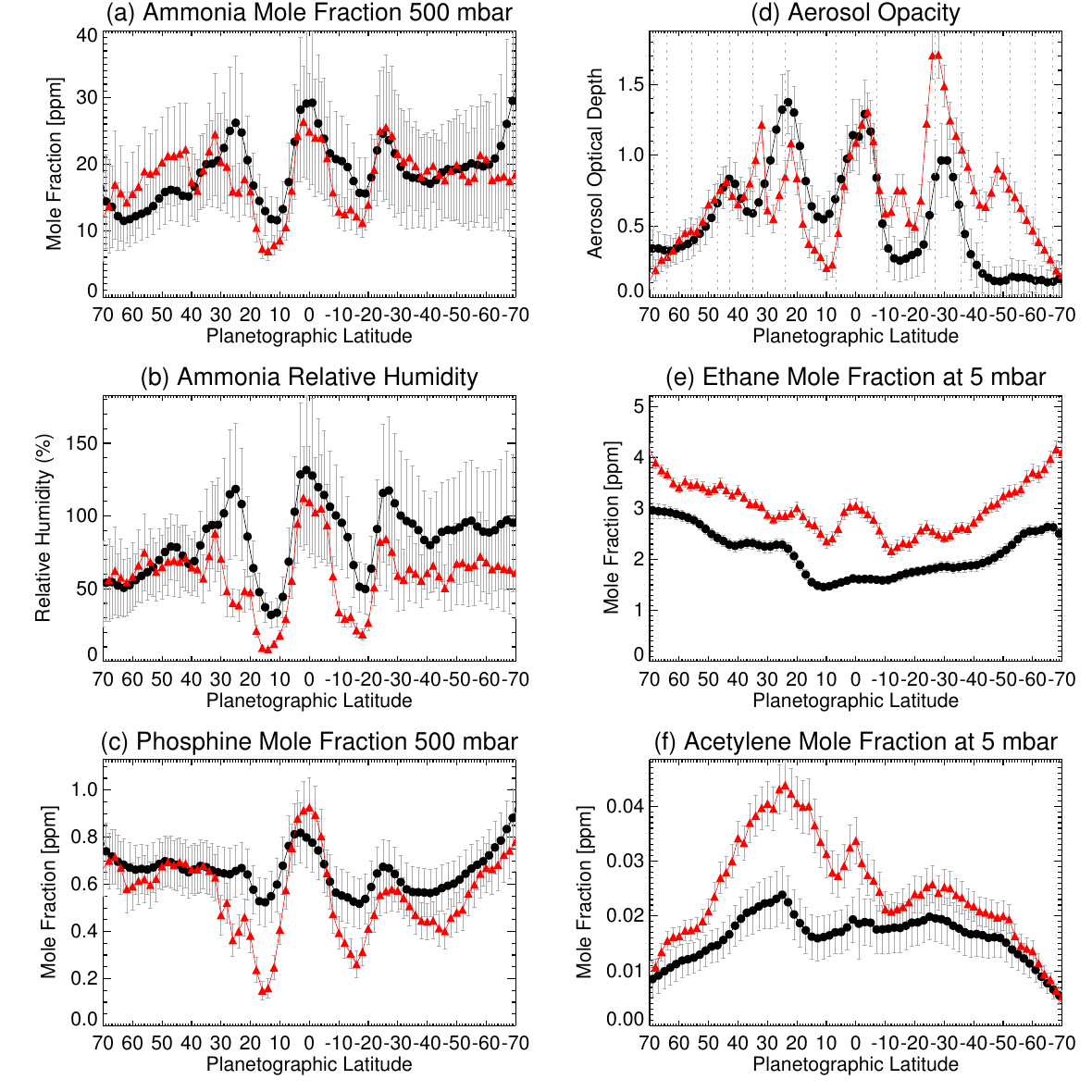}
\caption{Zonal-mean distributions of gaseous species and aerosols in the TEXES (black circle) and CIRS (red diamonds) datasets.  Ammonia (panel a), the ammonia relative humidity at 500 mbar (panel b) and phosphine (panel c) are plotted near the peak of the N-band contribution functions, as are ethane (panel e) and acetylene (panel f).  The aerosol opacity is the cumulative optical depth at 1 bar using our compact 800-mbar NH$_3$ ice cloud.  As in Fig. \ref{zonalT}, the error bars are from formal retrieval uncertainties and do not account for systematic offsets in absolute calibration.  Vertical dotted lines in pane (d) represent prograde jets as measured by \citet{03porco}.}
\label{zonal_gas}
\end{centering}
\end{figure*}

\begin{figure*}
\begin{centering}
\centerline{\includegraphics[angle=0,scale=.90]{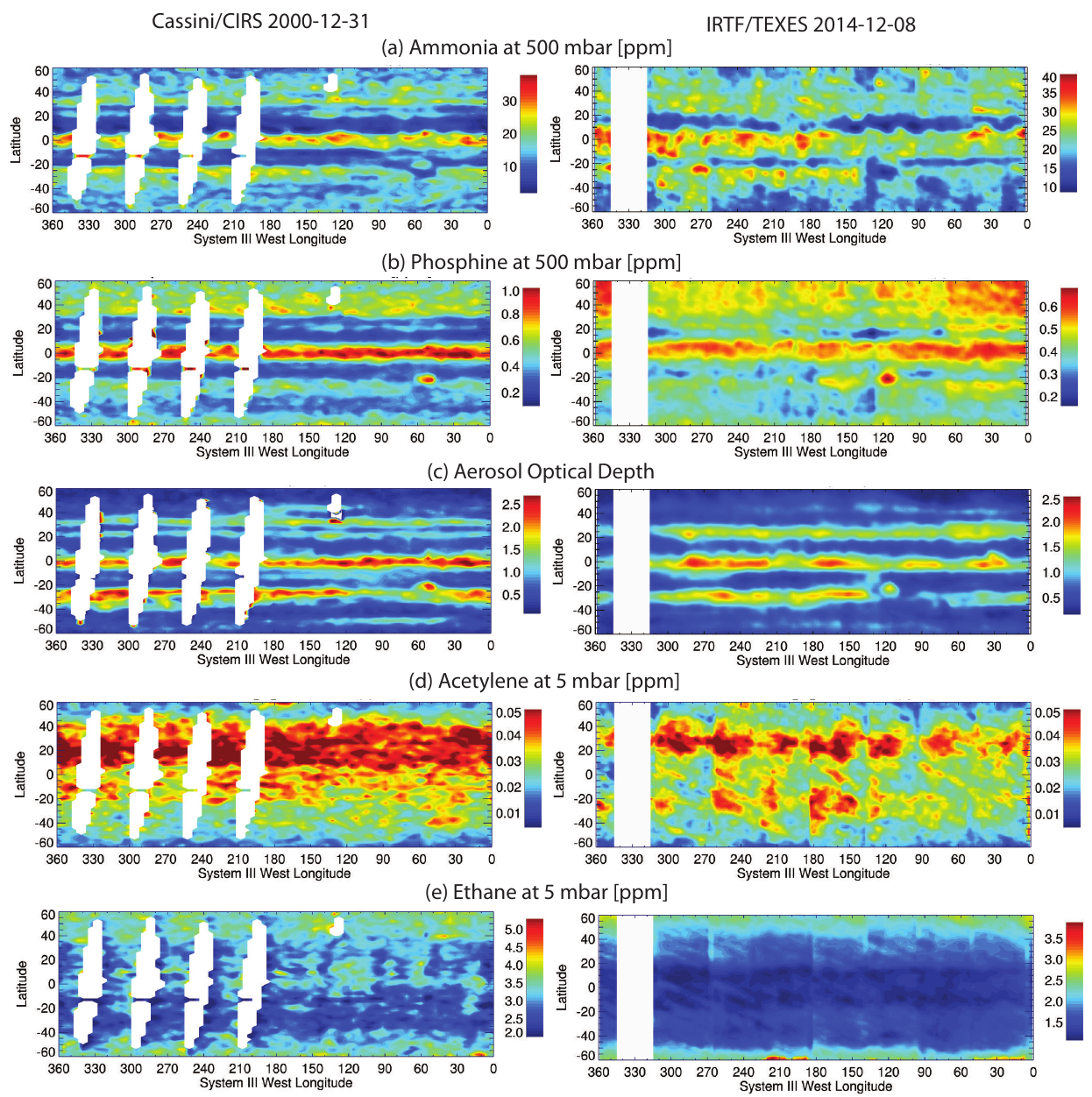}}
\caption{Spatially resolved maps of gases and aerosols derived from Cassini (December 2000) on the left and TEXES (December 2014) on the right.  A single scale factor was retrieved for each of the five categories.  Ammonia and phosphine are shown at 500 mbar, acetylene and ethane at 5 mbar, and the aerosol optical depth is a cumulative opacity based on a compact cloud at 800 mbar.  Note that the scale bars for each figure are not identical, given the different absolute abundances in Fig. \ref{zonal_gas}.  The Great Red Spot is near $50^\circ$W in 2000 and $120^\circ$W in 2014.  Missing data are shown as white gaps in each panel. }
\label{map_gas}
\end{centering}
\end{figure*}

The spatial distribution of jovian ammonia above the topmost clouds is expected to be governed by a balance between (i) vertical mixing from the deeper, well-mixed reservoir \citep{86depater, 04sault, 05showman}; (ii) condensation so that NH$_3$ follows a sub-saturated profile through the upper troposphere \citep{93carlson}; and (iii) photolytic destruction to generate dihydrazine, a potential contributor to Jupiter's upper tropospheric hazes \citep[e.g.,][]{04west}.  Previous studies of ammonia variability have come from a variety of sources, from centimetre wavelengths \citep[probing $p>1$ bar, e.g.,][]{01depater} to the thermal infrared in the 0.2-1.0 bar region \citep{00fouchet, 06achterberg, 10fletcher_grs} and near-infrared \citep[e.g.,]{98irwin}.  These studies suggest that the ammonia abundance is correlated with cloud opacity, with enhancements in zones and depletions in belts.  The CIRS and TEXES zonal-mean ammonia distribution and the corresponding relative humidity are plotted near the peak of the contribution function (500 mbar) in Fig. \ref{zonal_gas}a-b, and confirm the direct correlation with aerosol opacity.  The relative humidity varied from $\sim10$\% in the belts to $\sim125$\% in the equatorial zone, consistent with the distribution shown by \citet{06achterberg}.

The dominant characteristic of the NH$_3$ distribution in Fig. \ref{zonal_gas}a and Fig. \ref{map_gas}a is the factor of 3-5 contrast between the enhanced and supersaturated NH$_3$ of the EZ and the depletion over the SEB and NEB, demonstrating that the NH$_3$ abundance is anti-correlated with tropospheric temperatures.  This is the opposite of what we would expect from condensation alone (depletion in colder environments), and suggests the influence of dynamics.  500-mbar abundances range from $\sim5-15$ ppm over the NEB/SEB to $\sim30$ ppm at the equator, but belt/zone variations with smaller amplitudes are evident in the temperate domains (particularly in the NTropZ and STropZ).  These values are consistent with previous CIRS analysis by \citet{06achterberg} using the same dataset.  There is conflicting evidence for changes to the ammonia abundance in the polar domain, with CIRS results suggesting a gradual decline in abundance from $\pm45^\circ$ to $\pm75^\circ$ in both hemispheres, and TEXES results suggesting a slight upturn poleward of $\pm60^\circ$.  Given that sensitivity to tropospheric ammonia drops with increasing emission angle at these high latitudes, understanding the distribution of NH$_3$ at Jupiter's poles must likely await the arrival of the Juno spacecraft.

The low-latitude NH$_3$ contrasts could potentially be associated with Hadley-type circulation patterns \citep{86gierasch, 05showman}, with ascending air producing moist adiabats in the zones and transporting air from the deep reservoir of well-mixed NH$_3$ up to the cloud base.  It is interesting to note that the ammonia mole fractions deduced in the zones from CIRS and TEXES for $p>800$ mbar (i.e., below the NH$_3$ condensation cloud) are in the 400-500 ppm range in the equatorial zone, with spatial gradients similar to that seen at 500 mbar in Fig. \ref{zonal_gas}a.  These values are consistent with the maximum $\sim570$ ppm abundances inferred by the Galileo probe in the 10-bar region \citep{04wong}.  This is somewhat surprising, given that the formation of the NH$_4$SH cloud (expected to occur below the altitudes probed in the thermal infrared) would remove some of the NH$_3$ from the column.  Nevertheless, this supports the idea that large-scale upwelling can transport material from the $\sim8-12$ bar level all the way to the top-most condensation clouds in the zones.  Conversely, \citet{05showman} ruled out such direct upwelling on the basis of a global depletion of NH$_3$ at $p<4$ bar observed in radio observations \citep{01depater, 04sault}, even at the equator.  \citet{05showman} instead suggested that lateral mixing of dry air from belts and moist air from zones \citep{00ingersoll} created NH$_3$ abundances in the 1-4 bar regime that were much smaller than those in the deep reservoir.  Based on the thermal-IR inversions presented here, we do not need to invoke such a lateral circulation pattern at Jupiter's tropical latitudes.  Furthermore, we note that alternative profiles with deeper transition pressures near $p=1$ bar require even larger NH$_3$ abundances (by factors of 2-4) to reproduce the data (Fig. \ref{tropos_test}b).

The spatial distribution of NH$_3$ in Fig. \ref{map_gas}a shows that NH$_3$ does indeed form bands around the planet coinciding with the aerosol opacity, as well as significant longitudinal variability in abundance. The equatorial zone is not uniformly enhanced in NH$_3$, but both CIRS and TEXES shows peaks near $5^\circ$N that bias the zonal-mean distribution slightly northward of the equator.  These ammonia-rich plumes coincide with strong NH$_3$ depletions on the edge of the NEB that are most visible between $60-150^\circ$W in the TEXES map. We will explore their implications (and the complex NH$_3$ distribution associated with the GRS) in Section \ref{closeup}.  Both the zonal-mean and the NH$_3$ map reveal changes in the SEB between 2000 and 2014.  In 2000, the ammonia depletion over the SEB was broader in latitude, extending from $7-20^\circ$S.  But by 2014, the northern half of the SEB contained more NH$_3$, with the effect of shrinking the NH$_3$-depleted region into the southern part of the belt, $15-20^\circ$S.   A more complete time series is needed to assess whether these changes relate to the fade and revival cycle of the SEB, which recently occurred between 2009-2012 \citep{11fletcher_fade}.

\subsection{Phosphine}

Unlike NH$_3$, the distribution of PH$_3$ is not affected by condensation processes, and can instead be used to infer the strength of vertical mixing and the efficiency of photolytic destruction in Jupiter's upper troposphere. The phosphine distributions shown in Fig. \ref{zonal_gas}c confirm the findings previously reported by CIRS \citep{04irwin, 09fletcher_ph3, 10fletcher_grs}.  PH$_3$ is elevated over the EZ and depleted over the NEB and SEB, sensing the same Hadley-type circulation pattern as the NH$_3$ and aerosol distributions.  PH$_3$ shows a distinct secondary maximum near $30^\circ$S  (associated with the white STropZ) with a minimum near $45^\circ$S, that is not mirrored in the northern hemisphere.  In fact, $35-60^\circ$N represents a plateau in the zonal-mean PH$_3$ distribution that has no counterpart in the south.  This results in a decidedly asymmetric appearance in the spatial map (Fig. \ref{map_gas}b) with more PH$_3$ in the northern hemisphere.  This northern plateau and the southern contrasts are evident in both the TEXES and CIRS observations (late and early northern summer, respectively), hinting at an asymmetry that is stable over long periods of time.

The contrast between the enriched EZ and the depleted equatorial belts is more subdued in 2014, mirroring the smaller contrast in equatorial temperatures.  This is unlikely to be caused by differences in spatial resolution, given that these are comparable between the TEXES and CIRS datasets.  It could reflect the different weighting between the spectral fitting and the \textit{prior} in the TEXES and CIRS inversions with very different spectral resolutions, although Fig. \ref{scalefactors} does show real differences in zonally-averaged radiances at tropical latitudes.  If periodic increases in equatorial upwelling do serve to cool temperatures and enhance gases and aerosols, then it is plausible that the Cassini observations were closer to such an event.  Indeed, the 250-mbar time series of \citet{06simon} does suggest that equatorial contrasts were stronger in 2000 than in the early 1990s.  We suggest that the equatorial enhancement in PH$_3$ is therefore time variable.  Furthermore, unlike the ammonia updrafts that are localised in plumes in Fig. \ref{map_gas}a, the phosphine equatorial enhancement appears more uniform with longitude (although still with a slight bias towards the northern EZ), suggesting different processes controlling the abundances of these species.

At higher latitudes, the PH$_3$ abundances gradually increase poleward of $45^\circ$S and $60^\circ$N, being enhanced at high latitudes in both datasets.  This polar enhancement could either be driven by strong vertical mixing within the polar vortices, or due to enhanced photolytic shielding from Jupiter's stratospheric aerosols which increase in number density towards high latitudes \citep{13zhang_aer}, permitting PH$_3$ to have a longer lifetime before it is photochemically destroyed.  The poleward enhancement was first noted by \citet{90drossart} using ground-based 5-$\mu$m spectra acquired in 1987, which also suggests a long-lived stable PH$_3$ distribution at high latitudes.  Future 2D photochemical modelling efforts for tropospheric PH$_3$ will need to account for these spatial differences in aerosol shielding.

It is interesting to note that the PH$_3$ distributions inferred from the thermal-IR differ from those encountered in the deeper atmosphere at 5 $\mu$m.  Low spatial-resolution observations by Cassini/VIMS did not reveal strong contrasts between the equator and neighbouring belts, but they did observe the equator-to-pole rise in PH$_3$ \citep{15giles} that were identified by \citet{90drossart} at the same wavelengths.  Could this be a real change in the PH$_3$ distribution from above the clouds to the sub-cloud region?  If so, it could support the idea that vertical motions associated with the belt/zone structure are confined in altitude and may even switch directions at higher pressures.  This would complicate the picture of ammonia-rich air rising at the equator from the 10-bar region discussed above.  Phosphine mole fractions reported by 5-$\mu$m studies range from 0.7-1.1 ppm for $p>1$ bar depending on the choice of H$_2$O relative humidity \citep{15giles}.  As expected, these $p>1$ bar abundances are larger than the 500-mbar abundances reported in Fig. \ref{zonal_gas}c, but smaller than the $\sim$2 ppm deep abundance previously reported by CIRS \citep{04irwin, 09fletcher_ph3} - the source of this discrepancy between the 5-$\mu$m results and the 10-$\mu$m results is yet to be explained and combined modelling of both spectral ranges will be required.  Finally, PH$_3$ is elevated within the GRS in both the CIRS and TEXES data, to be discussed in Section \ref{closeup}.

\subsection{Aerosols}

Simulations of Jupiter's mid-infrared spectrum require the addition of tropospheric aerosol opacity in order to achieve adequate fits, but neither the spectral properties nor the vertical distributions of those aerosols are well-constrained (see Section \ref{sensitivity}).  It remains unclear whether the 800-mbar opacity is solely due to a condensate cloud (i.e., NH$_3$ ice), or due to hazes produced from NH$_3$/PH$_3$ photolysis in the upper troposphere (potentially mixed with hydrocarbons sedimenting from the stratosphere) \citep{04west}.  A mixture of these processes is likely, so we retain the general term `aerosol' in this section.  For consistency with previous studies \citep{98banfield, 98lara, 02irwin, 05matcheva, 04wong, 06achterberg, 09fletcher_ph3}, we assign the aerosol opacity to a compact layer near 800 mbar and use smooth absorption cross-sections representative of 10-$\mu$m radius NH$_3$ ice.  This altitude is consistent with the expected condensation altitude for $3\times$ solar-enriched NH$_3$ \citep{99atreya} and with our best-fitting transition pressure from the deep, well-mixed NH$_3$ reservoir at $p>800$ mbar.  We omit clouds in the deeper atmosphere at $p>1$ bar \citep{02irwin,04west} that are unlikely to be detected in mid-IR spectra.  Both CIRS and TEXES retrievals are consistent with cumulative optical depths (integrated for $p<1$ bar) of $\sim2-3$ in the zonal-mean (Fig. \ref{zonal_gas}d) and the full map (Fig \ref{map_gas}c).  

There is a strong anti-correlation between the aerosol opacity and the tropospheric temperatures, as expected for volatiles condensing in cooler atmospheric zones.  Fresh supply of volatiles via upwelling could contribute to the enhanced cloudiness of the zones, but is not required to qualitatively explain the distribution in Fig. \ref{map_gas}c.  Opacities in Jupiter's zones are approximately twice those found in the belts, reaching 1.0-1.5 in the zones compared to 0.1-0.6 in the cloud-free belts (although the absolute numbers are sensitive to the cloud parameters chosen).  There have been shifts in the relative opacities of the different regions between 2000 and 2014.  In 2014, the EZ had the highest opacity, the SEB was slightly more cloud-free than the NEB, local maxima in opacity occurred near $30^\circ$S (encompassing the STropZ and STZ with no distinct STB minimum), $24^\circ$N (the NTropZ) and $42^\circ$N (NNTZ).  In 2000 the STropZ was the cloudiest region near $27^\circ$ with a further zone near $50^\circ$S.  The northern hemisphere was rather different, with a local minimum associated with the NTB near $30^\circ$N and elevated opacity in the neighbouring zones - the NTropZ ($22^\circ$N), NTZ ($33^\circ$N) and NNTZ ($42^\circ$N).  The visibility and optical depth of these cloud bands in 2000 is similar to that derived by \citet{05matcheva} using the same CIRS dataset. The opacity of the belts and zones varies substantially over time in both the tropical and temperate domains, with some of the cloud-free belts disappearing entirely as occurred during the 2009-10 fade of the South Equatorial Belt (SEB) \citep{11fletcher_fade}.  Connecting these opacity changes to visible colour changes is a long-term goal of this TEXES programme.

A subtle asymmetry in the peak aerosol opacity is evident at the equator in Fig. \ref{zonal_gas}d, which tends to be more opaque in the $0-4^\circ$S region than in the $0-4^\circ$N region.  This asymmetry is evident in both the TEXES and the CIRS observations.  This is opposite to the asymmetry in the zonal distribution of NH$_3$, which is more enhanced north of the equator, and is possibly related to the differing dynamics on either side of the equator - the hotspots of the NEBs versus the chaotic turbulence of the SEBn in the vicinity of the Great Red Spot.

Poleward of $\pm50^\circ$ latitude, the tropospheric aerosol opacity drops away in both hemispheres.  This is not simply an effect of the geometry, as the retrieval requires significantly lower opacities than our prior (an optical depth of unity at 1 bar) at these high latitudes.  Condensation of NH$_3$ would not be any lower at the poles, as Jupiter's equator to pole temperature contrast is relatively small.  It is possible that the aerosols being sensed in this spectral range are of a photochemical origin (i.e., photolysis of NH$_3$ or PH$_3$), sedimenting downward from the upper troposphere, so we would expect them to be enhanced at low-latitudes (where photolysis rates are highest) and depleted at the poles, in the absence of latitudinal mixing.  This aerosol population is different from the polar aerosols sensed in the near-IR \citep{13zhang_aer}, which show a marked increase in opacity in the polar region.  Observations from the polar-orbiting Juno spacecraft should help to disentangle these different aerosol populations.

\subsection{Ethane and Acetylene}

The spatial distribution of stratospheric hydrocarbons depends on a delicate balance between photochemical production, loss and chemical transport (and potentially ion-neutral chemistry at high latitudes).  Cassini-derived zonal-mean hydrocarbon distributions \citep{04kunde, 07nixon, 10nixon, 13zhang} revealed a C$_2$H$_2$ abundance that decreased from equator to pole and a C$_2$H$_6$ abundance that was mostly uniform but with a moderate poleward enhancement.    Converse to the suggestions of chemical models \citep{05moses_jup}, the distributions of these two species appear to be decoupled.  The former suggests that the C$_2$H$_2$ distribution is controlled by insolation and has a lifetime shorter than the horizontal transport timescale.  The latter suggests that ethane's lifetime is longer than the transport timescale and therefore C$_2$H$_6$ could be influenced by an equator-to-pole circulation.  Figs.  \ref{zonal_gas}e-f and \ref{map_gas}d-e compare TEXES results to CIRS hydrocarbon distributions.  This is the first time that Jupiter's hydrocarbons have been derived using Cassini's closest approach data at 2.5-cm$^{-1}$ resolution \citep[previous studies of][used 0.5-cm$^{-1}$ observations at lower spatial resolution]{04kunde,07nixon,10nixon,13zhang}.  

A comparison of the CIRS and TEXES datasets in Figs. \ref{zonal_gas}e-f reveals equatorial maxima in both species in 2000 (coinciding with the warm 5-mbar airmass in Fig. \ref{zonalT}c) that were largely absent in 2014, suggesting that vertically-localised motions might be varying in strength in connection with the QQO wave pattern.  The spatial maps in Fig. \ref{map_gas}d-e indicate that low-latitude ethane was more homogeneous with longitude in 2014 than in 2000, and that subtle latitudinal contrasts could be seen in equatorial ethane in 2014.  At higher latitudes, ethane displays the equator-to-pole enhancement observed by previous authors, whereas acetylene is depleted.  The ethane enhancement is most notable poleward of $\pm50^\circ$ in both hemispheres (Fig. \ref{map_gas}e).  The presence of ethane at Jupiter's high latitudes, where photochemical production is expected to be lowest, could imply an equator-to-pole circulation with sinking polar air enhancing the polar C$_2$H$_6$ abundance in much the same way as it does on Saturn \citep{13sinclair, 15fletcher_poles}.  This two-cell pattern, one in each hemisphere, with air rising at the equator and falling at the poles, is consistent with the expected motions in the purely radiatively-forced model of \citet{90conrath}.  However, the thermal maps in Fig. \ref{mapT} reveal that there is more to this picture than radiative forcing alone.  Furthermore,  the influence of auroral-related ion-chemistry as a potential additional source of C$_2$H$_6$ (but not C$_2$H$_2$) at high latitudes is poorly understood, and could remove the need to invoke high-latitude subsidence to explain the ethane distribution \citep[e.g.,][]{16sinclair}.

Indeed, the most interesting result of this hydrocarbon comparison is the presence of mid-latitude maxima in hydrocarbon abundances in both 2000 and 2014 (Fig. \ref{zonal_gas}e-f).  The contrasts are more evident in the distribution of acetylene than ethane, which may be a result of the different vertical gradients of the two species:  C$_2$H$_2$ has the stronger vertical gradient \citep{05moses_jup} and will therefore show stronger abundance variations when subjected to localised vertical transport than C$_2$H$_6$.  These maxima are asymmetric, resulting in enhanced hydrocarbon abundances at northern mid-latitudes than in the south.  The asymmetry in acetylene appeared to be stronger in 2000, closer to northern summer solstice, and weaker in 2014 near to the northern autumnal equinox.  This is true even if we uniformly scale the TEXES hydrocarbon abundances to match the CIRS abundances.  The asymmetry was first noted by \citet{10nixon}, and given that Jupiter's northern summer coincides with orbital perihelion (and therefore higher insolation), it was suggested that this eccentricity might lead to enhanced production of the hydrocarbons in the north. However, Jupiter will spend less time at perihelion than at aphelion due to Kepler's 2nd law, implying that the annually-averaged hydrocarbon production rates should be approximately symmetric \citep[a suggestion borne out by recent photochemical modelling studies,][]{15hue_dps}.  Furthermore, the local maxima are near $30^\circ$N and $30^\circ$S, so are unlikely to be explained solely by radiative forcing due to Jupiter's small $3^\circ$ obliquity.  Stratospheric temperatures do vary with time \citep{94orton, 06simon} and should impact the hydrocarbon production rates \citep{15hue}, but the difference in the zonal-mean distributions revealed in Fig. \ref{zonal_gas}e-f is rather pronounced.  If seasonal differences in photochemistry are not to blame, how might we explain a sustained difference between the northern and southern stratospheres?

The global maps of ethane and acetylene (Fig. \ref{map_gas}d-e) reveal considerable longitudinal structure underlying the zonal averages in Fig. \ref{zonal_gas}.  Acetylene shows a rather different pattern to ethane, with considerably higher abundances between $10-40^\circ$N and lots of small-scale variability.  Longitudinal variations appeared to be stronger for C$_2$H$_2$ in 2014, although we remind the readers that the acetylene lines are measured in spectral regions strongly affected by telluric water vapour.  However, we speculate that the presence of small-scale variations in the acetylene distribution strongly suggest the influence of dynamics on these mid-latitude abundances, rather than simple radiative control.   

As described in Section \ref{stratos_wave}, there could be several underlying causes for thermal and compositional asymmetries between the two hemispheres.  Large-scale stratospheric circulation could explain the latitudinal contrasts and asymmetry, either as a single inter-hemispheric cell with rising air in the south, cross-equatorial transport and sinking in the north; or a two-cell system with equatorial upwelling, mid-latitude subsidence and seasonal/permanent differences in the strength of the downwelling causing the hemispheric asymmetry.  However, there may be no need to invoke such large-scale stratospheric transport.  Instead, the stratosphere could be influenced by Jupiter's asymmetric tropospheric dynamics - Jupiter's southern hemisphere features more large anticyclonic vortices and their associated turbulent wakes, whereas the zonal jet system in Jupiter's northern hemisphere decays more slowly with altitude.  It is intriguing to note that the stratospheric circulation models of \citet{90conrath} did show an annually-averaged temperature asymmetry between northern and southern mid-latitudes when mechanical forcings were included based on the observed cloud-top zonal flows.  The strength of vertical mixing might be weaker in the northern hemisphere, permitting long-term accumulation of hydrocarbons in the north.  Indeed, the south hosts some of the only spectroscopically-identifiable ammonia ice clouds on the planet \citep{02baines} and tantalising evidence of tropospheric H$_2$O ice from Voyager \citep{00simon} was apparently localised between $50^\circ$S and $20^\circ$N, again hinting at stronger vertical motions in the south and a more sluggish north. These facts point to a connection between the stratospheric circulation and the asymmetric tropospheric dynamics at cloud-level that has yet to be captured in combined chemistry-transport models.  

Alternatively, the presence of ephemeral wave activity in Jupiter's northern stratosphere may serve to mix the hydrocarbons down from their source regions at micro-bar pressures, and we are simply seeing a connection between wave activity and stratospheric composition.  In 2014 we find some of the largest acetylene abundances co-located with the strongest thermal contrasts associated with the stratospheric wave.  This wave activity may or may not be connected to the asymmetric tropospheric dynamics, and it is unclear whether the prevalence of waves is connected to Jupiter's seasons.  Disentangling these competing effects (photochemistry, large-scale transport, small-scale waves and mixing) will require long-term tracking of Jupiter's stratospheric hydrocarbon distributions, and particularly the mid-latitude asymmetry, over a full jovian year.  

\subsection{Winds}

\begin{figure*}
\begin{centering}
\centerline{\includegraphics[angle=0,scale=.80]{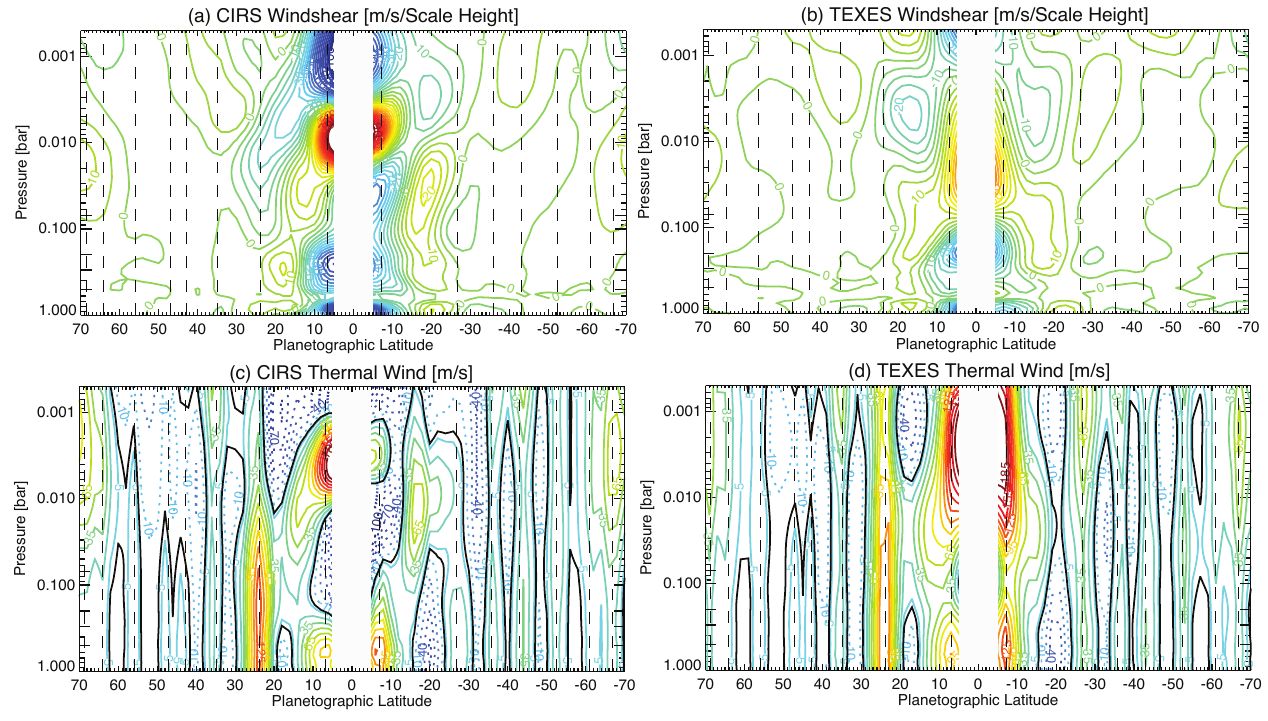}}
\caption{Zonal-mean cross-sections of the zonal windshear (top row) and thermal-wind (bottom row) from CIRS (left) and TEXES (right).  Contours of windshear are shown every 5 m/s/scale-height; contours of the thermal-wind are shown every 20 m/s.  The equatorial region is omitted as the thermal-wind equation greatly magnifies uncertainties in the winds at low latitudes.  Vertical dashed lines show the location of prograde zonal jets measured at the cloud-tops by \citet{03porco}. Solid black contours in panels c and d show the transition from retrograde flow (dotted contours) to prograde flow (solid contours).}
\label{winds}
\end{centering}
\end{figure*}

The discussion of asymmetries in PH$_3$ and the hydrocarbons suggest that there is some difference in vertical mixing processes or large-scale overturning between the northern and southern hemispheres.  This difference cannot be explained by radiative processes alone, but can be introduced via mechanical forcing based on Jupiter's zonal-wind field \citep[e.g.][]{90conrath}.  Latitudinal thermal contrasts measured by CIRS and TEXES in Fig. \ref{Tcontour} were used to investigate how the zonal jet system varies above the cloud-tops in Fig. \ref{winds}.  We used the latitudinal gradients in the CIRS and TEXES $T(p,\psi)$ cross sections (where $\psi$ is the latitude, p is the pressure) to calculate the two-dimensional wind-field $u(p,\psi)$ above the cloud tops via the thermal-wind equation\citep{87andrews}:
\begin{equation}
f\pderiv{u}{\ln \left( p \right)} = \frac{R}{a} \pderiv{T}{\psi} = R\pderiv{T}{y} = -fH\pderiv{u}{z}
\end{equation}
where $y$ is the north-south distance, $f$ is the Coriolis parameter $f=2\Omega sin(\psi)$ depends on the planetary angular velocity ($\Omega$), $R$ is the molar gas constant divided by the mean molar weight of Jupiter's atmosphere, $a(\psi)$ is the latitude-dependent planetary radius, and $H$ is the scale height.  We assume the Cassini-derived zonal velocities \citep{03porco} are measured at 500-mbar and integrate the thermal-wind equation with altitude.  We present the windshear per scale height and the thermal-wind cross-section in Fig. \ref{winds}.  

Propagation of temperature-retrieval uncertainties through the thermal-wind equation is complex given that latitudinal gradients magnify uncertainties.   \citet{16fletcher} used a Monte Carlo approach to demonstrate that uncertainties on the zonal winds from CIRS inversions grow with altitude and towards low latitudes, varying from $\sim10$ m/s near the mid-latitude tropopause to hundreds of m/s in the equatorial stratosphere.  Absolute winds shown in 2D $u(p,\psi)$ figures like Fig. \ref{winds} should therefore be viewed qualitatively, showing the switching directions of the winds and hemispheric asymmetries.  The TEXES and CIRS zonal-mean wind fields show similar properties, including oscillatory equatorial structure associated with Jupiter's QQO and an asymmetry between the mid-latitude jet systems.  

Section \ref{res_temp} demonstrated the stratospheric thermal contrasts associated with the QQO were more subdued in 2014 than 2000, and this is confirmed by Fig. \ref{winds}.  Both the CIRS and TEXES winds reveal alternating regions of positive and negative windshear above the prograde cloud-top jets at $\pm7^\circ$ latitude, with the strong positive shear near 10-mbar in 2000 leading to the fast prograde stratospheric jet identified by \citet{04flasar_jup}.  This windshear has evolved in the intervening 14 years so that this jet now covers a broader vertical range, coinciding with a weaker positive windshear in the 10-50 mbar region.  This confirms the suggestion \citep{06read_jup, 04flasar_jup} that the stratospheric jet is somewhat transient.  The reversal of the winds with altitude is exactly what is expected for the QQO.  However, as the vertical resolution of the CIRS and TEXES data differ due to the different spectral coverage and spectral resolution,  quantitative studies of the QQO temporal evolution will require long-term observations using the same instrument.

Finally, we note that mid-latitude ($\pm20$ to $\pm60^\circ$) wind fields are asymmetric between the two hemispheres.  In the south, prograde jets experience weaker shear and persist to low pressures (i.e., they are relatively barotropic), whereas the northern jets are divided into two categories - those that are barotropic between $20-40^\circ$N and those that vary substantially with altitude poleward of $40^\circ$N.  We observe these qualitative trends in both the CIRS and TEXES winds in Fig. \ref{winds}c-d, suggesting an asymmetry that remains stable with time.  A similar asymmetry was notable in the thermal-wind derivations of \citet{06read_jup}.  This asymmetric wind field could result in asymmetric forcing of the upper tropospheric and stratospheric circulation patterns, or the efficiency of vertical wave propagation and mixing, which would ultimately influence the distribution of PH$_3$ and the hydrocarbons.  However, we caution the reader that neither the CIRS nor TEXES windshear maps in Fig. \ref{winds}a-b has the latitudinal resolution to resolve all of the narrow belts and zones in Jupiter's temperate regions so that the shear may be underestimated (the shear only varies between $\pm5$ m/s/scale height outside of the tropics).  Nevertheless, Fig. \ref{winds} demonstrates that TEXES observations at multiple epochs could diagnose changes in the windshear with time, particularly those associated with belt/zone upheavals, the QQO, and long-term (seasonal?) changes in Jupiter's stratosphere.

\section{Discussion:  Discrete Regions}
\label{closeup}

\begin{figure*}
\begin{centering}
\includegraphics[angle=0,scale=.70]{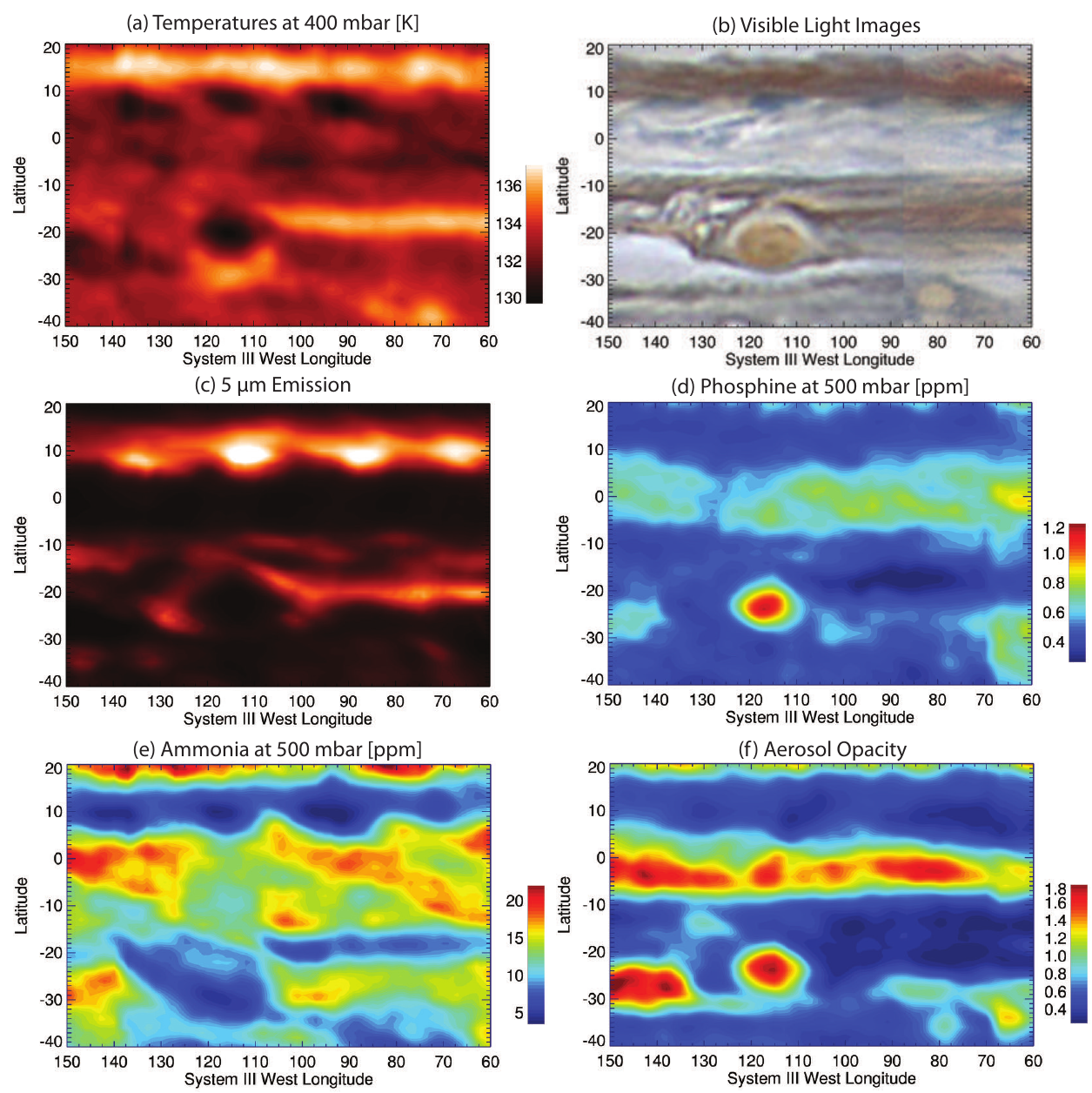}
\caption{Spatial maps of temperature, clouds and composition in the vicinity of Jupiter's Great Red Spot, Oval BA and equatorial plumes and hotspots, retrieved at the highest spatial resolution allowed by the TEXES dataset (a $1\times1^\circ$ grid, a factor of two improvement compared to the global maps in Fig. \ref{mapT} and \ref{map_gas}).}
\label{hires}
\end{centering}
\end{figure*}

Having demonstrated that temperatures, composition and aerosols retrieved from the TEXES and CIRS datasets are qualitatively and quantitatively consistent, we can use the retrieved TEXES maps to investigate the thermal structure and composition associated with discrete regions on Jupiter.  Fig. \ref{hires} compares the retrieved tropospheric temperature, ammonia, phosphine and aerosols to near-simultaneous images in visible light and at 5 $\mu$m, allowing us to focus on two regions - the Great Red Spot and the plumes and hotspots on the prograde NEBs jet.

\subsection{Great Red Spot and Oval BA}
These two red anticyclones were the two largest vortices on Jupiter in 2014, even though Oval BA appears to have paled in the years since its formation \citep{06simon_BA}.  Both are cold-core features, consistent with air rising and cooling adiabatically at the top of the anticyclone, although the spatial resolution of the IRTF does not resolve the cyclonic warm centre of the GRS that was discovered in higher-resolution imaging \citep{07choi, 10fletcher_grs}.  Both tropospheric aerosols and phosphine are notably enhanced within the GRS compared to the surroundings (Fig. \ref{hires}d-f, which leads to dark cloudy regions when viewed at 5-$\mu$m (Fig. \ref{hires}c).  A peripheral ring of subsidence is responsible for the low aerosol opacities in Fig. \ref{hires}f and the darker colours in the visible-light images, and may physically separate the internal red aerosols from the white aerosols of the STropZ.  The PH$_3$ and aerosol distributions are consistent with previous Cassini/CIRS analyses \citep{10fletcher_grs}, and marks a significant improvement compared to previous spectroscopic studies \citep{92griffith, 98lara}.  The latter study used IRTF/IRSHELL spectroscopy (the predecessor to TEXES) in 1991 \citep{98lara},but did not detect PH$_3$ variations across the GRS.

The spatial distribution of NH$_3$ across the Great Red Spot is complex \citep{92griffith, 96sada, 98lara, 06achterberg}.  CIRS and ground-based results previously suggested a north-south asymmetry, with ammonia elevated at the northern edge and depleted across the warm southern region that extended well into the STropZ and STB \citep{10fletcher_grs}.  This asymmetry is confirmed by the NH$_3$ map at 500 mbar in Fig. \ref{hires}e and by the CIRS map in Fig. \ref{map_gas}a.  Unlike phosphine and aerosols, ammonia is not enhanced within the GRS, meaning that the vortex structure does not stand out in the NH$_3$ map.  Furthermore, we do not see substantial enhancements in NH$_3$ in the turbulent wake to the northwest of the GRS, which is one of the few locations on the planet to host spectroscopically-identifiable NH$_3$ ice \citep{02baines}.  

At 400 mbar, Fig. \ref{hires} shows that the GRS has a warm southern periphery that coincides with depleted NH$_3$ gas and a break in the 5-$\mu$m-bright peripheral ring \citep[previously reported by][]{10fletcher_grs}.  However, this southern warm anomaly (SWA) does not apparently coincide with elevated aerosols at the $\sim800$-mbar level or any discrete features in the visible-light images, suggesting that a deeper aerosol population (possibly the NH$_4$SH cloud) is responsible for this feature. Both the SWA and the NH$_3$-enhancement north of the vortex could be related to the north-south `tilt' of the GRS deduced from visible cloud tracking \citep{02simon}.  Furthermore, the $\sim800$-mbar aerosol population (Fig. \ref{hires}f) appears to occupy a smaller area than this deeper cloud layer sensed at 5 $\mu$m (Fig. \ref{hires}c), possibly suggesting that the GRS area is larger at these deeper pressures.  Stratosphere-sensitive TEXES maps (Fig. \ref{mapT}a-b) confirm that the temperature and composition anomalies are restricted to the troposphere.  Finally, although the GRS temperatures, phosphine and aerosols do appear to be more circular in 2014 than 2000, consistent with the longitudinal shrinkage described by \citet{15simon}, we caution the reader that such an inter comparison should really be done with the same instrument.  Quantitative assessments of the GRS changes over time will be the subject of a future study.

Oval BA does show temperature and aerosol contrasts in Fig. \ref{hires}, with the cooler and high-opacity interior surrounded by a ring of peripheral warming due to subsidence, but it does not show contrast in PH$_3$ and NH$_3$ at this spatial resolution.  The warm periphery is co-located with darker clouds in the visible light images.  Smaller anticyclones, such as the white ovals near $40^\circ$S and the white oval embedded at the northern edge of the NEB near $280^\circ$W (known as `White Spot Z') are only barely visible as low-radiance features in Figs. \ref{texesmaps}-\ref{texesGRS} (i.e., cold-cored anticyclones), and these contrasts are insufficient to retrieve temperature or compositional variability associated with these features.  Spectroscopy at higher spatial resolutions will be required to diagnose their properties.

\subsection{Equatorial plumes and hotspots}


The TEXES zonal-means in Figs. \ref{zonalT}-\ref{zonal_gas} confirm that Jupiter's equatorial troposphere (between the prograde NEBs and SEBn jets at $7.1^\circ$N and $6.8^\circ$S) is characterised by cold temperatures and elevated PH$_3$, NH$_3$ and aerosol opacity.  Closer inspection of the equatorial region in Fig. \ref{hires} and Fig. \ref{map_gas} reveals some of the longitudinal variability associated with this equatorial enhancements.  The slight northward bias of the NH$_3$ zonal-mean is a result of a series of NH$_3$-rich regions between $0-7^\circ$N in the EZ, which have little counterpart in either the visible or infrared aerosol distributions.  These plumes appear to be to the south of the NEBs jet in Fig. \ref{hires}e, with corresponding NH$_3$-depleted regions to the northwest (centred on $10^\circ$N) in a regular wave pattern \citep[the so-called `hotspots' associated with the equatorial Rossby wave pattern,][]{90allison, 98ortiz, 99friedson}.  Surprisingly, the 400-mbar temperature map reveals that that depleted NH$_3$ regions are near to regions of cold temperatures of around 130 K \citep[consistent with the 470-mbar temperatures encountered by the Galileo probe as it entered one of these anomalous regions,][]{98seiff}.  Close inspection of the vertical temperature profiles suggests that these `cold anomalies' exist in the 300-600 mbar range.  

However, the cold anomalies are not precisely co-located with either the EZ plumes or the NEB hotspots.  Taking one plume-hotspot pair as an example, we find the strongest NH$_3$ depletion at $10^\circ$N, $118^\circ$W; the coldest temperatures at $8^\circ$N, $115^\circ$W; and the peak of an NH$_3$ plume at $2-4^\circ$N, $105^\circ$W.  As all TEXES channels used to perform these retrievals were taken within a few minutes of one another, longitudinal offsets due to zonal motion would have been negligible.  Comparing the temperature and ammonia maps with the 5-$\mu$m emission in Fig. \ref{hires}c, we observe that the cold anomalies and NH$_3$ dry spots sit slightly west of the brightest 5-$\mu$m emission ($9^\circ$N, $111^\circ$W), and that they are co-located with the visibly-dark albedo structures on the NEBs jet.  This longitudinal offset is consistent with the westward tilt with height of the bright emission features observed in Fig. \ref{texesNEB} (although such offsets were only apparent over a limited longitude range).  The equatorial NH$_3$ gas plumes are found in the same $2-7^\circ$N band of spectroscopically-identifiable NH$_3$ ice detected by Galileo/NIMS \citep[mean latitude of $6^\circ$N, approximately $1.5-3.0^\circ$ south of the bright hotspots,][]{02baines}, suggesting that the ice is condensing on powerful equatorial updrafts and that these updrafts cause the elevated zonal-mean NH$_3$ at the equator.  Interestingly, the westward offset between the NEB hotspots and the EZ plumes was also observed in the near-infrared by \citet{02baines}.  

This rather complex pattern of equatorial plumes and sinking hotspots confirms our earlier suggestions based on inspection of the TEXES radiance (Section \ref{image_inspection}) - the hotspots associated with the NEBs Rossby wave may be sheared westward as a function of altitude.  The zonal winds in Fig. \ref{winds} indicate strong westward shear on the eastward NEBs jet, so that channels sounding higher altitudes might see contrasts displaced westward compared to deeper-sounding channels.  Furthermore, the descent profile of the Galileo probe \citep{98atkinson} showed an increase in the prograde velocity from the 1-bar level to the 2-bar level of 40-50 m/s, consistent with hotspots at 5 $\mu$m (gaps in the NH$_4$SH cloud) being displaced eastward compared to those observed at 10 $\mu$m (gaps in the NH$_3$ cloud).  Vertical westward tilts are expected for equatorially-trapped Rossby waves but not for Kelvin waves \citep[e.g., Section 4.7 of][]{87andrews}.  Both wave types could be driven by large-scale convective motions near the equator, potentially associated with the ammonia-rich plumes identified here.  Higher spatial resolution simultaneous observations from larger telescopes will be necessary to confirm these longitudinal offsets, and temporal studies will be needed to understand the shear on the hotspots as a function of altitude.

\section{Conclusions}
\label{conclusion}

This paper reports on the development of a long-term mid-infrared Jupiter observing programme using the TEXES spectrograph on the IRTF.  The programme is designed to globally map Jupiter's temperatures, composition and aerosol opacity at frequent intervals preceding and during the Juno mission to Jupiter (2016-2018).   We use one example of a global TEXES map from December 2014 to show that the spatial and spectral resolution of this new archive of mid-infrared spectroscopy can surpass the quality of mid-IR maps acquired by Cassini during its close flyby of Jupiter in December 2000.  An optimal estimation spectral retrieval algorithm (NEMESIS) was used to simultaneously model eight TEXES spectral channels spanning the N and Q bands at moderate spectral resolution (0.06-0.5 cm$^{-1}$, $R\sim2000-12000$).  These inversions allow us to map temperatures and windshears in three dimensions, along with two-dimensional distributions of ammonia, phosphine, tropospheric aerosol opacity, ethane and acetylene for comparison with Jupiter's banded structure and discrete phenomena observed in visible-light imaging.  

Reduction and modelling of multiple spectral image cubes over two full nights of observing posed a significant challenge, primarily due to variations in the radiometric calibration as a function of time.  A comparison of the zonal-mean TEXES observations to those predicted based on Cassini's infrared observations suggested systematic offsets of 0-8 K in brightness temperature, depending on the channel.   In order to achieve realistic fits to all 8 channels, the spectra had to be globally scaled to match the Cassini predictions, assuming that global changes in Jupiter's temperatures have not occurred over the intervening 14 years.  The required scaling appears to be related to the TEXES calibration procedure (using the sky emission and a reference black card at the ambient temperature of the telescope) and is worst where the sky is most transparent.  Work is underway to better characterise the required radiometric corrections and whether they vary with time.  Unfortunately, the \textit{absolute} chemical abundances and their vertical gradients are strongly sensitive to these imposed scale factors, meaning that we can present only \textit{relative} spatial variations across Jupiter.  Nevertheless, those relative variations provide new insights into Jupiter's atmosphere, as outlined below.

\begin{enumerate}

\item \textbf{Equatorial plumes and hotspots:  }  Jupiter's equatorial troposphere is characterised by cold temperatures, elevated aerosol opacity and local maxima in ammonia and phosphine.  Global maps of these parameters reveal that ammonia enhancement occurs in localised plumes between the equator and the prograde NEBs jet at $7^\circ$N, which sit southeast of extreme NH$_3$ depletion regions coincident with cloud clearings in the 5-$\mu$m hotspots and dark regions in visible images.  Indeed, we demonstrate that the hotspots are detectable throughout the M and N bands (and possibly also in Q).  Upwelling in the plumes and subsidence in adjacent hotspots are consistent with the presence of an equatorial wave on the NEBs, and this is the first time that such activity has been studied in the mid-infrared.  Furthermore, the use of a broad spectral range sampling multiple altitudes suggests that the `hotspots' tilt westward with increasing altitude, consistent with the negative shear on the prograde NEBs.   Furthermore, PH$_3$ is also slightly enhanced in the northern region of the EZ (creating a north-south asymmetry in the equatorial PH$_3$ peak), although there is no direct correspondence with the NH$_3$ plumes.  The equatorial asymmetry, with plume activity dominating the northern EZ, is likely driven by the presence of the GRS in the SEB, preventing similar large-scale Rossby wave activity on the prograde SEBn.

\item \textbf{Tropospheric asymmetries: } Zonal-mean tropospheric temperatures and aerosol opacity varied between 2000 and 2014, potentially as a result of unpredictable upheavals of the banded structure \citep{95rogers, 08sanchez, 11fletcher_fade}, but despite these changes there appears to be a large-scale symmetry between the northern and southern hemispheres.  Between $\pm30^\circ$ latitude there is a good correspondence between cloudy zones (the EZ, STropZ and NTropZ) and elevated PH$_3$ and NH$_3$ gas in both the CIRS and TEXES observations, consistent with uplift in zones.  When the cloud-tracked zonal winds are extrapolated via the thermal-wind equation, we find that the differing 2D wind fields in the northern and southern hemispheres will mean that the efficiency of vertical mixing, wave propagation and mechanical forcing of higher-altitude circulations will differ between the two hemispheres.  This may be related to the other notable tropospheric asymmetry:  PH$_3$ is enhanced at northern temperate latitudes compared to southern temperate latitudes (a region we refer to as the `northern plateau'), and this enhancement was present in both 2000 and 2014 (early and late northern summer, respectively).  It is an open question as to whether this asymmetry is permanent or evolves with season, either due to enhanced photolytic shielding from aerosols/NH$_3$ or stronger vertical mixing in the north versus the south.  

\item \textbf{Stratospheric circulation: }  The equatorial stratosphere is controlled by the 4.2-year cycle of the QQO \citep{91leovy}, although the vertical oscillations of the temperature/wind fields and the contrast between the equator and $\pm10^\circ$ latitude were both more subdued in 2014 than in 2000, confirming the findings of \citet{06simon} that the amplitude of the oscillations changes over time.  5-mbar temperatures featured a local maximum at the equator in 2000 and 2014 (3.5 QQO cycles apart), 1-mbar temperatures featured a local minimum.  Away from the equator, two broad bands of warmer temperature exist between $\pm15-40^\circ$ latitude in the 1-5 mbar region.  The northern band appeared warmer than the southern band in both epochs, although the majority of the warming in 2014 was associated with a well-defined stratospheric wave ($15-20^\circ$ wavelength over a limited sector of the planet).  C$_2$H$_2$ was also elevated in the northern band and depleted towards both poles, whereas C$_2$H$_6$ shows a general equator-to-pole rise and a distinct 'plateau' near the northern mid-latitudes.  Both species are therefore enriched in the northern hemisphere in both epochs, and we speculate that this asymmetry is driven by dynamics rather than the small seasonal cycle.  As in the troposphere, this could result from a difference in the efficiency of vertical mixing between the northern and southern stratospheres, driven by the different zonal wind fields.  A greater propensity for stratospheric wave activity in the north could serve to mix the hydrocarbons down from their source regions at microbar pressure levels.   Indeed, both species show longitudinal variability that could be related to wave activity.  Alternatively, it may result from a large-scale interhemispheric transport from the south to the north during the summer season (i.e., advection, rather than mixing).  Distinguishing mid-latitude mixing from stratospheric advection will require tracking these hydrocarbon distributions over a full seasonal cycle.  

\item \textbf{Giant vortices: }  The spatial resolution of the IRTF is lower than that of 8-m facilities used in previous photometric imaging studies \citep{10fletcher_grs, 10depater}, but TEXES brings the substantial advantage of spatially-resolved spectroscopy.  This allows us to confirm the elevated PH$_3$ and aerosol opacity within the cold Great Red Spot, and to show that the distribution of NH$_3$ is more enhanced over the northern part of the GRS than over the south. Both Oval BA and the GRS have warm southern peripheries that coincide with high 5-$\mu$m opacity and a break in the peripheral rings of atmospheric subsidence.  These southern warm anomalies (SWAs) do not have counterparts in visible-light imaging (i.e., they do not appear to affect the upper tropospheric aerosols), and may be related to a north-south tilt in these giant anticyclones.  Smaller anticyclones, such as White Oval Z embedded in the NEB, show subtle temperature contrasts and cold cores, but no notable signatures in the composition or aerosol maps. 

\item \textbf{Polar Vortices: } Jupiter's polar vortices, poleward of the highest latitude jets near $66-68^\circ$ in both hemispheres, are challenging to observe from Earth due to the small obliquity.  Nevertheless, the atmospheric temperatures and composition undergo a distinct change at high latitudes.  Tropopause and lower stratospheric temperatures show a marked drop poleward of $\pm50^\circ$ suggestive of cold polar vortices, potentially due to enhanced radiative cooling from polar aerosols \citep{15zhang}.  However, tropospheric aerosol opacity reaches a minimum poleward of $\pm50^\circ$, whereas the population of stratospheric aerosols and hazes appear to increase \citep[e.g.,][]{13zhang_aer}, potentially in relation to the unusual chemistry occurring in proximity to the powerful jovian aurorae.  PH$_3$ shows a subtle enhancement that is more prominent in the north than in the south, acetylene is depleted poleward of $\pm50^\circ$ whereas C$_2$H$_6$ shows a strong enhancement, and there are no apparent changes in the distribution of NH$_3$ as we enter the polar region.  The polar ethane enhancement may be due to the downwelling branch of a large-scale equator-to-pole circulation pattern, superimposed onto the mid-latitude dynamic activity.  Alternatively, poorly-understood auroral-related ion chemistry could serve to enhance C$_2$H$_6$ (but not C$_2$H$_2$) at Jupiter's high latitudes.  The northern polar vortex also features the stratospheric hotspot driven by a combination of Joule heating, direct energy input and aerosol heating beneath the main auroral oval \citep{16sinclair}, with enhanced emission from all hydrocarbon features suggestive of high temperatures and (potentially) enhanced photochemical production there.  Higher spatial resolutions will ultimately be required to map thermochemical contrasts at the edges of Jupiter's polar vortices more precisely, combined with the eagerly anticipated results from the polar-orbiting Juno mission in 2016.

\end{enumerate}

The primary purpose of this paper has been to demonstrate the capabilities of TEXES for providing thermal-infrared scientific discoveries that can rival those of previous spacecraft missions.  The December 2014 observations are a snapshot of a larger infrared spectroscopic archive that will span from 2012 to the epoch of the Juno mission, allowing us to track belt/zone variability, changes to giant anticyclones, equatorial plume and wave phenomena, stratospheric circulation patterns and wave activity over the course of time.  We hope that the circulation patterns deduced from this long-term record will inform the next generation of numerical dynamical models and allow us to better understand the redistribution of material and energy within Jupiter's atmosphere.

\section*{Acknowledgments}

Fletcher was supported by a Royal Society Research Fellowship at the University of Leicester.  Fletcher, Greathouse, Orton and Giles were visiting astronomers at the Infrared Telescope Facility, which is operated by the University of Hawaii under Cooperative Agreement No. NNX-08AE38A with the National Aeronautics and Space Administration, Science Mission Directorate, Planetary Astronomy Program.  We recognise the significant cultural role of Mauna Kea within the indigenous Hawaiian community, and we appreciate the opportunity to conduct our Jupiter observations from this revered site.  The UK authors acknowledge the support of the Science and Technology Facilities Council (STFC). A portion of this work was performed by Orton and Sinclair at the Jet Propulsion Laboratory, California Institute of Technology, under a contract with NASA.  We thank Marco Vedovato of the Italian Amateur Astronomers Union for compiling the visible light images from I. Sharp, F. Fortunato, H. Einaga and T. Horiuchi to coincide with our TEXES programme.  We are extremely grateful to John Lacy and Matt Richter for their assistance in understanding the performance of the TEXES instrument and the uncertainties related to calibration.  We thank S. Guerlet, J. Moses, T. Fouchet and V. Hue for helpful comments and suggestions during this work, and Gordy Bjoraker and one anonymous reviewer for their critique of this manuscript.  This research used the ALICE High Performance Computing Facility at the University of Leicester.


%
%


\newpage
\appendix

\section{TEXES Observation Record in December 2014}
\label{datarecord}

\begin{table*}[p]
\caption{Observations on December 8th 2014.  'Lo' and 'Med' refer to the spectral resolution settings.}
\begin{center}
\begin{tabular}{|c|c|c|c|c|c|c|c|}
\hline
File ID & Date & Time [UT] & Wavenumber  & Mode & Total time  & Airmass & Longitude \\
& & & [cm$^{-1}$] & & [s] & & \\
\hline
 4011  & 14-12-08   & 10:19:36.100  &  1248  &  med	 & 	688.34  &  2.387   &  253.9\\						 
 4018  & 14-12-08   & 11:36:41.050  &  1248  &  med	 & 	680.34  &  1.472   &  300.4 \\						 
 4029  & 14-12-08   & 13:24:06.400  &  1248  &  med	 & 	680.34  &  1.079   & 5.4 \\						 
 4030  & 14-12-08   & 13:42:04.700  &  538   &  med	 & 	300.19  &  1.051   &  16.2 \\						 
 4031  & 14-12-08   & 13:48:15.300  &  587   &  med	 & 	318.19  &  1.043   &  19.9\\						 
 4032  & 14-12-08   & 13:54:33.100  &  744     &  med	 & 	318.19  &  1.035   & 23.7 \\						 
 4033  & 14-12-08   & 14:00:25.700  &  819     &  med	 & 	318.19  &  1.029   & 27.2  \\						 
 4034  & 14-12-08   & 14:03:38.150  &  819     &  med	 & 	318.19  &  1.026   & 29.2 \\						 
 4035  & 14-12-08   & 14:10:22.350  &  1248  &  med	 & 	354.22  &  1.021   & 33.3 \\						 
 4036  & 14-12-08   & 14:18:45.250  &  960     &  low	 & 	318.19  &  1.015   & 38.4 \\						 
 4037  & 14-12-08   & 14:24:19.850  &  901     &  low	 & 	318.19  &  1.012   & 41.7 \\						 
 4038  & 14-12-08   & 14:30:26.850  &  1161    &  low	 & 	318.19  &  1.009   & 45.4 \\						 
 4039  & 14-12-08   & 14:42:44.250  &  2137    &  med	 & 	680.34  &  1.005   & 52.8  \\						 
 4040  & 14-12-08   & 14:52:53.300  &  538.5   &  med	 & 	609.19  &  1.004   & 59.0 \\						 
 4041  & 14-12-08   & 15:00:48.250  &  587   &  med	 & 	318.19  &  1.004   &  63.7\\						 
 4042  & 14-12-08   & 15:05:59.500  &  587   &  med	 & 	318.19  &  1.005   & 66.9  \\						 
 4043  & 14-12-08   & 15:10:02.350  &  744     &  med	 & 	318.19  &  1.006   &  69.4\\						 
 4044  & 14-12-08   & 15:16:03.950  &  819     &  med	 & 	318.19  &  1.008   & 73.0 \\						 
 4046  & 14-12-08   & 15:26:03.600  &  1248  &  med	 & 	354.22  &  1.013   &  79.0\\						 
 4047  & 14-12-08   & 15:34:10.550  &  960     &  low	 & 	318.19  &  1.019   &  84.0 \\						 
 4048  & 14-12-08   & 15:39:52.300  &  901     &  low	 & 	318.19  &  1.023   &  87.4\\						 
 4049  & 14-12-08   & 15:46:38.900  &  1161    &  low	 & 	318.19  &  1.029   & 91.5 \\						 
 4051  & 14-12-08   & 15:58:06.100  &  2137    &  med	 & 	372.82  &  1.042   & 98.4  \\						 
 4052  & 14-12-08   & 16:05:40.050  &  538   &  med	 & 	336.79  &  1.051   & 103.0 \\						 
 4053  & 14-12-08   & 16:11:11.000  &  538   &  med	 & 	336.79  &  1.059   &  106.3\\						 
 4054  & 14-12-08   & 16:17:05.400  &  586   &  med	 & 	336.79  &  1.068   &  109.8\\						 
 4055  & 14-12-08   & 16:23:08.800  &  744     &  med	 & 	336.79  &  1.078   & 113.6 \\						 
 4056  & 14-12-08   & 16:29:10.400  &  819     &  med	 & 	336.79  &  1.090   &  117.3\\						 
 4057  & 14-12-08   & 16:35:55.700  &  1248  &  med	 & 	354.22  &  1.103   & 121.4 \\						 
 4058  & 14-12-08   & 16:44:38.400  &  960     &  low	 & 	313.39  &  1.122   & 126.6 \\						 
 4059  & 14-12-08   & 16:47:49.100  &  960     &  low	 & 	313.39  &  1.130   & 128.6 \\						 
 4060  & 14-12-08   & 16:53:18.301  &  901     &  low	 & 	313.39  &  1.144   & 131.9 \\						 
 4061  & 14-12-08   & 16:59:19.900  &  1161    &  low	 & 	313.39  &  1.160   & 135.5 \\						 
 4062  & 14-12-08   & 17:07:10.850  &  2137    &  med	 & 	372.82  &  1.183   &  140.2\\						 
 4063  & 14-12-08   & 17:14:31.650  &  538   &  med	 & 	318.19  &  1.207   & 144.6  \\						 
 4064  & 14-12-08   & 17:20:02.600  &  538   &  med	 & 	318.19  &  1.226   &  148.0\\								
 4065  & 14-12-08   & 17:25:44.400  &  587   &  med	 & 	318.19  &  1.247   &  151.2\\								
 4066  & 14-12-08   & 17:32:13.000  &  744     &  med	 & 	318.19  &  1.272   & 155.3 \\								
 4067  & 14-12-08   & 17:38:18.200  &  819     &  med	 & 	318.19  &  1.298   & 159.0 \\								
 4068  & 14-12-08   & 17:45:07.500  &  1248  &  med	 & 	354.22  &  1.329   &  163.1\\								
 4069  & 14-12-08   & 17:52:56.650  &  960     &  low	 & 	318.19  &  1.369   &  167.8\\								
 4070  & 14-12-08   & 17:58:45.600  &  901     &  low	 & 	318.19  &  1.400   &  171.4\\								
 4071  & 14-12-08   & 18:05:26.800  &  1161    &  low	 & 	318.19  &  1.440   &  175.4\\								
 4072  & 14-12-08   & 18:13:10.750  &  2137    &  med	 & 	372.82  &  1.489   & 180.2 \\								
 4073  & 14-12-08   & 18:20:28.250  &  538   &  med	 & 	318.19  &  1.541   &  184.5\\								
 4074  & 14-12-08   & 18:25:57.500  &  587   &  med	 & 	318.19  &  1.583   &  187.8\\								
 4075  & 14-12-08   & 18:31:53.650  &  744     &  med	 & 	318.19  &  1.632   &  191.4\\								
 4076  & 14-12-08   & 18:37:30.050  &  819     &  med	 & 	318.19  &  1.682   &  196.3\\								
 4077  & 14-12-08   & 18:44:40.050  &  960     &  low	 & 	318.19  &  1.753   &  199.2\\								
 4078  & 14-12-08   & 18:50:12.850  &  901     &  low	 & 	318.19  &  1.812   &  202.4\\								
 4079  & 14-12-08   & 18:56:37.850  &  1161    &  low	 & 	318.19  &  1.888   &  206.3\\								
 4080  & 14-12-08   & 19:04:32.800  &  2137    &  med	 & 	318.19  &  1.992   & 211.3 \\								
 4081  & 14-12-08   & 19:07:36.050  &  2137    &  med	 & 	372.82  &  2.036   &  213.0\\								

\hline

\end{tabular}
\end{center}
\label{data1}
\end{table*}%

\begin{table*}[p]
\caption{Observations on December 9th 2014.  'Low' and 'Med' refer to the spectral resolution settings.}
\begin{center}
\begin{tabular}{|c|c|c|c|c|c|c|c|}
\hline
File ID & Date & Time [UT] & Wavenumber  & Mode & Total time  & Airmass & Longitude \\
& & & [cm$^{-1}$] & & [s] & & \\
\hline

 5020  & 14-12-09   & 11:42:09.500  &  1248    &  med	 & 	354.22  &  1.414   &  90.7 \\						
 5037  & 14-12-09   & 13:29:45.100  &  1248    &  med	 & 	354.22  &  1.063   &  155.4 \\						
 5038  & 14-12-09   & 13:37:00.500  &  538   &  med	 & 	318.19  &  1.052   &  160.3 \\						
 5039  & 14-12-09   & 13:42:53.100  &  587   &  med	 & 	318.19  &  1.044   &  163.3 \\						
 5040  & 14-12-09   & 13:48:56.500  &  744     &  med	 & 	318.19  &  1.037   & 166.9  \\						
 5041  & 14-12-09   & 13:54:38.300  &  819     &  med	 & 	318.19  &  1.031   & 170.5 \\						
 5042  & 14-12-09   & 14:00:51.450  &  1248    &  med	 & 	354.22  &  1.025   &  174.2 \\						
 5043  & 14-12-09   & 14:08:55.800  &  960     &  low	 & 	318.19  &  1.019   & 179.0  \\						
 5044  & 14-12-09   & 14:14:30.500  &  901     &  low	 & 	318.19  &  1.015   &  182.6 \\						
 5045  & 14-12-09   & 14:20:28.500  &  1161    &  low	 & 	318.19  &  1.012   &  186.3 \\						
 5046  & 14-12-09   & 14:28:16.400  &  2137    &  med	 & 	372.82  &  1.008   & 191.1 \\						
 5047  & 14-12-09   & 14:35:36.600  &  538   &  med	 & 	318.19  &  1.006   & 195.3 \\						
 5048  & 14-12-09   & 14:41:09.450  &  587   &  med	 & 	318.19  &  1.005   &  199.0 \\						
 5049  & 14-12-09   & 14:47:03.850  &  744     &  med	 & 	318.19  &  1.004   &  202.6 \\						
 5050  & 14-12-09   & 14:52:38.450  &  819     &  med	 & 	318.19  &  1.004   &  205.6 \\						
 5051  & 14-12-09   & 14:59:08.050  &  1248    &  med	 & 	354.22  &  1.005   &  209.9 \\						
 5052  & 14-12-09   & 15:06:54.900  &  960     &  low	 & 	318.19  &  1.007   & 214.1  \\						
 5053  & 14-12-09   & 15:12:31.300  &  901     &  low	 & 	318.19  &  1.009   & 217.7  \\						
 5054  & 14-12-09   & 15:18:27.500  &  1161    &  low	 & 	318.19  &  1.011   &  221.3 \\						
 5055  & 14-12-09   & 15:23:49.500  &  1161    &  low	 & 	318.19  &  1.014   &  224.4 \\						
 5056  & 14-12-09   & 15:33:16.250  &  2137    &  med	 & 	372.82  &  1.021   & 230.4 \\						
 5057  & 14-12-09   & 15:40:30.400  &  538   &  med	 & 	336.79  &  1.027   &  234.6 \\						
 5058  & 14-12-09   & 15:46:19.350  &  587   &  med	 & 	336.79  &  1.033   &  238.3 \\						
 5059  & 14-12-09   & 15:52:26.350  &  744     &  med	 & 	318.19  &  1.040   & 241.9 \\						
 5060  & 14-12-09   & 15:58:04.600  &  819     &  med	 & 	318.19  &  1.046   &  245.5 \\						
 5061  & 14-12-09   & 16:04:22.650  &  1248    &  med	 & 	354.22  &  1.055   &  249.1 \\						
 5062  & 14-12-09   & 16:12:16.500  &  960     &  low	 & 	318.19  &  1.067   & 254.0  \\						
 5063  & 14-12-09   & 16:17:43.850  &  901     &  low	 & 	318.19  &  1.076   & 257.0 \\						
 5064  & 14-12-09   & 16:23:34.650  &  1161    &  low	 & 	318.19  &  1.086   &  260.6 \\						
 5065  & 14-12-09   & 16:31:02.700  &  2137    &  med	 & 	372.82  &  1.101   & 265.5  \\						
 5066  & 14-12-09   & 16:38:32.450  &  538   &  med	 & 	318.19  &  1.117   &  269.7\\						
 5067  & 14-12-09   & 16:44:19.650  &  587   &  med	 & 	318.19  &  1.131   &  273.3 \\						
 5068  & 14-12-09   & 16:50:30.200  &  744     &  med	 & 	318.19  &  1.147   &  277.0 \\						
 5069  & 14-12-09   & 16:56:12.000  &  819     &  med	 & 	318.19  &  1.162   &  280.6\\						
 5070  & 14-12-09   & 17:02:25.900  &  1248    &  med	 & 	354.22  &  1.181   &  284.2 \\						
 5076  & 14-12-09   & 18:02:32.250  &  2137    &  med	 & 	372.82  &  1.446   & 320.5 \\						
 5078  & 14-12-09   & 18:14:08.100  &  538   &  med	 & 	318.19  &  1.523   &  327.7 \\						
 5079  & 14-12-09   & 18:19:46.300  &  587   &  med	 & 	318.19  &  1.565   &  330.8 \\						
 5080  & 14-12-09   & 18:25:49.700  &  744     &  med	 & 	318.19  &  1.614   &  334.4 \\						
 5081  & 14-12-09   & 18:31:26.150  &  819     &  med	 & 	318.19  &  1.663   &  338.0 \\						
 5082  & 14-12-09   & 18:37:37.350  &  1248    &  med	 & 	354.22  &  1.721   &  341.6 \\						
 5086  & 14-12-09   & 19:03:34.200  &  2137    &  med	 & 	372.82  &  2.035   &  357.4 \\						
 5087  & 14-12-09   & 19:10:54.900  &  538   &  med	 & 	318.19  &  2.149   &  1.6 \\						
 5088  & 14-12-09   & 19:16:33.100  &  587   &  med	 & 	318.19  &  2.248   &  5.2 \\						
 5089  & 14-12-09   & 19:22:38.300  &  744     &  med	 & 	318.19  &  2.366   &  8.8 \\						
\hline

\end{tabular}
\end{center}
\label{data2}
\end{table*}%

\newpage

\bibliographystyle{elsarticle-harv}
\bibliography{references}

\begin{thebibliography}{135}
\expandafter\ifx\csname natexlab\endcsname\relax\def\natexlab#1{#1}\fi
\providecommand{\url}[1]{\texttt{#1}}
\providecommand{\href}[2]{#2}
\providecommand{\path}[1]{#1}
\providecommand{\DOIprefix}{doi:}
\providecommand{\ArXivprefix}{arXiv:}
\providecommand{\URLprefix}{URL: }
\providecommand{\Pubmedprefix}{pmid:}
\providecommand{\doi}[1]{\href{http://dx.doi.org/#1}{\path{#1}}}
\providecommand{\Pubmed}[1]{\href{pmid:#1}{\path{#1}}}
\providecommand{\bibinfo}[2]{#2}
\ifx\xfnm\relax \def\xfnm[#1]{\unskip,\space#1}\fi
\bibitem[{{Achterberg} et~al.(2006){Achterberg}, {Conrath} and
  {Gierasch}}]{06achterberg}
\bibinfo{author}{{Achterberg}, R.K.}, \bibinfo{author}{{Conrath}, B.J.},
  \bibinfo{author}{{Gierasch}, P.J.}, \bibinfo{year}{2006}.
\newblock \bibinfo{title}{{Cassini CIRS retrievals of Ammonia in Jupiter's
  Upper Troposphere}}.
\newblock \bibinfo{journal}{Icarus} \bibinfo{volume}{182},
  \bibinfo{pages}{169--180}.
\newblock \DOIprefix\doi{10.1016/j.icarus.2005.12.020}.
\bibitem[{{Allison}(1990)}]{90allison}
\bibinfo{author}{{Allison}, M.}, \bibinfo{year}{1990}.
\newblock \bibinfo{title}{{Planetary waves in Jupiter's equatorial
  atmosphere}}.
\newblock \bibinfo{journal}{Icarus} \bibinfo{volume}{83},
  \bibinfo{pages}{282--307}.
\newblock \DOIprefix\doi{10.1016/0019-1035(90)90069-L}.
\bibitem[{{Andrews} et~al.(1987){Andrews}, {Holton} and {Leovy}}]{87andrews}
\bibinfo{author}{{Andrews}, D.G.}, \bibinfo{author}{{Holton}, J.R.},
  \bibinfo{author}{{Leovy}, C.B.}, \bibinfo{year}{1987}.
\newblock \bibinfo{title}{{Middle atmosphere dynamics}}.
\newblock \bibinfo{publisher}{Academic Press, New York}.
\bibitem[{{Atkinson} et~al.(1998){Atkinson}, {Pollack} and
  {Seiff}}]{98atkinson}
\bibinfo{author}{{Atkinson}, D.H.}, \bibinfo{author}{{Pollack}, J.B.},
  \bibinfo{author}{{Seiff}, A.}, \bibinfo{year}{1998}.
\newblock \bibinfo{title}{{The Galileo probe Doppler wind experiment:
  Measurement of the deep zonal winds on Jupiter}}.
\newblock \bibinfo{journal}{J. Geophys. Res.} \bibinfo{volume}{103},
  \bibinfo{pages}{22911--22928}.
\newblock \DOIprefix\doi{10.1029/98JE00060}.
\bibitem[{{Atreya} et~al.(1999){Atreya}, {Wong}, {Owen}, {Mahaffy}, {Niemann},
  {de Pater}, {Drossart} and {Encrenaz}}]{99atreya}
\bibinfo{author}{{Atreya}, S.K.}, \bibinfo{author}{{Wong}, M.H.},
  \bibinfo{author}{{Owen}, T.C.}, \bibinfo{author}{{Mahaffy}, P.R.},
  \bibinfo{author}{{Niemann}, H.B.}, \bibinfo{author}{{de Pater}, I.},
  \bibinfo{author}{{Drossart}, P.}, \bibinfo{author}{{Encrenaz}, T.},
  \bibinfo{year}{1999}.
\newblock \bibinfo{title}{{A comparison of the atmospheres of Jupiter and
  Saturn: deep atmospheric composition, cloud structure, vertical mixing, and
  origin}}.
\newblock \bibinfo{journal}{Plan. \& Space Sci.} \bibinfo{volume}{47},
  \bibinfo{pages}{1243--1262}.
\bibitem[{Baines et~al.(2002)Baines, Carlson and Kamp}]{02baines}
\bibinfo{author}{Baines, K.}, \bibinfo{author}{Carlson, R.},
  \bibinfo{author}{Kamp, L.}, \bibinfo{year}{2002}.
\newblock \bibinfo{title}{{Fresh Ammonia Ice Clouds in Jupiter I. Spectroscopic
  Identification, Spatial Distribution, and Dynamical Implications}}.
\newblock \bibinfo{journal}{Icarus} \bibinfo{volume}{159},
  \bibinfo{pages}{74--94}.
\bibitem[{Banfield et~al.(1998)Banfield, Gierasch, Bell, Ustinov, Ingersoll,
  Vasavada, West and Belton}]{98banfield}
\bibinfo{author}{Banfield, D.}, \bibinfo{author}{Gierasch, P.},
  \bibinfo{author}{Bell, M.}, \bibinfo{author}{Ustinov, E.},
  \bibinfo{author}{Ingersoll, A.}, \bibinfo{author}{Vasavada, A.},
  \bibinfo{author}{West, R.}, \bibinfo{author}{Belton, M.},
  \bibinfo{year}{1998}.
\newblock \bibinfo{title}{{Jupiter's Cloud Structure from Galileo Imaging
  Data}}.
\newblock \bibinfo{journal}{Icarus} \bibinfo{volume}{135},
  \bibinfo{pages}{230--250}.
\bibitem[{{Bezard} et~al.(1997){Bezard}, {Griffith} and {Kelly}}]{97bezard_nh3}
\bibinfo{author}{{Bezard}, B.}, \bibinfo{author}{{Griffith}, C.A.},
  \bibinfo{author}{{Kelly}, D.M.}, \bibinfo{year}{1997}.
\newblock \bibinfo{title}{{Search for NH\_3 in Jupiter's Stratosphere Ten
  Months after SL9's Collision}}.
\newblock \bibinfo{journal}{Icarus} \bibinfo{volume}{125},
  \bibinfo{pages}{331--339}.
\newblock \DOIprefix\doi{10.1006/icar.1996.5615}.
\bibitem[{{Blass} et~al.(1987){Blass}, {Halsey} and {Jennings}}]{87blass}
\bibinfo{author}{{Blass}, W.E.}, \bibinfo{author}{{Halsey}, G.W.},
  \bibinfo{author}{{Jennings}, D.E.}, \bibinfo{year}{1987}.
\newblock \bibinfo{title}{{Self- and foreign-gas broadening of ethane lines
  determined from diode laser measurements at 12 microns}}.
\newblock \bibinfo{journal}{Journal of Quantitative Spectroscopy and Radiative
  Transfer} \bibinfo{volume}{38}, \bibinfo{pages}{183--184}.
\newblock \DOIprefix\doi{10.1016/0022-4073(87)90083-5}.
\bibitem[{Borysow and Frommhold(1986)}]{86borysow}
\bibinfo{author}{Borysow, A.}, \bibinfo{author}{Frommhold, L.},
  \bibinfo{year}{1986}.
\newblock \bibinfo{title}{Theoretical collision-induced rototranslational
  absorption spectra for the outer planets: {H}$_2$-{CH}$_4$ pairs}.
\newblock \bibinfo{journal}{Astrophys. J.} \bibinfo{volume}{304},
  \bibinfo{pages}{849--865}.
\bibitem[{Borysow and Frommhold(1987)}]{87borysow}
\bibinfo{author}{Borysow, A.}, \bibinfo{author}{Frommhold, L.},
  \bibinfo{year}{1987}.
\newblock \bibinfo{title}{Collision-induced rototranslational absorption
  spectra of {CH}$_4$-{CH}$_4$ pairs at temperatures from 50 to 300{K}}.
\newblock \bibinfo{journal}{Astrophys. J.} \bibinfo{volume}{318},
  \bibinfo{pages}{940--943}.
\bibitem[{{Borysow} et~al.(1988){Borysow}, {Frommhold} and
  {Birnbaum}}]{88borysow}
\bibinfo{author}{{Borysow}, J.}, \bibinfo{author}{{Frommhold}, L.},
  \bibinfo{author}{{Birnbaum}, G.}, \bibinfo{year}{1988}.
\newblock \bibinfo{title}{{Collison-induced rototranslational absorption
  spectra of H$_2$-He pairs at temperatures from 40 to 3000 K}}.
\newblock \bibinfo{journal}{Astrophys. J.} \bibinfo{volume}{326},
  \bibinfo{pages}{509--515}.
\newblock \DOIprefix\doi{10.1086/166112}.
\bibitem[{{Bouanich} et~al.(2003){Bouanich}, {Blanquet}, {Walrand} and
  {Lep{\`e}re}}]{03bouanich_c2h4}
\bibinfo{author}{{Bouanich}, J.P.}, \bibinfo{author}{{Blanquet}, G.},
  \bibinfo{author}{{Walrand}, J.}, \bibinfo{author}{{Lep{\`e}re}, M.},
  \bibinfo{year}{2003}.
\newblock \bibinfo{title}{{H$_{2}$-broadening in the {$\nu$}$_{7}$ band of
  ethylene by diode-laser spectroscopy}}.
\newblock \bibinfo{journal}{Journal of Molecular Spectroscopy}
  \bibinfo{volume}{218}, \bibinfo{pages}{22--27}.
\newblock \DOIprefix\doi{10.1016/S0022-2852(02)00034-6}.
\bibitem[{{Bouanich} et~al.(2004){Bouanich}, {Blanquet}, {Walrand} and
  {Lep{\`e}re}}]{04bouanich_c2h4}
\bibinfo{author}{{Bouanich}, J.P.}, \bibinfo{author}{{Blanquet}, G.},
  \bibinfo{author}{{Walrand}, J.}, \bibinfo{author}{{Lep{\`e}re}, M.},
  \bibinfo{year}{2004}.
\newblock \bibinfo{title}{{Hydrogen-broadening coefficients in the
  {$\nu$}$_{7}$ band of ethylene at low temperature}}.
\newblock \bibinfo{journal}{Journal of Molecular Spectroscopy}
  \bibinfo{volume}{227}, \bibinfo{pages}{172--179}.
\newblock \DOIprefix\doi{10.1016/j.jms.2004.06.001}.
\bibitem[{Bouanich et~al.(2004)Bouanich, Salem, Aroui, Walrand and
  Blanquet}]{04bouanich}
\bibinfo{author}{Bouanich, J.P.}, \bibinfo{author}{Salem, J.},
  \bibinfo{author}{Aroui, H.}, \bibinfo{author}{Walrand, J.},
  \bibinfo{author}{Blanquet, G.}, \bibinfo{year}{2004}.
\newblock \bibinfo{title}{{H}$_2$-broadening coefficients in the $\nu_2$ and
  $\nu_4$ bands of {PH}$_3$}.
\newblock \bibinfo{journal}{J. Quant. Spectro. Rad. Trans.}
  \bibinfo{volume}{84}, \bibinfo{pages}{195--205}.
\bibitem[{{Brown} et~al.(2003){Brown}, {Benner}, {Champion}, {Devi}, {Fejard},
  {Gamache}, {Gabard}, {Hilico}, {Lavorel}, {Loete}, {Mellau}, {Nikitin},
  {Pine}, {Predoi-Cross}, {Rinsland}, {Robert}, {Sams}, {Smith}, {Tashkun} and
  {Tyuterev}}]{03brown}
\bibinfo{author}{{Brown}, L.R.}, \bibinfo{author}{{Benner}, D.C.},
  \bibinfo{author}{{Champion}, J.P.}, \bibinfo{author}{{Devi}, V.M.},
  \bibinfo{author}{{Fejard}, L.}, \bibinfo{author}{{Gamache}, R.R.},
  \bibinfo{author}{{Gabard}, T.}, \bibinfo{author}{{Hilico}, J.C.},
  \bibinfo{author}{{Lavorel}, B.}, \bibinfo{author}{{Loete}, M.},
  \bibinfo{author}{{Mellau}, G.C.}, \bibinfo{author}{{Nikitin}, A.},
  \bibinfo{author}{{Pine}, A.S.}, \bibinfo{author}{{Predoi-Cross}, A.},
  \bibinfo{author}{{Rinsland}, C.P.}, \bibinfo{author}{{Robert}, O.},
  \bibinfo{author}{{Sams}, R.L.}, \bibinfo{author}{{Smith}, M.A.H.},
  \bibinfo{author}{{Tashkun}, S.A.}, \bibinfo{author}{{Tyuterev}, V.G.},
  \bibinfo{year}{2003}.
\newblock \bibinfo{title}{{Methane line parameters in HITRAN}}.
\newblock \bibinfo{journal}{Journal of Quantitative Spectroscopy and Radiative
  Transfer} \bibinfo{volume}{82}, \bibinfo{pages}{219--238}.
\bibitem[{{Brown} and {Peterson}(1994)}]{94brown}
\bibinfo{author}{{Brown}, L.R.}, \bibinfo{author}{{Peterson}, D.B.},
  \bibinfo{year}{1994}.
\newblock \bibinfo{title}{{An Empirical Expression for Linewidths of Ammonia
  from Far-Infrared Measurements}}.
\newblock \bibinfo{journal}{Journal of Molecular Spectroscopy}
  \bibinfo{volume}{168}, \bibinfo{pages}{593--606}.
\newblock \DOIprefix\doi{10.1006/jmsp.1994.1305}.
\bibitem[{Carlson et~al.(1993)Carlson, Lacis and Rossow}]{93carlson}
\bibinfo{author}{Carlson, B.}, \bibinfo{author}{Lacis, A.},
  \bibinfo{author}{Rossow, W.}, \bibinfo{year}{1993}.
\newblock \bibinfo{title}{Tropospheric gas composition and cloud structure of
  the jovian north equatorial belt}.
\newblock \bibinfo{journal}{Journal of Geophysical Research}
  \bibinfo{volume}{98}, \bibinfo{pages}{5251--5290}.
\bibitem[{{Carlson} et~al.(1992){Carlson}, {Lacis} and {Rossow}}]{92carlson}
\bibinfo{author}{{Carlson}, B.E.}, \bibinfo{author}{{Lacis}, A.A.},
  \bibinfo{author}{{Rossow}, W.B.}, \bibinfo{year}{1992}.
\newblock \bibinfo{title}{{Ortho-para-hydrogen equilibration on Jupiter}}.
\newblock \bibinfo{journal}{Astrophys. J.} \bibinfo{volume}{393},
  \bibinfo{pages}{357--372}.
\newblock \DOIprefix\doi{10.1086/171510}.
\bibitem[{{Cavali{\'e}} et~al.(2013){Cavali{\'e}}, {Feuchtgruber}, {Lellouch},
  {de Val-Borro}, {Jarchow}, {Moreno}, {Hartogh}, {Orton}, {Greathouse},
  {Billebaud}, {Dobrijevic}, {Lara}, {Gonz{\'a}lez} and {Sagawa}}]{13cavalie}
\bibinfo{author}{{Cavali{\'e}}, T.}, \bibinfo{author}{{Feuchtgruber}, H.},
  \bibinfo{author}{{Lellouch}, E.}, \bibinfo{author}{{de Val-Borro}, M.},
  \bibinfo{author}{{Jarchow}, C.}, \bibinfo{author}{{Moreno}, R.},
  \bibinfo{author}{{Hartogh}, P.}, \bibinfo{author}{{Orton}, G.},
  \bibinfo{author}{{Greathouse}, T.K.}, \bibinfo{author}{{Billebaud}, F.},
  \bibinfo{author}{{Dobrijevic}, M.}, \bibinfo{author}{{Lara}, L.M.},
  \bibinfo{author}{{Gonz{\'a}lez}, A.}, \bibinfo{author}{{Sagawa}, H.},
  \bibinfo{year}{2013}.
\newblock \bibinfo{title}{{Spatial distribution of water in the stratosphere of
  Jupiter from Herschel HIFI and PACS observations}}.
\newblock \bibinfo{journal}{Astron. Astrophys.} \bibinfo{volume}{553},
  \bibinfo{pages}{A21}.
\newblock \DOIprefix\doi{10.1051/0004-6361/201220797}.
\bibitem[{{Choi} et~al.(2007){Choi}, {Banfield}, {Gierasch} and
  {Showman}}]{07choi}
\bibinfo{author}{{Choi}, D.S.}, \bibinfo{author}{{Banfield}, D.},
  \bibinfo{author}{{Gierasch}, P.}, \bibinfo{author}{{Showman}, A.P.},
  \bibinfo{year}{2007}.
\newblock \bibinfo{title}{{Velocity and vorticity measurements of Jupiter's
  Great Red Spot using automated cloud feature tracking}}.
\newblock \bibinfo{journal}{Icarus} \bibinfo{volume}{188},
  \bibinfo{pages}{35--46}.
\newblock \DOIprefix\doi{10.1016/j.icarus.2006.10.037}.
\bibitem[{{Conrath} and {Gierasch}(1984)}]{84conrath}
\bibinfo{author}{{Conrath}, B.J.}, \bibinfo{author}{{Gierasch}, P.J.},
  \bibinfo{year}{1984}.
\newblock \bibinfo{title}{{Global variation of the para hydrogen fraction in
  Jupiter's atmosphere and implications for dynamics on the outer planets}}.
\newblock \bibinfo{journal}{Icarus} \bibinfo{volume}{57},
  \bibinfo{pages}{184--204}.
\newblock \DOIprefix\doi{10.1016/0019-1035(84)90065-4}.
\bibitem[{{Conrath} et~al.(1990){Conrath}, {Gierasch} and {Leroy}}]{90conrath}
\bibinfo{author}{{Conrath}, B.J.}, \bibinfo{author}{{Gierasch}, P.J.},
  \bibinfo{author}{{Leroy}, S.S.}, \bibinfo{year}{1990}.
\newblock \bibinfo{title}{{Temperature and circulation in the stratosphere of
  the outer planets}}.
\newblock \bibinfo{journal}{Icarus} \bibinfo{volume}{83},
  \bibinfo{pages}{255--281}.
\newblock \DOIprefix\doi{10.1016/0019-1035(90)90068-K}.
\bibitem[{{Conrath} et~al.(1998){Conrath}, {Gierasch} and
  {Ustinov}}]{98conrath}
\bibinfo{author}{{Conrath}, B.J.}, \bibinfo{author}{{Gierasch}, P.J.},
  \bibinfo{author}{{Ustinov}, E.A.}, \bibinfo{year}{1998}.
\newblock \bibinfo{title}{{Thermal Structure and Para Hydrogen Fraction on the
  Outer Planets from Voyager IRIS Measurements}}.
\newblock \bibinfo{journal}{Icarus} \bibinfo{volume}{135},
  \bibinfo{pages}{501--517}.
\newblock \DOIprefix\doi{10.1006/icar.1998.6000}.
\bibitem[{{Conrath} and {Pirraglia}(1983)}]{83conrath}
\bibinfo{author}{{Conrath}, B.J.}, \bibinfo{author}{{Pirraglia}, J.A.},
  \bibinfo{year}{1983}.
\newblock \bibinfo{title}{{Thermal structure of Saturn from Voyager infrared
  measurements - Implications for atmospheric dynamics}}.
\newblock \bibinfo{journal}{Icarus} \bibinfo{volume}{53},
  \bibinfo{pages}{286--292}.
\newblock \DOIprefix\doi{10.1016/0019-1035(83)90148-3}.
\bibitem[{{de Pater}(1986)}]{86depater}
\bibinfo{author}{{de Pater}, I.}, \bibinfo{year}{1986}.
\newblock \bibinfo{title}{{Jupiter's zone-belt structure at radio wavelengths.
  II - Comparison of observations with model atmosphere calculations}}.
\newblock \bibinfo{journal}{Icarus} \bibinfo{volume}{68},
  \bibinfo{pages}{344--365}.
\newblock \DOIprefix\doi{10.1016/0019-1035(86)90027-8}.
\bibitem[{{de Pater} et~al.(2001){de Pater}, {Dunn}, {Romani} and
  {Zahnle}}]{01depater}
\bibinfo{author}{{de Pater}, I.}, \bibinfo{author}{{Dunn}, D.},
  \bibinfo{author}{{Romani}, P.}, \bibinfo{author}{{Zahnle}, K.},
  \bibinfo{year}{2001}.
\newblock \bibinfo{title}{{Reconciling Galileo Probe Data and Ground-Based
  Radio Observations of Ammonia on Jupiter}}.
\newblock \bibinfo{journal}{Icarus} \bibinfo{volume}{149},
  \bibinfo{pages}{66--78}.
\newblock \DOIprefix\doi{10.1006/icar.2000.6527}.
\bibitem[{Deming et~al.(1989)Deming, Mumma, Espenak, Jennings, Kostiuk,
  Wiedemann, Loewenstein and Piscitelli}]{89deming}
\bibinfo{author}{Deming, D.}, \bibinfo{author}{Mumma, M.},
  \bibinfo{author}{Espenak, F.}, \bibinfo{author}{Jennings, D.},
  \bibinfo{author}{Kostiuk, T.}, \bibinfo{author}{Wiedemann, G.},
  \bibinfo{author}{Loewenstein, R.}, \bibinfo{author}{Piscitelli, J.},
  \bibinfo{year}{1989}.
\newblock \bibinfo{title}{A search for p-mode oscillations of jupiter-
  serendipitous observations of nonacoustic thermal wave structure}.
\newblock \bibinfo{journal}{Astrophysical Journal} \bibinfo{volume}{343},
  \bibinfo{pages}{456--467}.
\bibitem[{Deming et~al.(1997)Deming, Reuter, Jennings, Bjoraker, McCabe, Fast
  and Wiedemann}]{97deming}
\bibinfo{author}{Deming, D.}, \bibinfo{author}{Reuter, D.},
  \bibinfo{author}{Jennings, D.}, \bibinfo{author}{Bjoraker, G.},
  \bibinfo{author}{McCabe, G.}, \bibinfo{author}{Fast, K.},
  \bibinfo{author}{Wiedemann, G.}, \bibinfo{year}{1997}.
\newblock \bibinfo{title}{{Observations and Analysis of Longitudinal Thermal
  Waves on Jupiter}}.
\newblock \bibinfo{journal}{Icarus} \bibinfo{volume}{126},
  \bibinfo{pages}{301--312}.
\bibitem[{{Drossart} et~al.(1990){Drossart}, {Lellouch}, {Bezard}, {Maillard}
  and {Tarrogo}}]{90drossart}
\bibinfo{author}{{Drossart}, P.}, \bibinfo{author}{{Lellouch}, E.},
  \bibinfo{author}{{Bezard}, B.}, \bibinfo{author}{{Maillard}, J.P.},
  \bibinfo{author}{{Tarrogo}, G.}, \bibinfo{year}{1990}.
\newblock \bibinfo{title}{{Jupiter - Evidence for phosphine enhancement at high
  northern latitudes}}.
\newblock \bibinfo{journal}{Icarus} \bibinfo{volume}{83},
  \bibinfo{pages}{248--253}.
\newblock \DOIprefix\doi{10.1016/0019-1035(90)90018-5}.
\bibitem[{{Encrenaz} et~al.(2016){Encrenaz}, {Greathouse}, {Richter}, {DeWitt},
  {Drossart}, {Fouchet}, {Janssen}, {Gulkis}, {Orton}, {Bezard}, {Fletcher},
  {Giles} and {Atreya}}]{16encrenaz_texes}
\bibinfo{author}{{Encrenaz}, T.}, \bibinfo{author}{{Greathouse}, T.K.},
  \bibinfo{author}{{Richter}, M.J.}, \bibinfo{author}{{DeWitt}, C.},
  \bibinfo{author}{{Drossart}, P.}, \bibinfo{author}{{Fouchet}, T.},
  \bibinfo{author}{{Janssen}, M.A.}, \bibinfo{author}{{Gulkis}, S.},
  \bibinfo{author}{{Orton}, G.S.}, \bibinfo{author}{{Bezard}, B.},
  \bibinfo{author}{{Fletcher}, L.N.}, \bibinfo{author}{{Giles}, R.S.},
  \bibinfo{author}{{Atreya}, S.K.}, \bibinfo{year}{2016}.
\newblock \bibinfo{title}{{Monitoring Jovian dynamics using maps of NH$_3$ and
  PH$_3$}}.
\newblock \bibinfo{journal}{Astron. Astrophys., in prep.} .
\bibitem[{{Fast} et~al.(2011){Fast}, {Kostiuk}, {Livengood}, {Hewagama} and
  {Annen}}]{11fast}
\bibinfo{author}{{Fast}, K.E.}, \bibinfo{author}{{Kostiuk}, T.},
  \bibinfo{author}{{Livengood}, T.A.}, \bibinfo{author}{{Hewagama}, T.},
  \bibinfo{author}{{Annen}, J.}, \bibinfo{year}{2011}.
\newblock \bibinfo{title}{{Modification of Jupiter's stratosphere three weeks
  after the 2009 impact}}.
\newblock \bibinfo{journal}{Icarus} \bibinfo{volume}{213},
  \bibinfo{pages}{195--200}.
\newblock \DOIprefix\doi{10.1016/j.icarus.2011.02.008}.
\bibitem[{{Flasar} et~al.(1981){Flasar}, {Conrath}, {Pirraglia}, {Clark},
  {French} and {Gierasch}}]{81flasar}
\bibinfo{author}{{Flasar}, F.M.}, \bibinfo{author}{{Conrath}, B.J.},
  \bibinfo{author}{{Pirraglia}, J.}, \bibinfo{author}{{Clark}, P.C.},
  \bibinfo{author}{{French}, R.G.}, \bibinfo{author}{{Gierasch}, P.J.},
  \bibinfo{year}{1981}.
\newblock \bibinfo{title}{{Thermal structure and dynamics of the Jovian
  atmosphere. I - The Great Red Spot}}.
\newblock \bibinfo{journal}{J. Geophys. Res.} \bibinfo{volume}{86},
  \bibinfo{pages}{8759--8767}.
\newblock \DOIprefix\doi{10.1029/JA086iA10p08759}.
\bibitem[{{Flasar} et~al.(2004a){Flasar}, {Kunde}, {Abbas}, {Achterberg},
  {Ade}, {Barucci}, {B{\'e}zard}, {Bjoraker}, {Brasunas}, {Calcutt}, {Carlson},
  {C{\'e}sarsky}, {Conrath}, {Coradini}, {Courtin}, {Coustenis}, {Edberg},
  {Edgington}, {Ferrari}, {Fouchet}, {Gautier}, {Gierasch}, {Grossman},
  {Irwin}, {Jennings}, {Lellouch}, {Mamoutkine}, {Marten}, {Meyer}, {Nixon},
  {Orton}, {Owen}, {Pearl}, {Prang{\'e}}, {Raulin}, {Read}, {Romani},
  {Samuelson}, {Segura}, {Showalter}, {Simon-Miller}, {Smith}, {Spencer},
  {Spilker} and {Taylor}}]{04flasar}
\bibinfo{author}{{Flasar}, F.M.}, \bibinfo{author}{{Kunde}, V.G.},
  \bibinfo{author}{{Abbas}, M.M.}, \bibinfo{author}{{Achterberg}, R.K.},
  \bibinfo{author}{{Ade}, P.}, \bibinfo{author}{{Barucci}, A.},
  \bibinfo{author}{{B{\'e}zard}, B.}, \bibinfo{author}{{Bjoraker}, G.L.},
  \bibinfo{author}{{Brasunas}, J.C.}, \bibinfo{author}{{Calcutt}, S.},
  \bibinfo{author}{{Carlson}, R.}, \bibinfo{author}{{C{\'e}sarsky}, C.J.},
  \bibinfo{author}{{Conrath}, B.J.}, \bibinfo{author}{{Coradini}, A.},
  \bibinfo{author}{{Courtin}, R.}, \bibinfo{author}{{Coustenis}, A.},
  \bibinfo{author}{{Edberg}, S.}, \bibinfo{author}{{Edgington}, S.},
  \bibinfo{author}{{Ferrari}, C.}, \bibinfo{author}{{Fouchet}, T.},
  \bibinfo{author}{{Gautier}, D.}, \bibinfo{author}{{Gierasch}, P.J.},
  \bibinfo{author}{{Grossman}, K.}, \bibinfo{author}{{Irwin}, P.},
  \bibinfo{author}{{Jennings}, D.E.}, \bibinfo{author}{{Lellouch}, E.},
  \bibinfo{author}{{Mamoutkine}, A.A.}, \bibinfo{author}{{Marten}, A.},
  \bibinfo{author}{{Meyer}, J.P.}, \bibinfo{author}{{Nixon}, C.A.},
  \bibinfo{author}{{Orton}, G.S.}, \bibinfo{author}{{Owen}, T.C.},
  \bibinfo{author}{{Pearl}, J.C.}, \bibinfo{author}{{Prang{\'e}}, R.},
  \bibinfo{author}{{Raulin}, F.}, \bibinfo{author}{{Read}, P.L.},
  \bibinfo{author}{{Romani}, P.N.}, \bibinfo{author}{{Samuelson}, R.E.},
  \bibinfo{author}{{Segura}, M.E.}, \bibinfo{author}{{Showalter}, M.R.},
  \bibinfo{author}{{Simon-Miller}, A.A.}, \bibinfo{author}{{Smith}, M.D.},
  \bibinfo{author}{{Spencer}, J.R.}, \bibinfo{author}{{Spilker}, L.J.},
  \bibinfo{author}{{Taylor}, F.W.}, \bibinfo{year}{2004}a.
\newblock \bibinfo{title}{{Exploring The Saturn System In The Thermal Infrared:
  The Composite Infrared Spectrometer}}.
\newblock \bibinfo{journal}{Space Science Reviews} \bibinfo{volume}{115},
  \bibinfo{pages}{169--297}.
\newblock \DOIprefix\doi{10.1007/s11214-004-1454-9}.
\bibitem[{{Flasar} et~al.(2004b){Flasar}, {Kunde}, {Achterberg}, {Conrath},
  {Simon-Miller}, {Nixon}, {Gierasch}, {Romani}, {B{\'e}zard}, {Irwin},
  {Bjoraker}, {Brasunas}, {Jennings}, {Pearl}, {Smith}, {Orton}, {Spilker},
  {Carlson}, {Calcutt}, {Read}, {Taylor}, {Parrish}, {Barucci}, {Courtin},
  {Coustenis}, {Gautier}, {Lellouch}, {Marten}, {Prang{\'e}}, {Biraud},
  {Fouchet}, {Ferrari}, {Owen}, {Abbas}, {Samuelson}, {Raulin}, {Ade},
  {C{\'e}sarsky}, {Grossman} and {Coradini}}]{04flasar_jup}
\bibinfo{author}{{Flasar}, F.M.}, \bibinfo{author}{{Kunde}, V.G.},
  \bibinfo{author}{{Achterberg}, R.K.}, \bibinfo{author}{{Conrath}, B.J.},
  \bibinfo{author}{{Simon-Miller}, A.A.}, \bibinfo{author}{{Nixon}, C.A.},
  \bibinfo{author}{{Gierasch}, P.J.}, \bibinfo{author}{{Romani}, P.N.},
  \bibinfo{author}{{B{\'e}zard}, B.}, \bibinfo{author}{{Irwin}, P.},
  \bibinfo{author}{{Bjoraker}, G.L.}, \bibinfo{author}{{Brasunas}, J.C.},
  \bibinfo{author}{{Jennings}, D.E.}, \bibinfo{author}{{Pearl}, J.C.},
  \bibinfo{author}{{Smith}, M.D.}, \bibinfo{author}{{Orton}, G.S.},
  \bibinfo{author}{{Spilker}, L.J.}, \bibinfo{author}{{Carlson}, R.},
  \bibinfo{author}{{Calcutt}, S.B.}, \bibinfo{author}{{Read}, P.L.},
  \bibinfo{author}{{Taylor}, F.W.}, \bibinfo{author}{{Parrish}, P.},
  \bibinfo{author}{{Barucci}, A.}, \bibinfo{author}{{Courtin}, R.},
  \bibinfo{author}{{Coustenis}, A.}, \bibinfo{author}{{Gautier}, D.},
  \bibinfo{author}{{Lellouch}, E.}, \bibinfo{author}{{Marten}, A.},
  \bibinfo{author}{{Prang{\'e}}, R.}, \bibinfo{author}{{Biraud}, Y.},
  \bibinfo{author}{{Fouchet}, T.}, \bibinfo{author}{{Ferrari}, C.},
  \bibinfo{author}{{Owen}, T.C.}, \bibinfo{author}{{Abbas}, M.M.},
  \bibinfo{author}{{Samuelson}, R.E.}, \bibinfo{author}{{Raulin}, F.},
  \bibinfo{author}{{Ade}, P.}, \bibinfo{author}{{C{\'e}sarsky}, C.J.},
  \bibinfo{author}{{Grossman}, K.U.}, \bibinfo{author}{{Coradini}, A.},
  \bibinfo{year}{2004}b.
\newblock \bibinfo{title}{{An intense stratospheric jet on Jupiter}}.
\newblock \bibinfo{journal}{Nature} \bibinfo{volume}{427},
  \bibinfo{pages}{132--135}.
\bibitem[{{Fletcher} et~al.(2010a){Fletcher}, {Achterberg}, {Greathouse},
  {Orton}, {Conrath}, {Simon-Miller}, {Teanby}, {Guerlet}, {Irwin} and
  {Flasar}}]{10fletcher_seasons}
\bibinfo{author}{{Fletcher}, L.N.}, \bibinfo{author}{{Achterberg}, R.K.},
  \bibinfo{author}{{Greathouse}, T.K.}, \bibinfo{author}{{Orton}, G.S.},
  \bibinfo{author}{{Conrath}, B.J.}, \bibinfo{author}{{Simon-Miller}, A.A.},
  \bibinfo{author}{{Teanby}, N.}, \bibinfo{author}{{Guerlet}, S.},
  \bibinfo{author}{{Irwin}, P.G.J.}, \bibinfo{author}{{Flasar}, F.M.},
  \bibinfo{year}{2010}a.
\newblock \bibinfo{title}{{Seasonal change on Saturn from Cassini/CIRS
  observations, 2004-2009}}.
\newblock \bibinfo{journal}{Icarus} \bibinfo{volume}{208},
  \bibinfo{pages}{337--352}.
\newblock \DOIprefix\doi{10.1016/j.icarus.2010.01.022}.
\bibitem[{{Fletcher} et~al.(2014){Fletcher}, {Greathouse}, {Orton}, {Irwin},
  {Mousis}, {Sinclair} and {Giles}}]{14fletcher_texes}
\bibinfo{author}{{Fletcher}, L.N.}, \bibinfo{author}{{Greathouse}, T.K.},
  \bibinfo{author}{{Orton}, G.S.}, \bibinfo{author}{{Irwin}, P.G.J.},
  \bibinfo{author}{{Mousis}, O.}, \bibinfo{author}{{Sinclair}, J.A.},
  \bibinfo{author}{{Giles}, R.S.}, \bibinfo{year}{2014}.
\newblock \bibinfo{title}{{The origin of nitrogen on Jupiter and Saturn from
  the 15N/14N ratio}}.
\newblock \bibinfo{journal}{Icarus} \bibinfo{volume}{238},
  \bibinfo{pages}{170--190}.
\newblock \DOIprefix\doi{10.1016/j.icarus.2014.05.007},
  \href{http://arxiv.org/abs/1405.3800}{{\tt arXiv:1405.3800}}.
\bibitem[{{Fletcher} et~al.(2016){Fletcher}, {Irwin}, {Achterberg}, {Orton} and
  {Flasar}}]{16fletcher}
\bibinfo{author}{{Fletcher}, L.N.}, \bibinfo{author}{{Irwin}, P.G.J.},
  \bibinfo{author}{{Achterberg}, R.K.}, \bibinfo{author}{{Orton}, G.S.},
  \bibinfo{author}{{Flasar}, F.M.}, \bibinfo{year}{2016}.
\newblock \bibinfo{title}{{Seasonal variability of Saturn's tropospheric
  temperatures, winds and para-H$_{2}$ from Cassini far-IR spectroscopy}}.
\newblock \bibinfo{journal}{Icarus} \bibinfo{volume}{264},
  \bibinfo{pages}{137--159}.
\newblock \DOIprefix\doi{10.1016/j.icarus.2015.09.009},
  \href{http://arxiv.org/abs/1509.02281}{{\tt arXiv:1509.02281}}.
\bibitem[{{Fletcher} et~al.(2015){Fletcher}, {Irwin}, {Sinclair}, {Orton},
  {Giles}, {Hurley}, {Gorius}, {Achterberg}, {Hesman} and
  {Bjoraker}}]{15fletcher_poles}
\bibinfo{author}{{Fletcher}, L.N.}, \bibinfo{author}{{Irwin}, P.G.J.},
  \bibinfo{author}{{Sinclair}, J.A.}, \bibinfo{author}{{Orton}, G.S.},
  \bibinfo{author}{{Giles}, R.S.}, \bibinfo{author}{{Hurley}, J.},
  \bibinfo{author}{{Gorius}, N.}, \bibinfo{author}{{Achterberg}, R.K.},
  \bibinfo{author}{{Hesman}, B.E.}, \bibinfo{author}{{Bjoraker}, G.L.},
  \bibinfo{year}{2015}.
\newblock \bibinfo{title}{{Seasonal Evolution of Saturn's Polar Temperatures
  and Composition}}.
\newblock \bibinfo{journal}{Icarus} \bibinfo{volume}{250},
  \bibinfo{pages}{131--153}.
\bibitem[{{Fletcher} et~al.(2011a){Fletcher}, {Orton}, {de Pater}, {Edwards},
  {Yanamandra-Fisher}, {Hammel} and {Lisse}}]{11fletcher_trecs}
\bibinfo{author}{{Fletcher}, L.N.}, \bibinfo{author}{{Orton}, G.S.},
  \bibinfo{author}{{de Pater}, I.}, \bibinfo{author}{{Edwards}, M.},
  \bibinfo{author}{{Yanamandra-Fisher}, P.}, \bibinfo{author}{{Hammel}, H.B.},
  \bibinfo{author}{{Lisse}, C.M.}, \bibinfo{year}{2011}a.
\newblock \bibinfo{title}{{The Aftermath of the July 2009 Impact on Jupiter:
  Ammonia, Temperatures and Particulates from Gemini Thermal Infrared
  Spectroscopy}}.
\newblock \bibinfo{journal}{Icarus} \bibinfo{volume}{211},
  \bibinfo{pages}{568--586}.
\bibitem[{{Fletcher} et~al.(2010b){Fletcher}, {Orton}, {Mousis},
  {Yanamandra-Fisher}, {Parrish}, {Irwin}, {Fisher}, {Vanzi}, {Fujiyoshi},
  {Fuse}, {Simon-Miller}, {Edkins}, {Hayward} and {De Buizer}}]{10fletcher_grs}
\bibinfo{author}{{Fletcher}, L.N.}, \bibinfo{author}{{Orton}, G.S.},
  \bibinfo{author}{{Mousis}, O.}, \bibinfo{author}{{Yanamandra-Fisher}, P.},
  \bibinfo{author}{{Parrish}, P.D.}, \bibinfo{author}{{Irwin}, P.G.J.},
  \bibinfo{author}{{Fisher}, B.M.}, \bibinfo{author}{{Vanzi}, L.},
  \bibinfo{author}{{Fujiyoshi}, T.}, \bibinfo{author}{{Fuse}, T.},
  \bibinfo{author}{{Simon-Miller}, A.A.}, \bibinfo{author}{{Edkins}, E.},
  \bibinfo{author}{{Hayward}, T.L.}, \bibinfo{author}{{De Buizer}, J.},
  \bibinfo{year}{2010}b.
\newblock \bibinfo{title}{{Thermal structure and composition of Jupiter's Great
  Red Spot from high-resolution thermal imaging}}.
\newblock \bibinfo{journal}{Icarus} \bibinfo{volume}{208},
  \bibinfo{pages}{306--328}.
\newblock \DOIprefix\doi{10.1016/j.icarus.2010.01.005}.
\bibitem[{{Fletcher} et~al.(2011b){Fletcher}, {Orton}, {Rogers},
  {Simon-Miller}, {de Pater}, {Wong}, {Mousis}, {Irwin}, {Jacquesson} and
  {Yanamandra-Fisher}}]{11fletcher_fade}
\bibinfo{author}{{Fletcher}, L.N.}, \bibinfo{author}{{Orton}, G.S.},
  \bibinfo{author}{{Rogers}, J.H.}, \bibinfo{author}{{Simon-Miller}, A.A.},
  \bibinfo{author}{{de Pater}, I.}, \bibinfo{author}{{Wong}, M.H.},
  \bibinfo{author}{{Mousis}, O.}, \bibinfo{author}{{Irwin}, P.G.J.},
  \bibinfo{author}{{Jacquesson}, M.}, \bibinfo{author}{{Yanamandra-Fisher},
  P.A.}, \bibinfo{year}{2011}b.
\newblock \bibinfo{title}{{Jovian temperature and cloud variability during the
  2009-2010 fade of the South Equatorial Belt}}.
\newblock \bibinfo{journal}{Icarus} \bibinfo{volume}{213},
  \bibinfo{pages}{564--580}.
\newblock \DOIprefix\doi{10.1016/j.icarus.2011.03.007}.
\bibitem[{{Fletcher} et~al.(2009a){Fletcher}, {Orton}, {Teanby} and
  {Irwin}}]{09fletcher_ph3}
\bibinfo{author}{{Fletcher}, L.N.}, \bibinfo{author}{{Orton}, G.S.},
  \bibinfo{author}{{Teanby}, N.A.}, \bibinfo{author}{{Irwin}, P.G.J.},
  \bibinfo{year}{2009}a.
\newblock \bibinfo{title}{{Phosphine on Jupiter and Saturn from Cassini/CIRS}}.
\newblock \bibinfo{journal}{Icarus} \bibinfo{volume}{202},
  \bibinfo{pages}{543--564}.
\newblock \DOIprefix\doi{10.1016/j.icarus.2009.03.023}.
\bibitem[{{Fletcher} et~al.(2009b){Fletcher}, {Orton}, {Yanamandra-Fisher},
  {Fisher}, {Parrish} and {Irwin}}]{09fletcher_imaging}
\bibinfo{author}{{Fletcher}, L.N.}, \bibinfo{author}{{Orton}, G.S.},
  \bibinfo{author}{{Yanamandra-Fisher}, P.}, \bibinfo{author}{{Fisher}, B.M.},
  \bibinfo{author}{{Parrish}, P.D.}, \bibinfo{author}{{Irwin}, P.G.J.},
  \bibinfo{year}{2009}b.
\newblock \bibinfo{title}{{Retrievals of atmospheric variables on the gas
  giants from ground-based mid-infrared imaging}}.
\newblock \bibinfo{journal}{Icarus} \bibinfo{volume}{200},
  \bibinfo{pages}{154--175}.
\newblock \DOIprefix\doi{10.1016/j.icarus.2008.11.019}.
\bibitem[{{Fouchet} et~al.(2016){Fouchet}, {Greathouse}, {Spiga}, {Fletcher},
  {Guerlet}, {Leconte} and {Orton}}]{16fouchet_texes}
\bibinfo{author}{{Fouchet}, T.}, \bibinfo{author}{{Greathouse}, T.K.},
  \bibinfo{author}{{Spiga}, A.}, \bibinfo{author}{{Fletcher}, L.N.},
  \bibinfo{author}{{Guerlet}, S.}, \bibinfo{author}{{Leconte}, J.},
  \bibinfo{author}{{Orton}, G.S.}, \bibinfo{year}{2016}.
\newblock \bibinfo{title}{{Stratospheric aftermath of the 2010 Storm on Saturn
  as observed by the TEXES instrument. I. Temperature Structure}}.
\newblock \bibinfo{journal}{Icarus, submitted} .
\bibitem[{Fouchet et~al.(2008)Fouchet, Guerlet, Strobel, Simon-Miller,
  B{\'e}zard and Flasar}]{08fouchet}
\bibinfo{author}{Fouchet, T.}, \bibinfo{author}{Guerlet, S.},
  \bibinfo{author}{Strobel, D.}, \bibinfo{author}{Simon-Miller, A.},
  \bibinfo{author}{B{\'e}zard, B.}, \bibinfo{author}{Flasar, F.},
  \bibinfo{year}{2008}.
\newblock \bibinfo{title}{{An equatorial oscillation in Saturn's middle
  atmosphere.}}
\newblock \bibinfo{journal}{Nature} \bibinfo{volume}{453},
  \bibinfo{pages}{200--202}.
\bibitem[{{Fouchet} et~al.(2000){Fouchet}, {Lellouch}, {B{\'e}zard},
  {Encrenaz}, {Drossart}, {Feuchtgruber} and {de Graauw}}]{00fouchet}
\bibinfo{author}{{Fouchet}, T.}, \bibinfo{author}{{Lellouch}, E.},
  \bibinfo{author}{{B{\'e}zard}, B.}, \bibinfo{author}{{Encrenaz}, T.},
  \bibinfo{author}{{Drossart}, P.}, \bibinfo{author}{{Feuchtgruber}, H.},
  \bibinfo{author}{{de Graauw}, T.}, \bibinfo{year}{2000}.
\newblock \bibinfo{title}{{ISO-SWS Observations of Jupiter: Measurement of the
  Ammonia Tropospheric Profile and of the $^{15}$N/ $^{14}$N Isotopic Ratio}}.
\newblock \bibinfo{journal}{Icarus} \bibinfo{volume}{143},
  \bibinfo{pages}{223--243}.
\newblock \DOIprefix\doi{10.1006/icar.1999.6255},
  \href{http://arxiv.org/abs/astro-ph/9911257}{{\tt arXiv:astro-ph/9911257}}.
\bibitem[{{Friedson}(1999)}]{99friedson}
\bibinfo{author}{{Friedson}, A.J.}, \bibinfo{year}{1999}.
\newblock \bibinfo{title}{{New Observations and Modelling of a QBO-Like
  Oscillation in Jupiter's Stratosphere}}.
\newblock \bibinfo{journal}{Icarus} \bibinfo{volume}{137},
  \bibinfo{pages}{34--55}.
\newblock \DOIprefix\doi{10.1006/icar.1998.6038}.
\bibitem[{{Friedson}(2005)}]{05friedson}
\bibinfo{author}{{Friedson}, A.J.}, \bibinfo{year}{2005}.
\newblock \bibinfo{title}{{Water, ammonia, and H $_{2}$S mixing ratios in
  Jupiter's five-micron hot spots: A dynamical model}}.
\newblock \bibinfo{journal}{Icarus} \bibinfo{volume}{177},
  \bibinfo{pages}{1--17}.
\newblock \DOIprefix\doi{10.1016/j.icarus.2005.03.004}.
\bibitem[{{Geiss} and {Gloeckler}(2003)}]{03geiss}
\bibinfo{author}{{Geiss}, J.}, \bibinfo{author}{{Gloeckler}, G.},
  \bibinfo{year}{2003}.
\newblock \bibinfo{title}{{Isotopic Composition of H, HE and NE in the
  Protosolar Cloud}}.
\newblock \bibinfo{journal}{Space Sci. Rev.} \bibinfo{volume}{106},
  \bibinfo{pages}{3--18}.
\newblock \DOIprefix\doi{10.1023/A:1024651232758}.
\bibitem[{{Gierasch} et~al.(1986){Gierasch}, {Magalhaes} and
  {Conrath}}]{86gierasch}
\bibinfo{author}{{Gierasch}, P.J.}, \bibinfo{author}{{Magalhaes}, J.A.},
  \bibinfo{author}{{Conrath}, B.J.}, \bibinfo{year}{1986}.
\newblock \bibinfo{title}{{Zonal mean properties of Jupiter's upper troposphere
  from Voyager infrared observations}}.
\newblock \bibinfo{journal}{Icarus} \bibinfo{volume}{67},
  \bibinfo{pages}{456--483}.
\newblock \DOIprefix\doi{10.1016/0019-1035(86)90125-9}.
\bibitem[{{Giles} et~al.(2015){Giles}, {Fletcher} and {Irwin}}]{15giles}
\bibinfo{author}{{Giles}, R.S.}, \bibinfo{author}{{Fletcher}, L.N.},
  \bibinfo{author}{{Irwin}, P.G.J.}, \bibinfo{year}{2015}.
\newblock \bibinfo{title}{{Cloud structure and composition of Jupiter's
  troposphere from 5- {$\mu$} m Cassini VIMS spectroscopy}}.
\newblock \bibinfo{journal}{Icarus} \bibinfo{volume}{257},
  \bibinfo{pages}{457--470}.
\newblock \DOIprefix\doi{10.1016/j.icarus.2015.05.030},
  \href{http://arxiv.org/abs/1506.01608}{{\tt arXiv:1506.01608}}.
\bibitem[{{Goody} et~al.(1989){Goody}, {West}, {Chen} and {Crisp}}]{89goody_ck}
\bibinfo{author}{{Goody}, R.}, \bibinfo{author}{{West}, R.},
  \bibinfo{author}{{Chen}, L.}, \bibinfo{author}{{Crisp}, D.},
  \bibinfo{year}{1989}.
\newblock \bibinfo{title}{{The correlated-k method for radiation calculations
  in nonhomogeneous atmospheres}}.
\newblock \bibinfo{journal}{Journal of Quantitative Spectroscopy and Radiative
  Transfer} \bibinfo{volume}{42}, \bibinfo{pages}{539--550}.
\newblock \DOIprefix\doi{10.1016/0022-4073(89)90044-7}.
\bibitem[{{Greathouse} et~al.(2005){Greathouse}, {Lacy}, {B{\'e}zard}, {Moses},
  {Griffith} and {Richter}}]{05greathouse}
\bibinfo{author}{{Greathouse}, T.K.}, \bibinfo{author}{{Lacy}, J.H.},
  \bibinfo{author}{{B{\'e}zard}, B.}, \bibinfo{author}{{Moses}, J.I.},
  \bibinfo{author}{{Griffith}, C.A.}, \bibinfo{author}{{Richter}, M.J.},
  \bibinfo{year}{2005}.
\newblock \bibinfo{title}{{Meridional variations of temperature, C$_{2}$H$_{2}$
  and C$_{2}$H$_{6}$ abundances in Saturn's stratosphere at southern summer
  solstice}}.
\newblock \bibinfo{journal}{Icarus} \bibinfo{volume}{177},
  \bibinfo{pages}{18--31}.
\newblock \DOIprefix\doi{10.1016/j.icarus.2005.02.016}.
\bibitem[{Griffith et~al.(1992)Griffith, B{\'e}zard, Owen and
  Gautier}]{92griffith}
\bibinfo{author}{Griffith, C.}, \bibinfo{author}{B{\'e}zard, B.},
  \bibinfo{author}{Owen, T.}, \bibinfo{author}{Gautier, D.},
  \bibinfo{year}{1992}.
\newblock \bibinfo{title}{The tropospheric abundances of nh3 and ph3 in
  jupiter's great red spot, from voyager iris observations}.
\newblock \bibinfo{journal}{Icarus} \bibinfo{volume}{98},
  \bibinfo{pages}{82--93}.
\bibitem[{{Griffith} et~al.(2004){Griffith}, {B{\'e}zard}, {Greathouse},
  {Lellouch}, {Lacy}, {Kelly} and {Richter}}]{04griffith}
\bibinfo{author}{{Griffith}, C.A.}, \bibinfo{author}{{B{\'e}zard}, B.},
  \bibinfo{author}{{Greathouse}, T.}, \bibinfo{author}{{Lellouch}, E.},
  \bibinfo{author}{{Lacy}, J.}, \bibinfo{author}{{Kelly}, D.},
  \bibinfo{author}{{Richter}, M.J.}, \bibinfo{year}{2004}.
\newblock \bibinfo{title}{{Meridional transport of HCN from SL9 impacts on
  Jupiter}}.
\newblock \bibinfo{journal}{Icarus} \bibinfo{volume}{170},
  \bibinfo{pages}{58--69}.
\newblock \DOIprefix\doi{10.1016/j.icarus.2004.02.006}.
\bibitem[{{Griffith} et~al.(1997){Griffith}, {Bezard}, {Greathouse}, {Kelly},
  {Lacy} and {Noll}}]{97griffith}
\bibinfo{author}{{Griffith}, C.A.}, \bibinfo{author}{{Bezard}, B.},
  \bibinfo{author}{{Greathouse}, T.K.}, \bibinfo{author}{{Kelly}, D.M.},
  \bibinfo{author}{{Lacy}, J.H.}, \bibinfo{author}{{Noll}, K.S.},
  \bibinfo{year}{1997}.
\newblock \bibinfo{title}{{Thermal Infrared Imaging Spectroscopy of
  Shoemaker-Levy 9 Impact Sites: Spatial and Vertical Distributions of NH\_3,
  C\_2H\_4, and 10-{$\mu$}m Dust Emission}}.
\newblock \bibinfo{journal}{Icarus} \bibinfo{volume}{128},
  \bibinfo{pages}{275--293}.
\newblock \DOIprefix\doi{10.1006/icar.1997.5752}.
\bibitem[{{Guerlet} et~al.(2011){Guerlet}, {Fouchet}, {B{\'e}zard}, {Flasar}
  and {Simon-Miller}}]{11guerlet}
\bibinfo{author}{{Guerlet}, S.}, \bibinfo{author}{{Fouchet}, T.},
  \bibinfo{author}{{B{\'e}zard}, B.}, \bibinfo{author}{{Flasar}, F.M.},
  \bibinfo{author}{{Simon-Miller}, A.A.}, \bibinfo{year}{2011}.
\newblock \bibinfo{title}{{Evolution of the equatorial oscillation in Saturn's
  stratosphere between 2005 and 2010 from Cassini/CIRS limb data analysis}}.
\newblock \bibinfo{journal}{Geophysical Research Letters} \bibinfo{volume}{38},
  \bibinfo{pages}{9201}.
\newblock \DOIprefix\doi{10.1029/2011GL047192}.
\bibitem[{{Halsey} et~al.(1988){Halsey}, {Hillman}, {Nadler} and
  {Jennings}}]{88halsey}
\bibinfo{author}{{Halsey}, G.W.}, \bibinfo{author}{{Hillman}, J.J.},
  \bibinfo{author}{{Nadler}, S.}, \bibinfo{author}{{Jennings}, D.E.},
  \bibinfo{year}{1988}.
\newblock \bibinfo{title}{{Temperature dependence of the hydrogen-broadening
  coefficient for the nu 9 fundamental of ethane}}.
\newblock \bibinfo{journal}{Journal of Quantitative Spectroscopy and Radiative
  Transfer} \bibinfo{volume}{39}, \bibinfo{pages}{429--434}.
\newblock \DOIprefix\doi{10.1016/0022-4073(88)90087-8}.
\bibitem[{{Hanel} et~al.(1977){Hanel}, {Conrath}, {Kunde}, {Lowman}, {Maguire},
  {Pearl}, {Pirraglia}, {Gautier}, {Gierasch} and {Kumar}}]{77hanel}
\bibinfo{author}{{Hanel}, R.}, \bibinfo{author}{{Conrath}, B.},
  \bibinfo{author}{{Kunde}, V.}, \bibinfo{author}{{Lowman}, P.},
  \bibinfo{author}{{Maguire}, W.}, \bibinfo{author}{{Pearl}, J.},
  \bibinfo{author}{{Pirraglia}, J.}, \bibinfo{author}{{Gautier}, D.},
  \bibinfo{author}{{Gierasch}, P.}, \bibinfo{author}{{Kumar}, S.},
  \bibinfo{year}{1977}.
\newblock \bibinfo{title}{{The Voyager infrared spectroscopy and radiometry
  investigation}}.
\newblock \bibinfo{journal}{Space Science Reviews} \bibinfo{volume}{21},
  \bibinfo{pages}{129--157}.
\newblock \DOIprefix\doi{10.1007/BF00200848}.
\bibitem[{{Harrington} et~al.(2004){Harrington}, {de Pater}, {Brecht},
  {Deming}, {Meadows}, {Zahnle} and {Nicholson}}]{04harrington}
\bibinfo{author}{{Harrington}, J.}, \bibinfo{author}{{de Pater}, I.},
  \bibinfo{author}{{Brecht}, S.H.}, \bibinfo{author}{{Deming}, D.},
  \bibinfo{author}{{Meadows}, V.}, \bibinfo{author}{{Zahnle}, K.},
  \bibinfo{author}{{Nicholson}, P.D.}, \bibinfo{year}{2004}.
\newblock \bibinfo{title}{{Lessons from Shoemaker-Levy 9 about Jupiter and
  planetary impacts}}. \bibinfo{publisher}{Cambridge Planetary Science.
  Cambridge Univ. Press, New York}. chapter~\bibinfo{chapter}{8}.
\newblock pp. \bibinfo{pages}{159--184}.
\bibitem[{{Harrington} et~al.(1996){Harrington}, {Dowling} and
  {Baron}}]{96harrington}
\bibinfo{author}{{Harrington}, J.}, \bibinfo{author}{{Dowling}, T.E.},
  \bibinfo{author}{{Baron}, R.L.}, \bibinfo{year}{1996}.
\newblock \bibinfo{title}{{Jupiter's Tropospheric Thermal Emission. II. Power
  Spectrum Analysis and Wave Search}}.
\newblock \bibinfo{journal}{Icarus} \bibinfo{volume}{124},
  \bibinfo{pages}{32--44}.
\newblock \DOIprefix\doi{10.1006/icar.1996.0188}.
\bibitem[{{Hue} et~al.(2015a){Hue}, {Cavali{\'e}}, {Dobrijevic}, {Hersant} and
  {Greathouse}}]{15hue}
\bibinfo{author}{{Hue}, V.}, \bibinfo{author}{{Cavali{\'e}}, T.},
  \bibinfo{author}{{Dobrijevic}, M.}, \bibinfo{author}{{Hersant}, F.},
  \bibinfo{author}{{Greathouse}, T.K.}, \bibinfo{year}{2015}a.
\newblock \bibinfo{title}{{2D photochemical modeling of Saturn's stratosphere.
  Part I: Seasonal variation of atmospheric composition without meridional
  transport}}.
\newblock \bibinfo{journal}{Icarus} \bibinfo{volume}{257},
  \bibinfo{pages}{163--184}.
\newblock \DOIprefix\doi{10.1016/j.icarus.2015.04.001},
  \href{http://arxiv.org/abs/1504.02326}{{\tt arXiv:1504.02326}}.
\bibitem[{{Hue} et~al.(2015b){Hue}, {Hersant}, {Cavali{\'e}} and
  {Dobrijevic}}]{15hue_dps}
\bibinfo{author}{{Hue}, V.}, \bibinfo{author}{{Hersant}, F.},
  \bibinfo{author}{{Cavali{\'e}}, T.}, \bibinfo{author}{{Dobrijevic}, M.},
  \bibinfo{year}{2015}b.
\newblock \bibinfo{title}{{Photochemistry, mixing and transport in Jupiter's
  stratosphere constrained by Cassini}}, in: \bibinfo{booktitle}{AAS/Division
  for Planetary Sciences Meeting Abstracts}, p. \bibinfo{pages}{311.15}.
\bibitem[{{Ingersoll} et~al.(2000){Ingersoll}, J., D., {Vasavada} and {Galileo
  Imaging Team}}]{00ingersoll}
\bibinfo{author}{{Ingersoll}, A.P.}, \bibinfo{author}{J., G.P.},
  \bibinfo{author}{D., B.}, \bibinfo{author}{{Vasavada}, A.R.},
  \bibinfo{author}{{Galileo Imaging Team}}, \bibinfo{year}{2000}.
\newblock \bibinfo{title}{{Moist convection as an energy source for the
  large-scale motions in Jupiter's atmosphere}}.
\newblock \bibinfo{journal}{Nature} \bibinfo{volume}{403},
  \bibinfo{pages}{630--632}.
\newblock \DOIprefix\doi{10.1038/35001021}.
\bibitem[{Irwin et~al.(2008)Irwin, Teanby, de~Kok, Fletcher, Howett, Tsang,
  Wilson, Calcutt, Nixon and Parrish}]{08irwin}
\bibinfo{author}{Irwin, P.}, \bibinfo{author}{Teanby, N.},
  \bibinfo{author}{de~Kok, R.}, \bibinfo{author}{Fletcher, L.},
  \bibinfo{author}{Howett, C.}, \bibinfo{author}{Tsang, C.},
  \bibinfo{author}{Wilson, C.}, \bibinfo{author}{Calcutt, S.},
  \bibinfo{author}{Nixon, C.}, \bibinfo{author}{Parrish, P.},
  \bibinfo{year}{2008}.
\newblock \bibinfo{title}{{The NEMESIS planetary atmosphere radiative transfer
  and retrieval tool}}.
\newblock \bibinfo{journal}{Journal of Quantitative Spectroscopy and Radiative
  Transfer} \bibinfo{volume}{109}, \bibinfo{pages}{1136--1150}.
\bibitem[{{Irwin} and {Dyudina}(2002)}]{02irwin}
\bibinfo{author}{{Irwin}, P.G.J.}, \bibinfo{author}{{Dyudina}, U.},
  \bibinfo{year}{2002}.
\newblock \bibinfo{title}{{The Retrieval of Cloud Structure Maps in the
  Equatorial Region of Jupiter Using a Principal Component Analysis of
  Galileo/NIMS Data}}.
\newblock \bibinfo{journal}{Icarus} \bibinfo{volume}{156},
  \bibinfo{pages}{52--63}.
\newblock \DOIprefix\doi{10.1006/icar.2001.6773}.
\bibitem[{{Irwin} et~al.(2004){Irwin}, {Parrish}, {Fouchet}, {Calcutt},
  {Taylor}, {Simon-Miller} and {Nixon}}]{04irwin}
\bibinfo{author}{{Irwin}, P.G.J.}, \bibinfo{author}{{Parrish}, P.},
  \bibinfo{author}{{Fouchet}, T.}, \bibinfo{author}{{Calcutt}, S.B.},
  \bibinfo{author}{{Taylor}, F.W.}, \bibinfo{author}{{Simon-Miller}, A.A.},
  \bibinfo{author}{{Nixon}, C.A.}, \bibinfo{year}{2004}.
\newblock \bibinfo{title}{{Retrievals of jovian tropospheric phosphine from
  Cassini/CIRS}}.
\newblock \bibinfo{journal}{Icarus} \bibinfo{volume}{172},
  \bibinfo{pages}{37--49}.
\newblock \DOIprefix\doi{10.1016/j.icarus.2003.09.027}.
\bibitem[{{Irwin} et~al.(1998){Irwin}, {Weir}, {Smith}, {Taylor}, {Lambert},
  {Calcutt}, {Cameron-Smith}, {Carlson}, {Baines}, {Orton}, {Drossart},
  {Encrenaz} and {Roos-Serote}}]{98irwin}
\bibinfo{author}{{Irwin}, P.G.J.}, \bibinfo{author}{{Weir}, A.L.},
  \bibinfo{author}{{Smith}, S.E.}, \bibinfo{author}{{Taylor}, F.W.},
  \bibinfo{author}{{Lambert}, A.L.}, \bibinfo{author}{{Calcutt}, S.B.},
  \bibinfo{author}{{Cameron-Smith}, P.J.}, \bibinfo{author}{{Carlson}, R.W.},
  \bibinfo{author}{{Baines}, K.}, \bibinfo{author}{{Orton}, G.S.},
  \bibinfo{author}{{Drossart}, P.}, \bibinfo{author}{{Encrenaz}, T.},
  \bibinfo{author}{{Roos-Serote}, M.}, \bibinfo{year}{1998}.
\newblock \bibinfo{title}{{Cloud structure and atmospheric composition of
  Jupiter retrieved from Galileo near-infrared mapping spectrometer real-time
  spectra}}.
\newblock \bibinfo{journal}{J. Geophys. Res.} \bibinfo{volume}{103},
  \bibinfo{pages}{23001--23022}.
\newblock \DOIprefix\doi{10.1029/98JE00948}.
\bibitem[{{Jacquinet-Husson} et~al.(2011){Jacquinet-Husson}, {Crepeau},
  {Armante}, {Boutammine}, {Ch{\'e}din}, {Scott}, {Crevoisier}, {Capelle},
  {Boone}, {Poulet-Crovisier}, {Barbe}, {Campargue}, {Benner}, {Benilan},
  {B{\'e}zard}, {Boudon}, {Brown}, {Coudert}, {Coustenis}, {Dana}, {Devi},
  {Fally}, {Fayt}, {Flaud}, {Goldman}, {Herman}, {Harris}, {Jacquemart},
  {Jolly}, {Kleiner}, {Kleinb{\"o}hl}, {Kwabia-Tchana}, {Lavrentieva},
  {Lacome}, {Xu}, {Lyulin}, {Mandin}, {Maki}, {Mikhailenko}, {Miller},
  {Mishina}, {Moazzen-Ahmadi}, {M{\"u}ller}, {Nikitin}, {Orphal}, {Perevalov},
  {Perrin}, {Petkie}, {Predoi-Cross}, {Rinsland}, {Remedios}, {Rotger},
  {Smith}, {Sung}, {Tashkun}, {Tennyson}, {Toth}, {Vandaele} and {Vander
  Auwera}}]{11geisa}
\bibinfo{author}{{Jacquinet-Husson}, N.}, \bibinfo{author}{{Crepeau}, L.},
  \bibinfo{author}{{Armante}, R.}, \bibinfo{author}{{Boutammine}, C.},
  \bibinfo{author}{{Ch{\'e}din}, A.}, \bibinfo{author}{{Scott}, N.A.},
  \bibinfo{author}{{Crevoisier}, C.}, \bibinfo{author}{{Capelle}, V.},
  \bibinfo{author}{{Boone}, C.}, \bibinfo{author}{{Poulet-Crovisier}, N.},
  \bibinfo{author}{{Barbe}, A.}, \bibinfo{author}{{Campargue}, A.},
  \bibinfo{author}{{Benner}, D.C.}, \bibinfo{author}{{Benilan}, Y.},
  \bibinfo{author}{{B{\'e}zard}, B.}, \bibinfo{author}{{Boudon}, V.},
  \bibinfo{author}{{Brown}, L.R.}, \bibinfo{author}{{Coudert}, L.H.},
  \bibinfo{author}{{Coustenis}, A.}, \bibinfo{author}{{Dana}, V.},
  \bibinfo{author}{{Devi}, V.M.}, \bibinfo{author}{{Fally}, S.},
  \bibinfo{author}{{Fayt}, A.}, \bibinfo{author}{{Flaud}, J.M.},
  \bibinfo{author}{{Goldman}, A.}, \bibinfo{author}{{Herman}, M.},
  \bibinfo{author}{{Harris}, G.J.}, \bibinfo{author}{{Jacquemart}, D.},
  \bibinfo{author}{{Jolly}, A.}, \bibinfo{author}{{Kleiner}, I.},
  \bibinfo{author}{{Kleinb{\"o}hl}, A.}, \bibinfo{author}{{Kwabia-Tchana}, F.},
  \bibinfo{author}{{Lavrentieva}, N.}, \bibinfo{author}{{Lacome}, N.},
  \bibinfo{author}{{Xu}, L.H.}, \bibinfo{author}{{Lyulin}, O.M.},
  \bibinfo{author}{{Mandin}, J.Y.}, \bibinfo{author}{{Maki}, A.},
  \bibinfo{author}{{Mikhailenko}, S.}, \bibinfo{author}{{Miller}, C.E.},
  \bibinfo{author}{{Mishina}, T.}, \bibinfo{author}{{Moazzen-Ahmadi}, N.},
  \bibinfo{author}{{M{\"u}ller}, H.S.P.}, \bibinfo{author}{{Nikitin}, A.},
  \bibinfo{author}{{Orphal}, J.}, \bibinfo{author}{{Perevalov}, V.},
  \bibinfo{author}{{Perrin}, A.}, \bibinfo{author}{{Petkie}, D.T.},
  \bibinfo{author}{{Predoi-Cross}, A.}, \bibinfo{author}{{Rinsland}, C.P.},
  \bibinfo{author}{{Remedios}, J.J.}, \bibinfo{author}{{Rotger}, M.},
  \bibinfo{author}{{Smith}, M.A.H.}, \bibinfo{author}{{Sung}, K.},
  \bibinfo{author}{{Tashkun}, S.}, \bibinfo{author}{{Tennyson}, J.},
  \bibinfo{author}{{Toth}, R.A.}, \bibinfo{author}{{Vandaele}, A.C.},
  \bibinfo{author}{{Vander Auwera}, J.}, \bibinfo{year}{2011}.
\newblock \bibinfo{title}{{The 2009 edition of the GEISA spectroscopic
  database}}.
\newblock \bibinfo{journal}{Journal of Quantitative Spectroscopy and Radiative
  Transfer} \bibinfo{volume}{112}, \bibinfo{pages}{2395--2445}.
\newblock \DOIprefix\doi{10.1016/j.jqsrt.2011.06.004}.
\bibitem[{{Jacquinet-Husson} et~al.(2005){Jacquinet-Husson}, {Scott}, {Chedin},
  {Garceran}, {Armante}, {Chursin}, {Barbe}, {Birk}, {Brown}, {Camy-Peyret},
  {Claveau}, {Clerbaux}, {Coheur}, {Dana}, {Daumont}, {Debacker-Barilly},
  {Flaud}, {Goldman}, {Hamdouni}, {Hess}, {Jacquemart}, {Kopke}, {Mandin},
  {Massie}, {Mikhailenko}, {Nemtchinov}, {Nikitin}, {Newnham}, {Perrin},
  {Perevalov}, {Regalia-Jarlot}, {Rublev}, {Schreier}, {Schult}, {Smith},
  {Tashkun}, {Teffo}, {Toth}, {Tyuterev}, {Vander Auwera}, {Varanasi} and
  {Wagner}}]{05geisa}
\bibinfo{author}{{Jacquinet-Husson}, N.}, \bibinfo{author}{{Scott}, N.A.},
  \bibinfo{author}{{Chedin}, A.}, \bibinfo{author}{{Garceran}, K.},
  \bibinfo{author}{{Armante}, R.}, \bibinfo{author}{{Chursin}, A.A.},
  \bibinfo{author}{{Barbe}, A.}, \bibinfo{author}{{Birk}, M.},
  \bibinfo{author}{{Brown}, L.R.}, \bibinfo{author}{{Camy-Peyret}, C.},
  \bibinfo{author}{{Claveau}, C.}, \bibinfo{author}{{Clerbaux}, C.},
  \bibinfo{author}{{Coheur}, P.F.}, \bibinfo{author}{{Dana}, V.},
  \bibinfo{author}{{Daumont}, L.}, \bibinfo{author}{{Debacker-Barilly}, M.R.},
  \bibinfo{author}{{Flaud}, J.M.}, \bibinfo{author}{{Goldman}, A.},
  \bibinfo{author}{{Hamdouni}, A.}, \bibinfo{author}{{Hess}, M.},
  \bibinfo{author}{{Jacquemart}, D.}, \bibinfo{author}{{Kopke}, P.},
  \bibinfo{author}{{Mandin}, J.Y.}, \bibinfo{author}{{Massie}, S.},
  \bibinfo{author}{{Mikhailenko}, S.}, \bibinfo{author}{{Nemtchinov}, V.},
  \bibinfo{author}{{Nikitin}, A.}, \bibinfo{author}{{Newnham}, D.},
  \bibinfo{author}{{Perrin}, A.}, \bibinfo{author}{{Perevalov}, V.I.},
  \bibinfo{author}{{Regalia-Jarlot}, L.}, \bibinfo{author}{{Rublev}, A.},
  \bibinfo{author}{{Schreier}, F.}, \bibinfo{author}{{Schult}, I.},
  \bibinfo{author}{{Smith}, K.M.}, \bibinfo{author}{{Tashkun}, S.A.},
  \bibinfo{author}{{Teffo}, J.L.}, \bibinfo{author}{{Toth}, R.A.},
  \bibinfo{author}{{Tyuterev}, V.G.}, \bibinfo{author}{{Vander Auwera}, J.},
  \bibinfo{author}{{Varanasi}, P.}, \bibinfo{author}{{Wagner}, G.},
  \bibinfo{year}{2005}.
\newblock \bibinfo{title}{{The 2003 edition of the GEISA/IASI spectroscopic
  database}}.
\newblock \bibinfo{journal}{Journal of Quantitative Spectroscopy and Radiative
  Transfer} \bibinfo{volume}{95}, \bibinfo{pages}{429--467}.
\bibitem[{{Kleiner} et~al.(2003){Kleiner}, {Tarrago}, {Cottaz}, {Sagui},
  {Brown}, {Poynter}, {Pickett}, {Chen}, {Pearson}, {Sams}, {Blake},
  {Matsuura}, {Nemtchinov}, {Varanasi}, {Fusina} and {Di Lonardo}}]{03kleiner}
\bibinfo{author}{{Kleiner}, I.}, \bibinfo{author}{{Tarrago}, G.},
  \bibinfo{author}{{Cottaz}, C.}, \bibinfo{author}{{Sagui}, L.},
  \bibinfo{author}{{Brown}, L.R.}, \bibinfo{author}{{Poynter}, R.L.},
  \bibinfo{author}{{Pickett}, H.M.}, \bibinfo{author}{{Chen}, P.},
  \bibinfo{author}{{Pearson}, J.C.}, \bibinfo{author}{{Sams}, R.L.},
  \bibinfo{author}{{Blake}, G.A.}, \bibinfo{author}{{Matsuura}, S.},
  \bibinfo{author}{{Nemtchinov}, V.}, \bibinfo{author}{{Varanasi}, P.},
  \bibinfo{author}{{Fusina}, L.}, \bibinfo{author}{{Di Lonardo}, G.},
  \bibinfo{year}{2003}.
\newblock \bibinfo{title}{{NH$_3$ and PH$_3$ line parameters: the 2000 HITRAN
  update and new results}}.
\newblock \bibinfo{journal}{Journal of Quantitative Spectroscopy and Radiative
  Transfer} \bibinfo{volume}{82}, \bibinfo{pages}{293--312}.
\bibitem[{{Kostiuk} et~al.(1987){Kostiuk}, {Espenak}, {Mumma}, {Deming} and
  {Zipoy}}]{87kostiuk}
\bibinfo{author}{{Kostiuk}, T..}, \bibinfo{author}{{Espenak}, F.},
  \bibinfo{author}{{Mumma}, M.J.}, \bibinfo{author}{{Deming}, D.},
  \bibinfo{author}{{Zipoy}, D.}, \bibinfo{year}{1987}.
\newblock \bibinfo{title}{{Variability of ethane on Jupiter}}.
\newblock \bibinfo{journal}{Icarus} \bibinfo{volume}{72},
  \bibinfo{pages}{394--410}.
\newblock \DOIprefix\doi{10.1016/0019-1035(87)90182-5}.
\bibitem[{{Kostiuk} et~al.(1993){Kostiuk}, {Romani}, {Espenak} and
  {Livengood}}]{93kostiuk}
\bibinfo{author}{{Kostiuk}, T.}, \bibinfo{author}{{Romani}, P.},
  \bibinfo{author}{{Espenak}, F.}, \bibinfo{author}{{Livengood}, T.A.},
  \bibinfo{year}{1993}.
\newblock \bibinfo{title}{{Temperature and abundances in the Jovian auroral
  stratosphere. 2: Ethylene as a probe of the microbar region}}.
\newblock \bibinfo{journal}{Journal of Geophysical Research}
  \bibinfo{volume}{98}, \bibinfo{pages}{18}.
\newblock \DOIprefix\doi{10.1029/93JE01332}.
\bibitem[{{Kunde} et~al.(2004){Kunde}, {Flasar}, {Jennings}, {B{\'e}zard},
  {Strobel}, {Conrath}, {Nixon}, {Bjoraker}, {Romani}, {Achterberg},
  {Simon-Miller}, {Irwin}, {Brasunas}, {Pearl}, {Smith}, {Orton}, {Gierasch},
  {Spilker}, {Carlson}, {Mamoutkine}, {Calcutt}, {Read}, {Taylor}, {Fouchet},
  {Parrish}, {Barucci}, {Courtin}, {Coustenis}, {Gautier}, {Lellouch},
  {Marten}, {Prang{\'e}}, {Biraud}, {Ferrari}, {Owen}, {Abbas}, {Samuelson},
  {Raulin}, {Ade}, {C{\'e}sarsky}, {Grossman} and {Coradini}}]{04kunde}
\bibinfo{author}{{Kunde}, V.G.}, \bibinfo{author}{{Flasar}, F.M.},
  \bibinfo{author}{{Jennings}, D.E.}, \bibinfo{author}{{B{\'e}zard}, B.},
  \bibinfo{author}{{Strobel}, D.F.}, \bibinfo{author}{{Conrath}, B.J.},
  \bibinfo{author}{{Nixon}, C.A.}, \bibinfo{author}{{Bjoraker}, G.L.},
  \bibinfo{author}{{Romani}, P.N.}, \bibinfo{author}{{Achterberg}, R.K.},
  \bibinfo{author}{{Simon-Miller}, A.A.}, \bibinfo{author}{{Irwin}, P.},
  \bibinfo{author}{{Brasunas}, J.C.}, \bibinfo{author}{{Pearl}, J.C.},
  \bibinfo{author}{{Smith}, M.D.}, \bibinfo{author}{{Orton}, G.S.},
  \bibinfo{author}{{Gierasch}, P.J.}, \bibinfo{author}{{Spilker}, L.J.},
  \bibinfo{author}{{Carlson}, R.C.}, \bibinfo{author}{{Mamoutkine}, A.A.},
  \bibinfo{author}{{Calcutt}, S.B.}, \bibinfo{author}{{Read}, P.L.},
  \bibinfo{author}{{Taylor}, F.W.}, \bibinfo{author}{{Fouchet}, T.},
  \bibinfo{author}{{Parrish}, P.}, \bibinfo{author}{{Barucci}, A.},
  \bibinfo{author}{{Courtin}, R.}, \bibinfo{author}{{Coustenis}, A.},
  \bibinfo{author}{{Gautier}, D.}, \bibinfo{author}{{Lellouch}, E.},
  \bibinfo{author}{{Marten}, A.}, \bibinfo{author}{{Prang{\'e}}, R.},
  \bibinfo{author}{{Biraud}, Y.}, \bibinfo{author}{{Ferrari}, C.},
  \bibinfo{author}{{Owen}, T.C.}, \bibinfo{author}{{Abbas}, M.M.},
  \bibinfo{author}{{Samuelson}, R.E.}, \bibinfo{author}{{Raulin}, F.},
  \bibinfo{author}{{Ade}, P.}, \bibinfo{author}{{C{\'e}sarsky}, C.J.},
  \bibinfo{author}{{Grossman}, K.U.}, \bibinfo{author}{{Coradini}, A.},
  \bibinfo{year}{2004}.
\newblock \bibinfo{title}{{Jupiter's Atmospheric Composition from the Cassini
  Thermal Infrared Spectroscopy Experiment}}.
\newblock \bibinfo{journal}{Science} \bibinfo{volume}{305},
  \bibinfo{pages}{1582--1587}.
\newblock \DOIprefix\doi{10.1126/science.1100240}.
\bibitem[{{Lacis} and {Oinas}(1991)}]{91lacis}
\bibinfo{author}{{Lacis}, A.A.}, \bibinfo{author}{{Oinas}, V.},
  \bibinfo{year}{1991}.
\newblock \bibinfo{title}{{A description of the correlated-k distribution
  method for modelling nongray gaseous absorption, thermal emission, and
  multiple scattering in vertically inhomogeneous atmospheres}}.
\newblock \bibinfo{journal}{J. Geophys. Res.} \bibinfo{volume}{96},
  \bibinfo{pages}{9027--9064}.
\bibitem[{{Lacy} et~al.(1989){Lacy}, {Achtermann}, {Bruce}, {Lester}, {Arens},
  {Peck} and {Gaalema}}]{89lacy}
\bibinfo{author}{{Lacy}, J.H.}, \bibinfo{author}{{Achtermann}, J.M.},
  \bibinfo{author}{{Bruce}, D.E.}, \bibinfo{author}{{Lester}, D.F.},
  \bibinfo{author}{{Arens}, J.F.}, \bibinfo{author}{{Peck}, M.C.},
  \bibinfo{author}{{Gaalema}, S.D.}, \bibinfo{year}{1989}.
\newblock \bibinfo{title}{{IRSHELL: A Mid-Infrared Cryogenic Echelle
  Spectrograph}}.
\newblock \bibinfo{journal}{Publications of the Astronomical Society of the
  Pacific} \bibinfo{volume}{101}, \bibinfo{pages}{1166}.
\newblock \DOIprefix\doi{10.1086/132593}.
\bibitem[{{Lacy} et~al.(2002){Lacy}, {Richter}, {Greathouse}, {Jaffe} and
  {Zhu}}]{02lacy}
\bibinfo{author}{{Lacy}, J.H.}, \bibinfo{author}{{Richter}, M.J.},
  \bibinfo{author}{{Greathouse}, T.K.}, \bibinfo{author}{{Jaffe}, D.T.},
  \bibinfo{author}{{Zhu}, Q.}, \bibinfo{year}{2002}.
\newblock \bibinfo{title}{{TEXES: A Sensitive High-Resolution Grating
  Spectrograph for the Mid-Infrared}}.
\newblock \bibinfo{journal}{Publications of the Astronomical Society of the
  Pacific} \bibinfo{volume}{114}, \bibinfo{pages}{153--168}.
\newblock \DOIprefix\doi{10.1086/338730},
  \href{http://arxiv.org/abs/astro-ph/0110521}{{\tt arXiv:astro-ph/0110521}}.
\bibitem[{Lara et~al.(1998)Lara, B{\'e}zard, Griffith, Lacy and Owen}]{98lara}
\bibinfo{author}{Lara, L.}, \bibinfo{author}{B{\'e}zard, B.},
  \bibinfo{author}{Griffith, C.}, \bibinfo{author}{Lacy, J.},
  \bibinfo{author}{Owen, T.}, \bibinfo{year}{1998}.
\newblock \bibinfo{title}{High-resolution 10-micronmeter spectroscopy of
  ammonia and phosphine lines on jupiter}.
\newblock \bibinfo{journal}{Icarus} \bibinfo{volume}{131},
  \bibinfo{pages}{317--333}.
\bibitem[{{Leovy} et~al.(1991){Leovy}, {Friedson} and {Orton}}]{91leovy}
\bibinfo{author}{{Leovy}, C.B.}, \bibinfo{author}{{Friedson}, A.J.},
  \bibinfo{author}{{Orton}, G.S.}, \bibinfo{year}{1991}.
\newblock \bibinfo{title}{{The quasiquadrennial oscillation of Jupiter's
  equatorial stratosphere}}.
\newblock \bibinfo{journal}{Nature} \bibinfo{volume}{354},
  \bibinfo{pages}{380--382}.
\newblock \DOIprefix\doi{10.1038/354380a0}.
\bibitem[{{Levy} et~al.(1993){Levy}, {Lacome} and {Tarrago}}]{93levy}
\bibinfo{author}{{Levy}, A.}, \bibinfo{author}{{Lacome}, N.},
  \bibinfo{author}{{Tarrago}, G.}, \bibinfo{year}{1993}.
\newblock \bibinfo{title}{{Hydrogen- and helium-broadening of phosphine
  lines}}.
\newblock \bibinfo{journal}{Journal of Molecular Spectroscopy}
  \bibinfo{volume}{157}, \bibinfo{pages}{172--181}.
\bibitem[{{Li} et~al.(2006){Li}, {Ingersoll}, {Vasavada}, {Simon-Miller},
  {Achterberg}, {Ewald}, {Dyudina}, {Porco}, {West} and {Flasar}}]{06li}
\bibinfo{author}{{Li}, L.}, \bibinfo{author}{{Ingersoll}, A.P.},
  \bibinfo{author}{{Vasavada}, A.R.}, \bibinfo{author}{{Simon-Miller}, A.A.},
  \bibinfo{author}{{Achterberg}, R.K.}, \bibinfo{author}{{Ewald}, S.P.},
  \bibinfo{author}{{Dyudina}, U.A.}, \bibinfo{author}{{Porco}, C.C.},
  \bibinfo{author}{{West}, R.A.}, \bibinfo{author}{{Flasar}, F.M.},
  \bibinfo{year}{2006}.
\newblock \bibinfo{title}{{Waves in Jupiter's atmosphere observed by the
  Cassini ISS and CIRS instruments}}.
\newblock \bibinfo{journal}{Icarus} \bibinfo{volume}{185},
  \bibinfo{pages}{416--429}.
\newblock \DOIprefix\doi{10.1016/j.icarus.2006.08.005}.
\bibitem[{Lindal(1992)}]{92lindal_nep}
\bibinfo{author}{Lindal, G.}, \bibinfo{year}{1992}.
\newblock \bibinfo{title}{The atmosphere of neptune- an analysis of radio
  occultation data acquired with voyager 2}.
\newblock \bibinfo{journal}{Astronomical Journal} \bibinfo{volume}{103},
  \bibinfo{pages}{967--982}.
\bibitem[{{Livengood} et~al.(1993){Livengood}, {Kostiuk} and
  {Espenak}}]{93livengood}
\bibinfo{author}{{Livengood}, T.A.}, \bibinfo{author}{{Kostiuk}, T.},
  \bibinfo{author}{{Espenak}, F.}, \bibinfo{year}{1993}.
\newblock \bibinfo{title}{{Temperature and abundances in the Jovian auroral
  stratosphere. 1: Ethane as a probe of the millibar region}}.
\newblock \bibinfo{journal}{Journal of Geophysical Research}
  \bibinfo{volume}{98}, \bibinfo{pages}{18}.
\newblock \DOIprefix\doi{10.1029/93JE01043}.
\bibitem[{Lord(1992)}]{92lord}
\bibinfo{author}{Lord, S.}, \bibinfo{year}{1992}.
\newblock \bibinfo{title}{{A New Software Tool for Computing Earth's
  Atmospheric Transmission of Near-and Far-Infrared Radiation}}.
\newblock \bibinfo{journal}{NASA Technical Memorandum}
  \bibinfo{volume}{103957}, \bibinfo{pages}{94035--1000}.
\bibitem[{{Magalhaes} et~al.(1989){Magalhaes}, {Weir}, {Conrath}, {Gierasch}
  and {Leroy}}]{89magalhaes}
\bibinfo{author}{{Magalhaes}, J.A.}, \bibinfo{author}{{Weir}, A.L.},
  \bibinfo{author}{{Conrath}, B.J.}, \bibinfo{author}{{Gierasch}, P.J.},
  \bibinfo{author}{{Leroy}, S.S.}, \bibinfo{year}{1989}.
\newblock \bibinfo{title}{{Slowly moving thermal features on Jupiter}}.
\newblock \bibinfo{journal}{Nature} \bibinfo{volume}{337},
  \bibinfo{pages}{444--447}.
\newblock \DOIprefix\doi{10.1038/337444a0}.
\bibitem[{{Margolis}(1993)}]{93margolis}
\bibinfo{author}{{Margolis}, J.S.}, \bibinfo{year}{1993}.
\newblock \bibinfo{title}{{Measurement of hydrogen-broadened methane lines in
  the {$\nu$}$_{4}$ band at 296 and 200K.}}
\newblock \bibinfo{journal}{Journal of Quantitative Spectroscopy and Radiative
  Transfer} \bibinfo{volume}{50}, \bibinfo{pages}{431--441}.
\newblock \DOIprefix\doi{10.1016/0022-4073(93)90073-Q}.
\bibitem[{{Martonchik} et~al.(1984){Martonchik}, {Orton} and
  {Appleby}}]{84martonchik}
\bibinfo{author}{{Martonchik}, J.V.}, \bibinfo{author}{{Orton}, G.S.},
  \bibinfo{author}{{Appleby}, J.F.}, \bibinfo{year}{1984}.
\newblock \bibinfo{title}{{Optical properties of NH3 ice from the far infrared
  to the near ultraviolet}}.
\newblock \bibinfo{journal}{Applied Optics} \bibinfo{volume}{23},
  \bibinfo{pages}{541--547}.
\bibitem[{Matcheva et~al.(2005)Matcheva, Conrath, Gierasch and
  Flasar}]{05matcheva}
\bibinfo{author}{Matcheva, K.}, \bibinfo{author}{Conrath, B.},
  \bibinfo{author}{Gierasch, P.}, \bibinfo{author}{Flasar, F.},
  \bibinfo{year}{2005}.
\newblock \bibinfo{title}{{The cloud structure of the jovian atmosphere as seen
  by the Cassini/CIRS experiment}}.
\newblock \bibinfo{journal}{Icarus} \bibinfo{volume}{179},
  \bibinfo{pages}{432--448}.
\bibitem[{{Moses} et~al.(2005){Moses}, {Fouchet}, {B{\'e}zard}, {Gladstone},
  {Lellouch} and {Feuchtgruber}}]{05moses_jup}
\bibinfo{author}{{Moses}, J.I.}, \bibinfo{author}{{Fouchet}, T.},
  \bibinfo{author}{{B{\'e}zard}, B.}, \bibinfo{author}{{Gladstone}, G.R.},
  \bibinfo{author}{{Lellouch}, E.}, \bibinfo{author}{{Feuchtgruber}, H.},
  \bibinfo{year}{2005}.
\newblock \bibinfo{title}{{Photochemistry and diffusion in Jupiter's
  stratosphere: Constraints from ISO observations and comparisons with other
  giant planets}}.
\newblock \bibinfo{journal}{Journal of Geophysical Research (Planets)}
  \bibinfo{volume}{110}, \bibinfo{pages}{E08001}.
\newblock \DOIprefix\doi{10.1029/2005JE002411}.
\bibitem[{{Niemann} et~al.(1998){Niemann}, {Atreya}, {Carignan}, {Donahue},
  {Haberman}, {Harpold}, {Hartle}, {Hunten}, {Kasprzak}, {Mahaffy}, {Owen} and
  {Way}}]{98niemann}
\bibinfo{author}{{Niemann}, H.B.}, \bibinfo{author}{{Atreya}, S.K.},
  \bibinfo{author}{{Carignan}, G.R.}, \bibinfo{author}{{Donahue}, T.M.},
  \bibinfo{author}{{Haberman}, J.A.}, \bibinfo{author}{{Harpold}, D.N.},
  \bibinfo{author}{{Hartle}, R.E.}, \bibinfo{author}{{Hunten}, D.M.},
  \bibinfo{author}{{Kasprzak}, W.T.}, \bibinfo{author}{{Mahaffy}, P.R.},
  \bibinfo{author}{{Owen}, T.C.}, \bibinfo{author}{{Way}, S.H.},
  \bibinfo{year}{1998}.
\newblock \bibinfo{title}{{The composition of the Jovian atmosphere as
  determined by the Galileo probe mass spectrometer}}.
\newblock \bibinfo{journal}{J. Geophys. Res.} \bibinfo{volume}{103},
  \bibinfo{pages}{22831--22846}.
\newblock \DOIprefix\doi{10.1029/98JE01050}.
\bibitem[{{Nixon} et~al.(2007){Nixon}, {Achterberg}, {Conrath}, {Irwin},
  {Teanby}, {Fouchet}, {Parrish}, {Romani}, {Abbas}, {Leclair}, {Strobel},
  {Simon-Miller}, {Jennings}, {Flasar} and {Kunde}}]{07nixon}
\bibinfo{author}{{Nixon}, C.A.}, \bibinfo{author}{{Achterberg}, R.K.},
  \bibinfo{author}{{Conrath}, B.J.}, \bibinfo{author}{{Irwin}, P.G.J.},
  \bibinfo{author}{{Teanby}, N.A.}, \bibinfo{author}{{Fouchet}, T.},
  \bibinfo{author}{{Parrish}, P.D.}, \bibinfo{author}{{Romani}, P.N.},
  \bibinfo{author}{{Abbas}, M.}, \bibinfo{author}{{Leclair}, A.},
  \bibinfo{author}{{Strobel}, D.}, \bibinfo{author}{{Simon-Miller}, A.A.},
  \bibinfo{author}{{Jennings}, D.J.}, \bibinfo{author}{{Flasar}, F.M.},
  \bibinfo{author}{{Kunde}, V.G.}, \bibinfo{year}{2007}.
\newblock \bibinfo{title}{{Meridional variations of C$_{2}$H$_{2}$ and
  C$_{2}$H$_{6}$ in Jupiter's atmosphere from Cassini CIRS infrared spectra}}.
\newblock \bibinfo{journal}{Icarus} \bibinfo{volume}{188},
  \bibinfo{pages}{47--71}.
\newblock \DOIprefix\doi{10.1016/j.icarus.2006.11.016}.
\bibitem[{{Nixon} et~al.(2010){Nixon}, {Achterberg}, {Romani}, {Allen},
  {Zhang}, {Teanby}, {Irwin} and {Flasar}}]{10nixon}
\bibinfo{author}{{Nixon}, C.A.}, \bibinfo{author}{{Achterberg}, R.K.},
  \bibinfo{author}{{Romani}, P.N.}, \bibinfo{author}{{Allen}, M.},
  \bibinfo{author}{{Zhang}, X.}, \bibinfo{author}{{Teanby}, N.A.},
  \bibinfo{author}{{Irwin}, P.G.J.}, \bibinfo{author}{{Flasar}, F.M.},
  \bibinfo{year}{2010}.
\newblock \bibinfo{title}{{Abundances of Jupiter's trace hydrocarbons from
  Voyager and Cassini}}.
\newblock \bibinfo{journal}{Plan. \& Space Sci.} \bibinfo{volume}{58},
  \bibinfo{pages}{1667--1680}.
\newblock \DOIprefix\doi{10.1016/j.pss.2010.05.008},
  \href{http://arxiv.org/abs/1005.3959}{{\tt arXiv:1005.3959}}.
\bibitem[{{Ortiz} et~al.(1998){Ortiz}, {Orton}, {Friedson}, {Stewart}, {Fisher}
  and {Spencer}}]{98ortiz}
\bibinfo{author}{{Ortiz}, J.L.}, \bibinfo{author}{{Orton}, G.S.},
  \bibinfo{author}{{Friedson}, A.J.}, \bibinfo{author}{{Stewart}, S.T.},
  \bibinfo{author}{{Fisher}, B.M.}, \bibinfo{author}{{Spencer}, J.R.},
  \bibinfo{year}{1998}.
\newblock \bibinfo{title}{{Evolution and persistence of 5-{$\mu$}m hot spots at
  the Galileo probe entry latitude}}.
\newblock \bibinfo{journal}{J. Geophys. Res.} \bibinfo{volume}{103},
  \bibinfo{pages}{23051--23069}.
\newblock \DOIprefix\doi{10.1029/98JE00696}.
\bibitem[{{Orton} et~al.(1996){Orton}, {Ortiz}, {Baines}, {Bjoraker},
  {Carsenty}, {Colas}, {Dayal}, {Deming}, {Drossart}, {Frappa}, {Friedson},
  {Goguen}, {Golisch}, {Griep}, {Hernandez}, {Hoffmann}, {Jennings},
  {Kaminski}, {Kuhn}, {Laques}, {Limaye}, {Lin}, {Lecacheux}, {Martin},
  {McCabe}, {Momary}, {Parker}, {Puetter}, {Ressler}, {Reyes}, {Sada},
  {Spencer}, {Spitale}, {Stewart}, {Varsik}, {Warell}, {Wild},
  {Yanamandra-Fisher}, {Fazio}, {Hora} and {Deutsch}}]{96orton}
\bibinfo{author}{{Orton}, G.}, \bibinfo{author}{{Ortiz}, J.L.},
  \bibinfo{author}{{Baines}, K.}, \bibinfo{author}{{Bjoraker}, G.},
  \bibinfo{author}{{Carsenty}, U.}, \bibinfo{author}{{Colas}, F.},
  \bibinfo{author}{{Dayal}, A.}, \bibinfo{author}{{Deming}, D.},
  \bibinfo{author}{{Drossart}, P.}, \bibinfo{author}{{Frappa}, E.},
  \bibinfo{author}{{Friedson}, J.}, \bibinfo{author}{{Goguen}, J.},
  \bibinfo{author}{{Golisch}, W.}, \bibinfo{author}{{Griep}, D.},
  \bibinfo{author}{{Hernandez}, C.}, \bibinfo{author}{{Hoffmann}, W.},
  \bibinfo{author}{{Jennings}, D.}, \bibinfo{author}{{Kaminski}, C.},
  \bibinfo{author}{{Kuhn}, J.}, \bibinfo{author}{{Laques}, P.},
  \bibinfo{author}{{Limaye}, S.}, \bibinfo{author}{{Lin}, H.},
  \bibinfo{author}{{Lecacheux}, J.}, \bibinfo{author}{{Martin}, T.},
  \bibinfo{author}{{McCabe}, G.}, \bibinfo{author}{{Momary}, T.},
  \bibinfo{author}{{Parker}, D.}, \bibinfo{author}{{Puetter}, R.},
  \bibinfo{author}{{Ressler}, M.}, \bibinfo{author}{{Reyes}, G.},
  \bibinfo{author}{{Sada}, P.}, \bibinfo{author}{{Spencer}, J.},
  \bibinfo{author}{{Spitale}, J.}, \bibinfo{author}{{Stewart}, S.},
  \bibinfo{author}{{Varsik}, J.}, \bibinfo{author}{{Warell}, J.},
  \bibinfo{author}{{Wild}, W.}, \bibinfo{author}{{Yanamandra-Fisher}, P.},
  \bibinfo{author}{{Fazio}, G.}, \bibinfo{author}{{Hora}, J.},
  \bibinfo{author}{{Deutsch}, L.}, \bibinfo{year}{1996}.
\newblock \bibinfo{title}{{Earth-based observations of the Galileo probe entry
  site}}.
\newblock \bibinfo{journal}{Science} \bibinfo{volume}{272},
  \bibinfo{pages}{839--840}.
\bibitem[{Orton et~al.(2008)Orton, Yanamandra-Fisher, Fisher, Friedson,
  Parrish, Nelson, Bauermeister, Fletcher, Gezari, Varosi, Tokunaga, Caldwell,
  Baines, Hora, Ressler, Fujiyoshi, Fuse, Hagopian, Martin, Bergstralh, Howett,
  Hoffmann, Deutsch, Van~Cleve, Noe, Adams, Kassis and
  Tollestrup}]{08orton_qxo}
\bibinfo{author}{Orton, G.}, \bibinfo{author}{Yanamandra-Fisher, P.},
  \bibinfo{author}{Fisher, B.}, \bibinfo{author}{Friedson, A.},
  \bibinfo{author}{Parrish, P.}, \bibinfo{author}{Nelson, J.},
  \bibinfo{author}{Bauermeister, A.}, \bibinfo{author}{Fletcher, L.},
  \bibinfo{author}{Gezari, D.}, \bibinfo{author}{Varosi, F.},
  \bibinfo{author}{Tokunaga, A.}, \bibinfo{author}{Caldwell, J.},
  \bibinfo{author}{Baines, K.}, \bibinfo{author}{Hora, J.},
  \bibinfo{author}{Ressler, M.}, \bibinfo{author}{Fujiyoshi, T.},
  \bibinfo{author}{Fuse, T.}, \bibinfo{author}{Hagopian, H.},
  \bibinfo{author}{Martin, T.}, \bibinfo{author}{Bergstralh, J.},
  \bibinfo{author}{Howett, C.}, \bibinfo{author}{Hoffmann, W.},
  \bibinfo{author}{Deutsch, L.}, \bibinfo{author}{Van~Cleve, J.},
  \bibinfo{author}{Noe, E.}, \bibinfo{author}{Adams, J.},
  \bibinfo{author}{Kassis, M.}, \bibinfo{author}{Tollestrup, E.},
  \bibinfo{year}{2008}.
\newblock \bibinfo{title}{{Semi-annual oscillations in Saturn's low-latitude
  stratospheric temperatures}}.
\newblock \bibinfo{journal}{Nature} \bibinfo{volume}{453},
  \bibinfo{pages}{196--198}.
\bibitem[{{Orton} et~al.(1998){Orton}, {Fisher}, {Baines}, {Stewart},
  {Friedson}, {Ortiz}, {Marinova}, {Ressler}, {Dayal}, {Hoffmann}, {Hora},
  {Hinkley}, {Krishnan}, {Masanovic}, {Tesic}, {Tziolas} and
  {Parija}}]{98orton}
\bibinfo{author}{{Orton}, G.S.}, \bibinfo{author}{{Fisher}, B.M.},
  \bibinfo{author}{{Baines}, K.H.}, \bibinfo{author}{{Stewart}, S.T.},
  \bibinfo{author}{{Friedson}, A.J.}, \bibinfo{author}{{Ortiz}, J.L.},
  \bibinfo{author}{{Marinova}, M.}, \bibinfo{author}{{Ressler}, M.},
  \bibinfo{author}{{Dayal}, A.}, \bibinfo{author}{{Hoffmann}, W.},
  \bibinfo{author}{{Hora}, J.}, \bibinfo{author}{{Hinkley}, S.},
  \bibinfo{author}{{Krishnan}, V.}, \bibinfo{author}{{Masanovic}, M.},
  \bibinfo{author}{{Tesic}, J.}, \bibinfo{author}{{Tziolas}, A.},
  \bibinfo{author}{{Parija}, K.C.}, \bibinfo{year}{1998}.
\newblock \bibinfo{title}{{Characteristics of the Galileo probe entry site from
  Earth-based remote sensing observations}}.
\newblock \bibinfo{journal}{J. Geophys. Res.} \bibinfo{volume}{103},
  \bibinfo{pages}{22791--22814}.
\newblock \DOIprefix\doi{10.1029/98JE02380}.
\bibitem[{{Orton} et~al.(1991){Orton}, {Friedson}, {Caldwell}, {Hammel},
  {Baines}, {Bergstralh}, {Martin}, {Malcom}, {West}, {Golisch}, {Griep},
  {Kaminski}, {Tokunaga}, {Baron} and {Shure}}]{91orton}
\bibinfo{author}{{Orton}, G.S.}, \bibinfo{author}{{Friedson}, A.J.},
  \bibinfo{author}{{Caldwell}, J.}, \bibinfo{author}{{Hammel}, H.B.},
  \bibinfo{author}{{Baines}, K.H.}, \bibinfo{author}{{Bergstralh}, J.T.},
  \bibinfo{author}{{Martin}, T.Z.}, \bibinfo{author}{{Malcom}, M.E.},
  \bibinfo{author}{{West}, R.A.}, \bibinfo{author}{{Golisch}, W.F.},
  \bibinfo{author}{{Griep}, D.M.}, \bibinfo{author}{{Kaminski}, C.D.},
  \bibinfo{author}{{Tokunaga}, A.T.}, \bibinfo{author}{{Baron}, R.},
  \bibinfo{author}{{Shure}, M.}, \bibinfo{year}{1991}.
\newblock \bibinfo{title}{{Thermal maps of Jupiter - Spatial organization and
  time dependence of stratospheric temperatures, 1980 to 1990}}.
\newblock \bibinfo{journal}{Science} \bibinfo{volume}{252},
  \bibinfo{pages}{537--542}.
\bibitem[{{Orton} et~al.(1994){Orton}, {Friedson}, {Yanamandra-Fisher},
  {Caldwell}, {Hammel}, {Baines}, {Bergstralh}, {Martin}, {West}, {Veeder},
  {Lynch}, {Russell}, {Malcom}, {Golisch}, {Griep}, {Kaminski}, {Tokunaga},
  {Herbst} and {Shure}}]{94orton}
\bibinfo{author}{{Orton}, G.S.}, \bibinfo{author}{{Friedson}, A.J.},
  \bibinfo{author}{{Yanamandra-Fisher}, P.A.}, \bibinfo{author}{{Caldwell},
  J.}, \bibinfo{author}{{Hammel}, H.B.}, \bibinfo{author}{{Baines}, K.H.},
  \bibinfo{author}{{Bergstralh}, J.T.}, \bibinfo{author}{{Martin}, T.Z.},
  \bibinfo{author}{{West}, R.A.}, \bibinfo{author}{{Veeder}, Jr., G.J.},
  \bibinfo{author}{{Lynch}, D.K.}, \bibinfo{author}{{Russell}, R.},
  \bibinfo{author}{{Malcom}, M.E.}, \bibinfo{author}{{Golisch}, W.F.},
  \bibinfo{author}{{Griep}, D.M.}, \bibinfo{author}{{Kaminski}, C.D.},
  \bibinfo{author}{{Tokunaga}, A.T.}, \bibinfo{author}{{Herbst}, T.},
  \bibinfo{author}{{Shure}, M.}, \bibinfo{year}{1994}.
\newblock \bibinfo{title}{{Spatial Organization and Time Dependence of
  Jupiter's Tropospheric Temperatures, 1980-1993}}.
\newblock \bibinfo{journal}{Science} \bibinfo{volume}{265},
  \bibinfo{pages}{625--631}.
\newblock \DOIprefix\doi{10.1126/science.265.5172.625}.
\bibitem[{{Orton} et~al.(2007){Orton}, {Gustafsson}, {Burgdorf} and
  {Meadows}}]{07orton}
\bibinfo{author}{{Orton}, G.S.}, \bibinfo{author}{{Gustafsson}, M.},
  \bibinfo{author}{{Burgdorf}, M.}, \bibinfo{author}{{Meadows}, V.},
  \bibinfo{year}{2007}.
\newblock \bibinfo{title}{{Revised Ab Initio Models for H$_2$-H$_2$ Collision
  Induced Absorption at Low Temperatures}}.
\newblock \bibinfo{journal}{Icarus} \bibinfo{volume}{189},
  \bibinfo{pages}{544--549}.
\bibitem[{{Orton} et~al.(2015){Orton}, {Sinclair}, {Fletcher}, {Fujiyoshi},
  {Yanamandra-Fisher}, {Rogers}, {Irwin}, {Greathouse}, {Seede}, {Simon},
  {Nguyen} and {Lai}}]{15orton_dps}
\bibinfo{author}{{Orton}, G.S.}, \bibinfo{author}{{Sinclair}, J.},
  \bibinfo{author}{{Fletcher}, L.}, \bibinfo{author}{{Fujiyoshi}, T.},
  \bibinfo{author}{{Yanamandra-Fisher}, P.}, \bibinfo{author}{{Rogers}, J.},
  \bibinfo{author}{{Irwin}, P.}, \bibinfo{author}{{Greathouse}, T.},
  \bibinfo{author}{{Seede}, R.}, \bibinfo{author}{{Simon}, J.},
  \bibinfo{author}{{Nguyen}, M.}, \bibinfo{author}{{Lai}, M.},
  \bibinfo{year}{2015}.
\newblock \bibinfo{title}{{Are Brown Barges the Deserts of the Upper Jovian
  Atmosphere?}}, in: \bibinfo{booktitle}{AAS/Division for Planetary Sciences
  Meeting Abstracts}, p. \bibinfo{pages}{502.04}.
\bibitem[{{Owen} et~al.(2001){Owen}, {Mahaffy}, {Niemann}, {Atreya} and
  {Wong}}]{01owen}
\bibinfo{author}{{Owen}, T.}, \bibinfo{author}{{Mahaffy}, P.R.},
  \bibinfo{author}{{Niemann}, H.B.}, \bibinfo{author}{{Atreya}, S.},
  \bibinfo{author}{{Wong}, M.}, \bibinfo{year}{2001}.
\newblock \bibinfo{title}{{Protosolar Nitrogen}}.
\newblock \bibinfo{journal}{Astrophys. J. Letters} \bibinfo{volume}{553},
  \bibinfo{pages}{L77--L79}.
\newblock \DOIprefix\doi{10.1086/320501}.
\bibitem[{de~Pater et~al.(2010)de~Pater, Fletcher, Perez-Hoyos, Hammel, Orton,
  Wong, Luszcz-Cook, Sanchez-Lavega and Boslough}]{10depater}
\bibinfo{author}{de~Pater, I.}, \bibinfo{author}{Fletcher, L.N.},
  \bibinfo{author}{Perez-Hoyos, S.}, \bibinfo{author}{Hammel, H.B.},
  \bibinfo{author}{Orton, G.S.}, \bibinfo{author}{Wong, M.H.},
  \bibinfo{author}{Luszcz-Cook, S.}, \bibinfo{author}{Sanchez-Lavega, A.},
  \bibinfo{author}{Boslough, M.}, \bibinfo{year}{2010}.
\newblock \bibinfo{title}{{A Multi-Wavelength Study of the 2009 Impact on
  Jupiter: Comparison of High Resolution Images from Gemini, Keck and HST}}.
\newblock \bibinfo{journal}{Icarus, in press} .
\bibitem[{{Porco} et~al.(2003){Porco}, {West}, {McEwen}, {Del Genio},
  {Ingersoll}, {Thomas}, {Squyres}, {Dones}, {Murray}, {Johnson}, {Burns},
  {Brahic}, {Neukum}, {Veverka}, {Barbara}, {Denk}, {Evans}, {Ferrier},
  {Geissler}, {Helfenstein}, {Roatsch}, {Throop}, {Tiscareno} and
  {Vasavada}}]{03porco}
\bibinfo{author}{{Porco}, C.C.}, \bibinfo{author}{{West}, R.A.},
  \bibinfo{author}{{McEwen}, A.}, \bibinfo{author}{{Del Genio}, A.D.},
  \bibinfo{author}{{Ingersoll}, A.P.}, \bibinfo{author}{{Thomas}, P.},
  \bibinfo{author}{{Squyres}, S.}, \bibinfo{author}{{Dones}, L.},
  \bibinfo{author}{{Murray}, C.D.}, \bibinfo{author}{{Johnson}, T.V.},
  \bibinfo{author}{{Burns}, J.A.}, \bibinfo{author}{{Brahic}, A.},
  \bibinfo{author}{{Neukum}, G.}, \bibinfo{author}{{Veverka}, J.},
  \bibinfo{author}{{Barbara}, J.M.}, \bibinfo{author}{{Denk}, T.},
  \bibinfo{author}{{Evans}, M.}, \bibinfo{author}{{Ferrier}, J.J.},
  \bibinfo{author}{{Geissler}, P.}, \bibinfo{author}{{Helfenstein}, P.},
  \bibinfo{author}{{Roatsch}, T.}, \bibinfo{author}{{Throop}, H.},
  \bibinfo{author}{{Tiscareno}, M.}, \bibinfo{author}{{Vasavada}, A.R.},
  \bibinfo{year}{2003}.
\newblock \bibinfo{title}{{Cassini Imaging of Jupiter's Atmosphere, Satellites,
  and Rings}}.
\newblock \bibinfo{journal}{Science} \bibinfo{volume}{299},
  \bibinfo{pages}{1541--1547}.
\newblock \DOIprefix\doi{10.1126/science.1079462}.
\bibitem[{Read et~al.(2006a)Read, Gierasch and Conrath}]{06read_grs}
\bibinfo{author}{Read, P.}, \bibinfo{author}{Gierasch, P.},
  \bibinfo{author}{Conrath, B.}, \bibinfo{year}{2006}a.
\newblock \bibinfo{title}{{Mapping potential-vorticity dynamics on Jupiter. II:
  The Great Red Spot from Voyager 1 and 2 data}}.
\newblock \bibinfo{journal}{Quarterly Journal of the Royal Meteorological
  Society} \bibinfo{volume}{132}, \bibinfo{pages}{1605--1625}.
\bibitem[{Read et~al.(2006b)Read, Gierasch, Conrath, Simon-Miller, Fouchet and
  Yamazaki}]{06read_jup}
\bibinfo{author}{Read, P.}, \bibinfo{author}{Gierasch, P.},
  \bibinfo{author}{Conrath, B.}, \bibinfo{author}{Simon-Miller, A.},
  \bibinfo{author}{Fouchet, T.}, \bibinfo{author}{Yamazaki, Y.},
  \bibinfo{year}{2006}b.
\newblock \bibinfo{title}{{Mapping potential-vorticity dynamics on Jupiter. I:
  Zonal-mean circulation from Cassini and Voyager 1 data}}.
\newblock \bibinfo{journal}{Q. J. R. Meteorol. Soc.} \bibinfo{volume}{132},
  \bibinfo{pages}{1577--1603}.
\bibitem[{{Reuter} and {Sirota}(1994)}]{94reuter}
\bibinfo{author}{{Reuter}, D.C.}, \bibinfo{author}{{Sirota}, J.M.},
  \bibinfo{year}{1994}.
\newblock \bibinfo{title}{{Line strength and self-broadening coefficient of the
  pure rotational S(1) quadrupole line in H2}}.
\newblock \bibinfo{journal}{{Astrophys. J.}} \bibinfo{volume}{428},
  \bibinfo{pages}{L77--L79}.
\newblock \DOIprefix\doi{10.1086/187397}.
\bibitem[{{Rodgers}(2000)}]{00rodgers}
\bibinfo{author}{{Rodgers}, C.D.}, \bibinfo{year}{2000}.
\newblock \bibinfo{title}{{Inverse Methods for Atmospheric Remote Sounding:
  Theory and Practice}}.
\newblock \bibinfo{publisher}{World Scientific}.
\bibitem[{Rogers(1995)}]{95rogers}
\bibinfo{author}{Rogers, J.}, \bibinfo{year}{1995}.
\newblock \bibinfo{title}{{The Giant Planet Jupiter}}.
\newblock \bibinfo{publisher}{Cambridge University Press}.
\bibitem[{{Romani}(1996)}]{96romani}
\bibinfo{author}{{Romani}, P.N.}, \bibinfo{year}{1996}.
\newblock \bibinfo{title}{{Recent Rate Constant and Product Measurements of the
  Reactions C$_{2}$H$_{3}$+ H$_{2}$ and C$_{2}$H$_{3}$+ H - Importance for
  Photochemical Modeling of Hydrocarbons on Jupiter}}.
\newblock \bibinfo{journal}{Icarus} \bibinfo{volume}{122},
  \bibinfo{pages}{233--241}.
\newblock \DOIprefix\doi{10.1006/icar.1996.0122}.
\bibitem[{{Rothman} et~al.(2013){Rothman}, {Gordon}, {Babikov}, {Barbe}, {Chris
  Benner}, {Bernath}, {Birk}, {Bizzocchi}, {Boudon}, {Brown}, {Campargue},
  {Chance}, {Cohen}, {Coudert}, {Devi}, {Drouin}, {Fayt}, {Flaud}, {Gamache},
  {Harrison}, {Hartmann}, {Hill}, {Hodges}, {Jacquemart}, {Jolly}, {Lamouroux},
  {Le Roy}, {Li}, {Long}, {Lyulin}, {Mackie}, {Massie}, {Mikhailenko},
  {M{\"u}ller}, {Naumenko}, {Nikitin}, {Orphal}, {Perevalov}, {Perrin},
  {Polovtseva}, {Richard}, {Smith}, {Starikova}, {Sung}, {Tashkun}, {Tennyson},
  {Toon}, {Tyuterev} and {Wagner}}]{13rothman}
\bibinfo{author}{{Rothman}, L.S.}, \bibinfo{author}{{Gordon}, I.E.},
  \bibinfo{author}{{Babikov}, Y.}, \bibinfo{author}{{Barbe}, A.},
  \bibinfo{author}{{Chris Benner}, D.}, \bibinfo{author}{{Bernath}, P.F.},
  \bibinfo{author}{{Birk}, M.}, \bibinfo{author}{{Bizzocchi}, L.},
  \bibinfo{author}{{Boudon}, V.}, \bibinfo{author}{{Brown}, L.R.},
  \bibinfo{author}{{Campargue}, A.}, \bibinfo{author}{{Chance}, K.},
  \bibinfo{author}{{Cohen}, E.A.}, \bibinfo{author}{{Coudert}, L.H.},
  \bibinfo{author}{{Devi}, V.M.}, \bibinfo{author}{{Drouin}, B.J.},
  \bibinfo{author}{{Fayt}, A.}, \bibinfo{author}{{Flaud}, J.M.},
  \bibinfo{author}{{Gamache}, R.R.}, \bibinfo{author}{{Harrison}, J.J.},
  \bibinfo{author}{{Hartmann}, J.M.}, \bibinfo{author}{{Hill}, C.},
  \bibinfo{author}{{Hodges}, J.T.}, \bibinfo{author}{{Jacquemart}, D.},
  \bibinfo{author}{{Jolly}, A.}, \bibinfo{author}{{Lamouroux}, J.},
  \bibinfo{author}{{Le Roy}, R.J.}, \bibinfo{author}{{Li}, G.},
  \bibinfo{author}{{Long}, D.A.}, \bibinfo{author}{{Lyulin}, O.M.},
  \bibinfo{author}{{Mackie}, C.J.}, \bibinfo{author}{{Massie}, S.T.},
  \bibinfo{author}{{Mikhailenko}, S.}, \bibinfo{author}{{M{\"u}ller}, H.S.P.},
  \bibinfo{author}{{Naumenko}, O.V.}, \bibinfo{author}{{Nikitin}, A.V.},
  \bibinfo{author}{{Orphal}, J.}, \bibinfo{author}{{Perevalov}, V.},
  \bibinfo{author}{{Perrin}, A.}, \bibinfo{author}{{Polovtseva}, E.R.},
  \bibinfo{author}{{Richard}, C.}, \bibinfo{author}{{Smith}, M.A.H.},
  \bibinfo{author}{{Starikova}, E.}, \bibinfo{author}{{Sung}, K.},
  \bibinfo{author}{{Tashkun}, S.}, \bibinfo{author}{{Tennyson}, J.},
  \bibinfo{author}{{Toon}, G.C.}, \bibinfo{author}{{Tyuterev}, V.G.},
  \bibinfo{author}{{Wagner}, G.}, \bibinfo{year}{2013}.
\newblock \bibinfo{title}{{The HITRAN2012 molecular spectroscopic database}}.
\newblock \bibinfo{journal}{J. Quant. Spectro. Rad. Trans.}
  \bibinfo{volume}{130}, \bibinfo{pages}{4--50}.
\newblock \DOIprefix\doi{10.1016/j.jqsrt.2013.07.002}.
\bibitem[{{Sada} et~al.(1996){Sada}, {Beebe} and {Conrath}}]{96sada}
\bibinfo{author}{{Sada}, P.V.}, \bibinfo{author}{{Beebe}, R.F.},
  \bibinfo{author}{{Conrath}, B.J.}, \bibinfo{year}{1996}.
\newblock \bibinfo{title}{{Comparison of the Structure and Dynamics of
  Jupiter's Great Red SPOT between the Voyager 1 and 2 Encounters}}.
\newblock \bibinfo{journal}{Icarus} \bibinfo{volume}{119},
  \bibinfo{pages}{311--335}.
\newblock \DOIprefix\doi{10.1006/icar.1996.0022}.
\bibitem[{{Sada} et~al.(1998){Sada}, {Bjoraker}, {Jennings}, {McCabe} and
  {Romani}}]{98sada}
\bibinfo{author}{{Sada}, P.V.}, \bibinfo{author}{{Bjoraker}, G.L.},
  \bibinfo{author}{{Jennings}, D.E.}, \bibinfo{author}{{McCabe}, G.H.},
  \bibinfo{author}{{Romani}, P.N.}, \bibinfo{year}{1998}.
\newblock \bibinfo{title}{{Observations of CH $_{4}$, C $_{2}$H $_{6}$, and C
  $_{2}$H $_{2}$in the Stratosphere of Jupiter}}.
\newblock \bibinfo{journal}{Icarus} \bibinfo{volume}{136},
  \bibinfo{pages}{192--201}.
\newblock \DOIprefix\doi{10.1006/icar.1998.6021}.
\bibitem[{{Salem} et~al.(2004){Salem}, {Bouanich}, {Walrand}, {Aroui} and
  {Blanquet}}]{04salem}
\bibinfo{author}{{Salem}, J.}, \bibinfo{author}{{Bouanich}, J.P.},
  \bibinfo{author}{{Walrand}, J.}, \bibinfo{author}{{Aroui}, H.},
  \bibinfo{author}{{Blanquet}, G.}, \bibinfo{year}{2004}.
\newblock \bibinfo{title}{{Hydrogen line broadening in the $\nu_{2}$ and
  $\nu_{4}$ bands of phosphine at low temperature}}.
\newblock \bibinfo{journal}{Journal of Molecular Spectroscopy}
  \bibinfo{volume}{228}, \bibinfo{pages}{23--30}.
\newblock \DOIprefix\doi{10.1016/j.jms.2004.06.015}.
\bibitem[{{S{\'a}nchez-Lavega} et~al.(2008){S{\'a}nchez-Lavega}, {Orton},
  {Hueso}, {Garc{\'{\i}}a-Melendo}, {P{\'e}rez-Hoyos}, {Simon-Miller}, {Rojas},
  {G{\'o}mez}, {Yanamandra-Fisher}, {Fletcher}, {Joels}, {Kemerer}, {Hora},
  {Karkoschka}, {de Pater}, {Wong}, {Marcus}, {Pinilla-Alonso}, {Carvalho},
  {Go}, {Parker}, {Salway}, {Valimberti}, {Wesley} and {Pujic}}]{08sanchez}
\bibinfo{author}{{S{\'a}nchez-Lavega}, A.}, \bibinfo{author}{{Orton}, G.S.},
  \bibinfo{author}{{Hueso}, R.}, \bibinfo{author}{{Garc{\'{\i}}a-Melendo}, E.},
  \bibinfo{author}{{P{\'e}rez-Hoyos}, S.}, \bibinfo{author}{{Simon-Miller},
  A.}, \bibinfo{author}{{Rojas}, J.F.}, \bibinfo{author}{{G{\'o}mez}, J.M.},
  \bibinfo{author}{{Yanamandra-Fisher}, P.}, \bibinfo{author}{{Fletcher}, L.},
  \bibinfo{author}{{Joels}, J.}, \bibinfo{author}{{Kemerer}, J.},
  \bibinfo{author}{{Hora}, J.}, \bibinfo{author}{{Karkoschka}, E.},
  \bibinfo{author}{{de Pater}, I.}, \bibinfo{author}{{Wong}, M.H.},
  \bibinfo{author}{{Marcus}, P.S.}, \bibinfo{author}{{Pinilla-Alonso}, N.},
  \bibinfo{author}{{Carvalho}, F.}, \bibinfo{author}{{Go}, C.},
  \bibinfo{author}{{Parker}, D.}, \bibinfo{author}{{Salway}, M.},
  \bibinfo{author}{{Valimberti}, M.}, \bibinfo{author}{{Wesley}, A.},
  \bibinfo{author}{{Pujic}, Z.}, \bibinfo{year}{2008}.
\newblock \bibinfo{title}{{Depth of a strong jovian jet from a planetary-scale
  disturbance driven by storms}}.
\newblock \bibinfo{journal}{Nature} \bibinfo{volume}{451},
  \bibinfo{pages}{437--440}.
\newblock \DOIprefix\doi{10.1038/nature06533}.
\bibitem[{{Sault} et~al.(2004){Sault}, {Engel} and {de Pater}}]{04sault}
\bibinfo{author}{{Sault}, R.J.}, \bibinfo{author}{{Engel}, C.},
  \bibinfo{author}{{de Pater}, I.}, \bibinfo{year}{2004}.
\newblock \bibinfo{title}{{Longitude-resolved imaging of Jupiter at
  {$\lambda$}=2 cm}}.
\newblock \bibinfo{journal}{Icarus} \bibinfo{volume}{168},
  \bibinfo{pages}{336--343}.
\newblock \DOIprefix\doi{10.1016/j.icarus.2003.11.014},
  \href{http://arxiv.org/abs/astro-ph/0612769}{{\tt arXiv:astro-ph/0612769}}.
\bibitem[{{Schinder} et~al.(2011){Schinder}, {Flasar}, {Marouf}, {French},
  {McGhee}, {Kliore}, {Rappaport}, {Barbinis}, {Fleischman} and
  {Anabtawi}}]{11schinder}
\bibinfo{author}{{Schinder}, P.J.}, \bibinfo{author}{{Flasar}, F.M.},
  \bibinfo{author}{{Marouf}, E.A.}, \bibinfo{author}{{French}, R.G.},
  \bibinfo{author}{{McGhee}, C.A.}, \bibinfo{author}{{Kliore}, A.J.},
  \bibinfo{author}{{Rappaport}, N.J.}, \bibinfo{author}{{Barbinis}, E.},
  \bibinfo{author}{{Fleischman}, D.}, \bibinfo{author}{{Anabtawi}, A.},
  \bibinfo{year}{2011}.
\newblock \bibinfo{title}{{Saturn's equatorial oscillation: Evidence of
  descending thermal structure from Cassini radio occultations}}.
\newblock \bibinfo{journal}{Geophys. Res. Lett.} \bibinfo{volume}{38},
  \bibinfo{pages}{8205}.
\newblock \DOIprefix\doi{10.1029/2011GL047191}.
\bibitem[{{Seiff} et~al.(1998){Seiff}, {Kirk}, {Knight}, {Young}, {Mihalov},
  {Young}, {Milos}, {Schubert}, {Blanchard} and {Atkinson}}]{98seiff}
\bibinfo{author}{{Seiff}, A.}, \bibinfo{author}{{Kirk}, D.B.},
  \bibinfo{author}{{Knight}, T.C.D.}, \bibinfo{author}{{Young}, R.E.},
  \bibinfo{author}{{Mihalov}, J.D.}, \bibinfo{author}{{Young}, L.A.},
  \bibinfo{author}{{Milos}, F.S.}, \bibinfo{author}{{Schubert}, G.},
  \bibinfo{author}{{Blanchard}, R.C.}, \bibinfo{author}{{Atkinson}, D.},
  \bibinfo{year}{1998}.
\newblock \bibinfo{title}{{Thermal structure of Jupiter's atmosphere near the
  edge of a 5-{$\mu$}m hot spot in the north equatorial belt}}.
\newblock \bibinfo{journal}{J. Geophys. Res.} \bibinfo{volume}{103},
  \bibinfo{pages}{22857--22890}.
\newblock \DOIprefix\doi{10.1029/98JE01766}.
\bibitem[{{Showman} and {de Pater}(2005)}]{05showman}
\bibinfo{author}{{Showman}, A.P.}, \bibinfo{author}{{de Pater}, I.},
  \bibinfo{year}{2005}.
\newblock \bibinfo{title}{{Dynamical implications of Jupiter's tropospheric
  ammonia abundance}}.
\newblock \bibinfo{journal}{Icarus} \bibinfo{volume}{174},
  \bibinfo{pages}{192--204}.
\newblock \DOIprefix\doi{10.1016/j.icarus.2004.10.004}.
\bibitem[{{Showman} and {Dowling}(2000)}]{00showman}
\bibinfo{author}{{Showman}, A.P.}, \bibinfo{author}{{Dowling}, T.E.},
  \bibinfo{year}{2000}.
\newblock \bibinfo{title}{{Nonlinear Simulations of Jupiter's 5-Micron Hot
  Spots}}.
\newblock \bibinfo{journal}{Science} \bibinfo{volume}{289},
  \bibinfo{pages}{1737--1740}.
\newblock \DOIprefix\doi{10.1126/science.289.5485.1737}.
\bibitem[{{Simon} et~al.(2015){Simon}, {Wong} and {Orton}}]{15simon}
\bibinfo{author}{{Simon}, A.A.}, \bibinfo{author}{{Wong}, M.H.},
  \bibinfo{author}{{Orton}, G.S.}, \bibinfo{year}{2015}.
\newblock \bibinfo{title}{{First Results from the Hubble OPAL Program: Jupiter
  in 2015}}.
\newblock \bibinfo{journal}{ApJ Letters} \bibinfo{volume}{812},
  \bibinfo{pages}{55}.
\newblock \DOIprefix\doi{10.1088/0004-637X/812/1/55}.
\bibitem[{Simon-Miller et~al.(2002)Simon-Miller, Gierasch, Beebe, Conrath,
  Flasar and Achterberg}]{02simon}
\bibinfo{author}{Simon-Miller, A.}, \bibinfo{author}{Gierasch, P.},
  \bibinfo{author}{Beebe, R.}, \bibinfo{author}{Conrath, B.},
  \bibinfo{author}{Flasar, F.}, \bibinfo{author}{Achterberg, R.},
  \bibinfo{year}{2002}.
\newblock \bibinfo{title}{{New Observational Results Concerning Jupiter's Great
  Red Spot}}.
\newblock \bibinfo{journal}{Icarus} \bibinfo{volume}{158},
  \bibinfo{pages}{249--266}.
\bibitem[{{Simon-Miller} et~al.(2006a){Simon-Miller}, {Chanover}, {Orton},
  {Sussman}, {Tsavaris} and {Karkoschka}}]{06simon_BA}
\bibinfo{author}{{Simon-Miller}, A.A.}, \bibinfo{author}{{Chanover}, N.J.},
  \bibinfo{author}{{Orton}, G.S.}, \bibinfo{author}{{Sussman}, M.},
  \bibinfo{author}{{Tsavaris}, I.G.}, \bibinfo{author}{{Karkoschka}, E.},
  \bibinfo{year}{2006}a.
\newblock \bibinfo{title}{{Jupiter's White Oval turns red}}.
\newblock \bibinfo{journal}{Icarus} \bibinfo{volume}{185},
  \bibinfo{pages}{558--562}.
\newblock \DOIprefix\doi{10.1016/j.icarus.2006.08.002}.
\bibitem[{{Simon-Miller} et~al.(2000){Simon-Miller}, {Conrath}, {Gierasch} and
  {Beebe}}]{00simon}
\bibinfo{author}{{Simon-Miller}, A.A.}, \bibinfo{author}{{Conrath}, B.},
  \bibinfo{author}{{Gierasch}, P.J.}, \bibinfo{author}{{Beebe}, R.F.},
  \bibinfo{year}{2000}.
\newblock \bibinfo{title}{{A detection of water ice on Jupiter with Voyager
  IRIS}}.
\newblock \bibinfo{journal}{Icarus} \bibinfo{volume}{145},
  \bibinfo{pages}{454--461}.
\newblock \DOIprefix\doi{10.1006/icar.2000.6359}.
\bibitem[{{Simon-Miller} et~al.(2006b){Simon-Miller}, {Conrath}, {Gierasch},
  {Orton}, {Achterberg}, {Flasar} and {Fisher}}]{06simon}
\bibinfo{author}{{Simon-Miller}, A.A.}, \bibinfo{author}{{Conrath}, B.J.},
  \bibinfo{author}{{Gierasch}, P.J.}, \bibinfo{author}{{Orton}, G.S.},
  \bibinfo{author}{{Achterberg}, R.K.}, \bibinfo{author}{{Flasar}, F.M.},
  \bibinfo{author}{{Fisher}, B.M.}, \bibinfo{year}{2006}b.
\newblock \bibinfo{title}{{Jupiter's atmospheric temperatures: From Voyager
  IRIS to Cassini CIRS}}.
\newblock \bibinfo{journal}{Icarus} \bibinfo{volume}{180},
  \bibinfo{pages}{98--112}.
\newblock \DOIprefix\doi{10.1016/j.icarus.2005.07.019}.
\bibitem[{{Sinclair} et~al.(2015){Sinclair}, {Orton}, {Greathouse}, {Fletcher}
  and {Irwin}}]{15sinclair_dps}
\bibinfo{author}{{Sinclair}, J.}, \bibinfo{author}{{Orton}, G.},
  \bibinfo{author}{{Greathouse}, T.}, \bibinfo{author}{{Fletcher}, L.},
  \bibinfo{author}{{Irwin}, P.}, \bibinfo{year}{2015}.
\newblock \bibinfo{title}{{Jupiter's auroral-related thermal infrared emission
  from IRTF-TEXES}}, in: \bibinfo{booktitle}{AAS/Division for Planetary
  Sciences Meeting Abstracts}, p. \bibinfo{pages}{311.13}.
\bibitem[{{Sinclair} et~al.(2013){Sinclair}, {Irwin}, {Fletcher}, {Moses},
  {Greathouse}, {Friedson}, {Hesman}, {Hurley} and {Merlet}}]{13sinclair}
\bibinfo{author}{{Sinclair}, J.A.}, \bibinfo{author}{{Irwin}, P.G.J.},
  \bibinfo{author}{{Fletcher}, L.N.}, \bibinfo{author}{{Moses}, J.I.},
  \bibinfo{author}{{Greathouse}, T.K.}, \bibinfo{author}{{Friedson}, A.J.},
  \bibinfo{author}{{Hesman}, B.}, \bibinfo{author}{{Hurley}, J.},
  \bibinfo{author}{{Merlet}, C.}, \bibinfo{year}{2013}.
\newblock \bibinfo{title}{{Seasonal variations of temperature, acetylene and
  ethane in Saturn's atmosphere from 2005 to 2010, as observed by
  Cassini-CIRS}}.
\newblock \bibinfo{journal}{Icarus} \bibinfo{volume}{225},
  \bibinfo{pages}{257--271}.
\newblock \DOIprefix\doi{10.1016/j.icarus.2013.03.011}.
\bibitem[{{Sinclair} et~al.(2016){Sinclair}, {Orton}, {Greathouse}, {Fletcher}
  and {Irwin}}]{16sinclair}
\bibinfo{author}{{Sinclair}, J.A.}, \bibinfo{author}{{Orton}, G.S.},
  \bibinfo{author}{{Greathouse}, T.K.}, \bibinfo{author}{{Fletcher}, L.N.},
  \bibinfo{author}{{Irwin}, P.G.J.}, \bibinfo{year}{2016}.
\newblock \bibinfo{title}{{Jupiter's auroral-related heating and chemistry I: a
  re-analysis of Voyager-IRIS and Cassini-CIRS spectra}}.
\newblock \bibinfo{journal}{Icarus, in prep} .
\bibitem[{{Vander Auwera} et~al.(2007){Vander Auwera}, {Moazzen-Ahmadi} and
  {Flaud}}]{07vander}
\bibinfo{author}{{Vander Auwera}, J.}, \bibinfo{author}{{Moazzen-Ahmadi}, N.},
  \bibinfo{author}{{Flaud}, J.}, \bibinfo{year}{2007}.
\newblock \bibinfo{title}{{Toward an Accurate Database for the 12 $\mu$m Region
  of the Ethane Spectrum}}.
\newblock \bibinfo{journal}{Astrophysical Journal} \bibinfo{volume}{662},
  \bibinfo{pages}{750--757}.
\newblock \DOIprefix\doi{10.1086/515567}.
\bibitem[{{Varanasi}(1992)}]{92varanasi}
\bibinfo{author}{{Varanasi}, P.}, \bibinfo{year}{1992}.
\newblock \bibinfo{title}{{Intensity and linewidth measurements in the
  13.7-micron fundamental bands of (C-12)2H2 and (C-12)(C-13)H2 at planetary
  atmospheric temperatures}}.
\newblock \bibinfo{journal}{Journal of Quantitative Spectroscopy and Radiative
  Transfer} \bibinfo{volume}{47}, \bibinfo{pages}{263--274}.
\newblock \DOIprefix\doi{10.1016/0022-4073(92)90145-T}.
\bibitem[{West et~al.(2004)West, Baines, Friedson, Banfield, Ragent and
  Taylor}]{04west}
\bibinfo{author}{West, R.}, \bibinfo{author}{Baines, K.},
  \bibinfo{author}{Friedson, A.}, \bibinfo{author}{Banfield, D.},
  \bibinfo{author}{Ragent, B.}, \bibinfo{author}{Taylor, F.},
  \bibinfo{year}{2004}.
\newblock \bibinfo{title}{In Jupiter. The Planet, Satellites and
  Magnetosphere}. chapter \bibinfo{chapter}{Jovian Clouds and Hazes}.
\newblock pp. \bibinfo{pages}{79--104}.
\bibitem[{{Wong} et~al.(2004){Wong}, {Bjoraker}, {Smith}, {Flasar} and
  {Nixon}}]{04wong}
\bibinfo{author}{{Wong}, M.H.}, \bibinfo{author}{{Bjoraker}, G.L.},
  \bibinfo{author}{{Smith}, M.D.}, \bibinfo{author}{{Flasar}, F.M.},
  \bibinfo{author}{{Nixon}, C.A.}, \bibinfo{year}{2004}.
\newblock \bibinfo{title}{{Identification of the 10-{$\mu$}m ammonia ice
  feature on Jupiter}}.
\newblock \bibinfo{journal}{Plan. \& Space Sci.} \bibinfo{volume}{52},
  \bibinfo{pages}{385--395}.
\newblock \DOIprefix\doi{10.1016/j.pss.2003.06.005}.
\bibitem[{{Zhang} et~al.(2013a){Zhang}, {Nixon}, {Shia}, {West}, {Irwin},
  {Yelle}, {Allen} and {Yung}}]{13zhang}
\bibinfo{author}{{Zhang}, X.}, \bibinfo{author}{{Nixon}, C.A.},
  \bibinfo{author}{{Shia}, R.L.}, \bibinfo{author}{{West}, R.A.},
  \bibinfo{author}{{Irwin}, P.G.J.}, \bibinfo{author}{{Yelle}, R.V.},
  \bibinfo{author}{{Allen}, M.A.}, \bibinfo{author}{{Yung}, Y.L.},
  \bibinfo{year}{2013}a.
\newblock \bibinfo{title}{{Radiative forcing of the stratosphere of Jupiter,
  Part I: Atmospheric cooling rates from Voyager to Cassini}}.
\newblock \bibinfo{journal}{Plan. \& Space Sci.} \bibinfo{volume}{88},
  \bibinfo{pages}{3--25}.
\newblock \DOIprefix\doi{10.1016/j.pss.2013.07.005}.
\bibitem[{{Zhang} et~al.(2013b){Zhang}, {West}, {Banfield} and
  {Yung}}]{13zhang_aer}
\bibinfo{author}{{Zhang}, X.}, \bibinfo{author}{{West}, R.A.},
  \bibinfo{author}{{Banfield}, D.}, \bibinfo{author}{{Yung}, Y.L.},
  \bibinfo{year}{2013}b.
\newblock \bibinfo{title}{{Stratospheric aerosols on Jupiter from Cassini
  observations}}.
\newblock \bibinfo{journal}{Icarus} \bibinfo{volume}{226},
  \bibinfo{pages}{159--171}.
\newblock \DOIprefix\doi{10.1016/j.icarus.2013.05.020}.
\bibitem[{{Zhang} et~al.(2015){Zhang}, {West}, {Irwin}, {Nixon} and
  {Yung}}]{15zhang}
\bibinfo{author}{{Zhang}, X.}, \bibinfo{author}{{West}, R.A.},
  \bibinfo{author}{{Irwin}, P.G.J.}, \bibinfo{author}{{Nixon}, C.A.},
  \bibinfo{author}{{Yung}, Y.L.}, \bibinfo{year}{2015}.
\newblock \bibinfo{title}{{Aerosol influence on energy balance of the middle
  atmosphere of Jupiter}}.
\newblock \bibinfo{journal}{Nature Communications} \bibinfo{volume}{6},
  \bibinfo{pages}{10231}.
\newblock \DOIprefix\doi{10.1038/ncomms10231}.

\end{thebibliography}







\end{document}